\RequirePackage{fix-cm}
\documentclass[smallextended]{svjour3} 
\smartqed 
\usepackage{graphicx}
\usepackage{booktabs}
\usepackage[table]{xcolor}
\usepackage{tabularray}
\UseTblrLibrary{booktabs}
\UseTblrLibrary{varwidth} 
\usepackage{tablefootnote} 
\usepackage{colortbl}
\definecolor{myyellow1}{RGB}{255,255,230}
\definecolor{myyellow2}{RGB}{255,253,210}
\definecolor{myyellow3}{RGB}{255,250,190}
\definecolor{myyellow4}{RGB}{255,245,160}
\definecolor{myyellow5}{RGB}{255,240,130}
\definecolor{myyellow6}{RGB}{255,235,100}
\definecolor{myyellow7}{RGB}{255,225,70}
\definecolor{myyellow8}{RGB}{255,210,40}
\definecolor{myyellow9}{RGB}{255,190,20}
\definecolor{myyellow10}{RGB}{240,170,0}

\definecolor{myblue1}{RGB}{220,235,255}
\definecolor{myblue2}{RGB}{200,225,255}
\definecolor{myblue3}{RGB}{180,210,255}
\definecolor{myblue4}{RGB}{160,190,255}
\definecolor{myblue5}{RGB}{140,170,255}
\definecolor{myblue6}{RGB}{120,150,240}
\definecolor{myblue7}{RGB}{100,130,220}
\definecolor{myblue8}{RGB}{80,110,200}
\definecolor{myblue9}{RGB}{60,90,180}
\definecolor{myblue10}{RGB}{40,70,160}

\definecolor{lfd-yellow}{RGB}{240,170,0}
\definecolor{lfd-blue}{RGB}{31,120,180}
\definecolor{lfd-green}{RGB}{0,147,113}
\definecolor{lfd-lilac}{RGB}{190,174,212}
\definecolor{lfd-gray}{RGB}{153,153,153}

\newcommand{\colorcell}[2]{%
  \ifdim #1 pt < -0.9pt
    \cellcolor{myyellow10}{\ensuremath{\color{black}#2}}%
  \else\ifdim #1 pt < -0.7pt
    \cellcolor{myyellow9}{\ensuremath{\color{black}#2}}%
  \else\ifdim #1 pt < -0.5pt
    \cellcolor{myyellow8}{\ensuremath{\color{black}#2}}%
  \else\ifdim #1 pt < -0.3pt
    \cellcolor{myyellow7}{\ensuremath{\color{black}#2}}%
  \else\ifdim #1 pt < -0.1pt
    \cellcolor{myyellow6}{\ensuremath{\color{black}#2}}%
  \else\ifdim #1 pt < 0pt
    \cellcolor{myyellow3}{\ensuremath{\color{black}#2}}%
  \else\ifdim #1 pt = 0pt
    \cellcolor{myyellow1}{\ensuremath{\color{black}#2}}%
  \else\ifdim #1 pt < 0.1pt
    \cellcolor{myblue1}{\ensuremath{\color{black}#2}}%
  \else\ifdim #1 pt < 0.3pt
    \cellcolor{myblue2}{\ensuremath{\color{black}#2}}%
  \else\ifdim #1 pt < 0.4pt
    \cellcolor{myblue3}{\ensuremath{\color{black}#2}}%
  \else\ifdim #1 pt < 0.5pt
    \cellcolor{myblue4}{\ensuremath{\color{white}#2}}%
  \else\ifdim #1 pt < 0.6pt
    \cellcolor{myblue5}{\ensuremath{\color{white}#2}}%
  \else\ifdim #1 pt < 0.7pt
    \cellcolor{myblue6}{\ensuremath{\color{white}#2}}%
  \else\ifdim #1 pt < 0.8pt
    \cellcolor{myblue7}{\ensuremath{\color{white}#2}}%
  \else\ifdim #1 pt < 0.9pt
    \cellcolor{myblue8}{\ensuremath{\color{white}#2}}%
  \else\ifdim #1 pt < 1pt
    \cellcolor{myblue9}{\ensuremath{\color{white}#2}}%
  \else\ifdim #1 pt < 1.1pt
    \cellcolor{myblue10}{\ensuremath{\color{white}#2}}%
  \else
    \cellcolor{myblue10}{\ensuremath{\color{white}#2}}%
  \fi\fi\fi\fi\fi\fi\fi\fi\fi\fi\fi\fi\fi\fi\fi\fi\fi
}
\usepackage{multirow}
\usepackage{hyperref}
\usepackage{amssymb}
\usepackage{mathtools}
\usepackage{natbib}
\usepackage{subcaption}
\usepackage{caption}
\usepackage{xspace}
\usepackage[most]{tcolorbox}
\tcbset{
  colback=gray!15,  
  colframe=gray!15,
  boxrule=0pt,
  arc=0pt,
  left=6pt,
  right=6pt,
  top=6pt,
  bottom=6pt,
  enhanced
}

\newtcolorbox{answerbox}[1][]{}

\journalname{}

\begin{document}

\title{Oops!\dots I did it again}

\subtitle{Analysing and Handling Conclusion (In-)Stability in Socio-Technical Software Engineering}


\author{Nicole Hoess \and
        Carlos Paradis \and 
        Rick Kazman \and
        Wolfgang Mauerer
}


\institute{Nicole Hoess \at
              Technical University of Applied Sciences Regensburg \\
              Tel.: +49-941-943-1333 \\
              \email{nicole.hoess@othr.de}
           \and
           Carlos Paradis \at
              Independent researcher\\
              \email{cvas@acm.org}
           \and
           Rick Kazman \at
              University of Hawaii at Mānoa \\
              \email{kazman@hawaii.edu}
           \and
           Wolfgang Mauerer \at
              Technical University of Applied Sciences Regensburg \\
              Siemens AG, Foundational Technology\\
              Tel.: +49-941-943-9702 \\
              \email{wolfgang.mauerer@othr.de}   
}

\date{Revised preprint.}

\newcommand{\etal}{\emph{et al.}\xspace}

\maketitle

\begin{abstract} 
\mbox{}\par
\paragraph{Context:} Mining software repositories is a popular means to gain insights into a software project's evolution, monitor project health, support decisions and derive best practices. Tools supporting the mining process are commonly applied by researchers and practitioners, but their limitations and agreement are often not well understood.
\paragraph{Objective:} This study investigates some threats to validity in complex tool pipelines for evolutionary socio-technical software analyses. We evaluate the tools' agreement in terms of data, study outcomes and conclusions for the same research questions to derive actionable advice for researchers and practitioners.
\paragraph{Method:} We conduct a lightweight literature review to select \emph{three} studies on collaboration and coordination, software maintenance and software quality from high-ranked venues, which we formally replicate with \emph{four} independent, systematically selected mining tools to quantitatively and qualitatively compare the extracted data, analysis results and conclusions.
\paragraph{Results:} We find and summarise numerous technical details in tool design and implementation which accumulate along the complex mining pipelines and can cause substantial differences in the extracted baseline data, its derivatives, subsequent results of statistical analyses and, under specific circumstances, conclusions.
\paragraph{Conclusions:} Users should evaluate mining tools and their limitations carefully to scope the validity of their conclusions, for instance with the checklist we provide. Researchers and tool authors can further reduce uncertainty through reproduction packages and comparative studies in other MSR fields following our methodology.

\mbox{}\par
\keywords{Mining software repositories \and Software analysis \and Tools 
\and Replication \and Validity \and Conclusion stability}
\end{abstract}

\section{Introduction}
\label{intro}

Software repositories, managed by version-control systems (VCS) such as Git, record a project's entire development history, making them a valuable data source for empirical studies. Mining software repositories (MSR) has therefore grown into a relevant field in software engineering. While earlier studies focused on extracting and representing information from repositories, for instance to analyse software quality~\citep{sliwerski_when_2005}, activity and coordination patterns~\citep{dinh-trong_freebsd_2005}, studies soon grew more advanced, constructing developer networks~\citep{bird_mining_2006} and modeling statistical relationships~\citep{fernandez-ramil_what_2009}. Today, studies leverage advanced machine learning to support developers in specific tasks, such as recommending libraries~\citep{he_multi-metric_2021}. As research questions grow more complex, so do the methods needed to answer them.

Tools automating data extraction and processing are therefore essential for comprehensive analyses at scale. However, implementing automated pipelines requires many design and implementation decisions to be made. For instance, there are numerous degrees of freedom in choosing data sources, APIs, algorithms, data schemes, filters and metrics~\citep{paradis_building_2022}. During implementation, researchers and practitioners must carefully evaluate for which decisions they provide configuration options and for which ones they implement defaults. Given the many seemingly minor technical details, lots of these decisions are made implicitly based on assumptions of the tool author.

Nowadays, a plethora of tools automates different steps of the analysis pipeline. For instance, tools such as \textsc{Codeface}~\citep{joblin_developer_2015}, \textsc{git2net}~\citep{gote_git2net_2019}, \textsc{GrimoireLab}~\citep{duenas_grimoirelab_2021}, \textsc{Kaiaulu}~\citep{paradis_building_2022} and \textsc{SmartSHARK}~\citep{trautsch_adressing_2016} extract data from VCS, issue trackers, mailing lists, chats, Q\&A platforms and map, unify and visualise data to incorporate even more useful information into analyses and models. Automated retrieval and processing of such large amounts of diverse data is obviously a complex problem, where validation of correctness is sometimes not feasible. For instance, we \emph{cannot} simply determine if all timestamps captured in the Git log are correct, or if developers using different aliases in Git and chats over years really belong to the same person. Due to this limitation, data extracted by tools is often trusted without further verification, giving the misleading impression of a \emph{ground truth} for subsequent analyses.

Although several threats in mining pipelines have been studied in previous research~\citep{bird_promises_2009, kalliamvakou_promises_2014} and methods were proposed to overcome them~\citep{saarimaki_towards_2022}, there may be more uncertainties introduced by the implicit assumptions inherently made by tool developers. Ideally, this would not pose a big problem, as robust research processes, results and conclusions should not depend on the specific implementation of an extraction and analysis process. Furthermore, tools with a defined set of extraction and analytical capabilities should ideally allow for substitution by another tool, and thereby overcome problems such as unavailable artefacts, scripts and dependencies, which researchers often face during reproduction and replication~\citep{hermann_community_2020, gonzalez-barahona_revisiting_2023, abualhaija_replication_2024}.

In previous work, we found that even for two very similar tools following the same data extraction and analysis process, small technical differences accumulate along the pipeline and in sum can have a substantial impact on the resulting data~\citep{hoess_does_2025}. However, we did not evaluate the actual \emph{impact} of the entirety of potential and observed discrepancies along a pipeline on the high-level \emph{conclusions} of studies. Although very few other studies address the stability of analysis results through replication~\citep{eng_replicating_2022}, an extensive evaluation of the impact of tool choice on conclusion stability is still missing. If equally valid fine-grained implementation details and assumptions can influence the overarching statement made by a study, this would raise serious concerns about the generalisability of many impactful findings within socio-technical software engineering.

To better understand the relevance and consequences of threats in software repository mining, we build on our technical comparison and adjustment of the two mining tools \textsc{Codeface} and \textsc{Kaiaulu} and investigate the impact of the observed discrepancies on empirical study conclusions from a higher level of abstraction by replicating three influential studies from the last decade with four different tools. 

We deliberately scope our investigation to socio-technical studies that analyse developer collaboration, coordination, productivity, and software quality primarily from VCS data, rather than from other artefacts such as issues or pull requests. This focus lets us compare tools in depth on a coherent class of studies. Correspondingly, our findings apply to this socio-technical, VCS-centred setting and should not be generalised beyond it.

The results of this study are two-fold: (1) The high-level \emph{conclusions} of two of three studies are stable across four independent mining tools. This is positive for the community, as it demonstrates that conclusions \emph{can} be robust against tool variations if studies make very broad statements. (2) Despite this, the underlying \emph{results} are \emph{not stable} throughout all replications: In several cases, metric values differ substantially. In one study, these differences are large enough to change the overall
\emph{conclusion} of the replication, indicating that tools are \emph{not} readily interchangeable. As uncertainty accumulates across each extraction, processing, and analysis step, even minor technical inconsistencies can substantially influence the resulting conclusions. Since such inconsistencies are easy to overlook and hard to detect, we provide an actionable checklist that helps tool authors and users anticipate and address them based on our lessons learned.\\

In summary, we make the following contributions:
\begin{itemize}
    \item We perform a literature review of seven high-ranked venues to identify tools and popular topics frequently addressed by software repository mining to evaluate the potential fields affected by discrepancies.
    \item We compare the analysis pipelines and results of four similar mining tools and summarise the most relevant causes of discrepancies in the baseline data.
    \item We evaluate the impact of such discrepancies on the stability of the actual study \emph{conclusions} by replicating the central analyses of three studies from three different fields with data extracted by the four tools, simulating a tool switch. We assess \emph{result} and \emph{conclusion} stability both qualitatively and quantitatively.
    \item We summarise lessons learned and recommendations regarding the most important tool threats to which researchers and practitioners must pay particular attention, including an actionable checklist guiding them through future work.
    \item We provide a replication package containing all data, code and supplementary artefacts. As we cannot cover all areas of software repository mining within the scope of this study, other researchers can use our methods and the annotated literature material for similar, future investigations.
\end{itemize}

The study is an extension of our previous work~\citep{hoess_does_2025}. The major extensions in this paper are as follows:
\begin{itemize}
    \item We include two additional mining tools -- git2net and GrimoireLab -- in our baseline data comparison.
    \item We perform the literature review as described above for a more structured evaluation of the \emph{impact} of the threats identified in our previous work.
    \item As a major extension of the baseline data comparison in our previous work, we now formally replicate \emph{three} carefully chosen studies heavily relying on data extracted by mining tools to investigate conclusion stability across tools. Using four independent mining tools, this results in twelve new replications.
    \item We extend our list of lessons learned and recommendations by novel aspects found in the replications and compile an extensive checklist table as actionable advice.
\end{itemize}

The rest of this work is structured as follows: Section \ref{related-work} provides an overview of related studies exploring the validity of repository mining techniques, reproducibility and replicability. Section \ref{methodology} describes our study design and research questions, starting with a literature review, the setup for the baseline data comparison and our replication procedure. Section \ref{results} presents the results of the literature review, the baseline data comparison and each replication study to answer our research questions. In section \ref{discussion}, we discuss the implications of our findings on research and practice. We reflect on these findings to provide lessons learned and a checklist for future studies, before we conclude the paper in section \ref{conclusion}. Section \ref{reproduction-package} refers to data availability and our reproduction package.

\section{Related Work}
\label{related-work}

Threats to validity in software repository mining have been identified and explored in numerous previous studies. Related studies primarily focused on the validity of specific analysis steps in the mining pipeline and compared competing approaches of their implementation. Other threats have been found in replication studies.

\paragraph{Validity and threats} The most influential work on pitfalls in software repository mining has been conducted by Bird~\etal~\citeyearpar{bird_promises_2009}, whose study on the promises and perils of mining git addresses possible traceability issues. Its findings were later extended by a study evaluating threats caused by peculiar properties of GitHub~\citep{kalliamvakou_promises_2014}. 
Further studies explore threats and their overcoming when mining unstructured data~\citep{bavota_mining_2016} or working with time series data~\citep{zheng_method_2015, moonen_what_2018, flint_pitfalls_2022, saarimaki_towards_2022}. In addition, the impact of untracked entity changes was studied in the context of refactoring~\citep{hora_assessing_2018}.

Particularly for socio-technical aspects, Zhu~\etal~\citeyearpar{zhu_empirical_2019} study the use of multiple developer identities in open-source projects, emphasising the relevance of identity merging when calculating metrics.
Nia~\etal~\citeyearpar{nia_validity_2010} explore the impact of pitfalls in e-mail network construction, resulting in missing edges or edges out of temporal order. Their results show that metrics such as node centrality are stable despite the changes in topology. Other studies explore developer perception of collaboration and team structure expressed by such networks~\citep{meneely_socio-technical_2011, joblin_developer_2015}.

Further studies investigate threats in data labeling~\citep{tantithamthavorn_impact_2015, herbold_fine-grained_2022, song_validity_2023, guo_estimating_2024}, often specific to defect prediction~\citep{chowdhury_method-level_2024}. 
For empirical studies, Härtel~\etal~\citeyearpar{hartel_operationalizing_2022, hartel_operationalizing_2023} propose simulation-based testing to evaluate the validity of statistical methods. Tu~\etal~\citeyearpar{tu_be_2018} investigate data leakage in predictive models due to time-related problems in issue tracking data and, similar to our study, show the impact of such threats by reproducing and analysing three prior studies. Other studies explore popularity bias in recommender systems~\citep{nguyen_dealing_2023}
and methods to detect and handle endogeneity in statistical methods~\citep{graf-vlachy_cleaning_2024}.
While previous studies of validity and threats focused on methods for specific analysis steps in isolation, our study focuses on their interaction and propagation along the entire tool pipeline.

\paragraph{Method and tool comparisons} Several studies investigate the interchangeability of tools in other fields of empirical software engineering, such as software composition analysis~\citep{zhao_software_2023} and architecture recovery~\citep{schneider_comparison_2025}. Lefever~\etal~\citeyearpar{lefever_lack_2021} compare the results obtained by commercial and open-source tools for technical debt detection, finding that they disagree even for very common, basic measures such as lines of code (LOC). To the best of our knowledge, no similar studies exist for tools extracting evolutionary aspects such as time series from the version-control history. 

However, some studies highlighted differentiating aspects of specific steps in mining pipelines. For instance, studies propose and compare heuristics for developer identity matching~\citep{goeminne_comparison_2013, amreen_alfaa_2020}. 
Bertoncello~\etal~\citeyearpar{bertoncello_pull_2020} 
find that pull-requests are more accurate for measuring contributions than commits to distinguish core and casual developers. 
Joblin~\etal~\citeyearpar{joblin_developer_2015, joblin_classifying_2017} investigate differences in collaboration network construction methods, community detection and core developer classification metrics and evaluate correspondence with the real perception of developers.
Several authors study the agreement of developer networks~\citep{tymchuk_collaboration_2014, panichella_how_2014} and communities~\citep{bock_measuring_2021} constructed from data extracted from different communication and collaboration sources. Tymchuk~\etal~\citeyearpar{tymchuk_collaboration_2014} find that the combination of these channels is essential to obtain a comprehensive view of a project.

For studies using machine learning subsequent to mining, comparisons across different models and datasets are more common. For instance, Mahadi~\etal~\citeyearpar{mahadi_conclusion_2021} evaluate stability of conclusions from classifiers predicting whether discussions are design-related on different datasets, finding that conclusion stability across domains and datasets is poor and degrees of freedom in the analysis pipeline are high. This problem is likely to also apply to the data extraction itself. Therefore, our work contributes an evaluation of conclusion stability of different tools for this purpose.

\paragraph{Replication studies}

Studies missing technical details as described above can impair both, the possibility of \emph{reproduction}, referring to the repetition of an experiment by using the exact same setup, and \emph{replication}, referring to the independent repetition of an experiment using a different setup~\citep{mauerer_beyond_2022, gonzalez-barahona_revisiting_2023}, possibly by the same team (\emph{internal} replication) or a different team (\emph{external} replication)~\citep{shepperd_role_2018} with operational or conceptual changes~\citep{gomez_understanding_2014}. As few replication studies exist in the field of software repository mining, the consequences of deviations in mining setups have not yet been fully evaluated.

The limited availability of replication studies in empirical software engineering is discussed as a threat since decades~\citep{basili_building_1999,robles_replicating_2010}. In general, advances in reproducibility engineering contributed notably to the improvement of study reproducibility in the last years~\citep{gonzalez-barahona_revisiting_2023}. For instance, González-Barahone~\etal\citeyearpar{gonzalez-barahona_reproducibility_2012} propose a general process with related elements suitable for most mining studies to improve reproducibility. The elements include data sources, their transformations, any extraction, processing and analysing methodology and study results. Mauerer~\etal~\citeyearpar{mauerer_beyond_2022} present best practices for reproduction packages built as self-contained docker images. \emph{Self-contained} means the image contains all artefacts, such as data, code and scripts, as well as the execution environment with all dependencies. A single dispatcher, such as a Makefile, then allows others to execute the entire pipeline from data processing to the final paper. Other approaches propose dedicated platforms designed for reproduction purposes~\citep{ghezzi_replicating_2013, trautsch_addressing_2018}. Despite these advances, \emph{replications} of mining studies yet remain challenging, as artefacts and implementations are often not published at all, only in parts, or in form of a diverse~\citep{trautsch_addressing_2018, liang_can_2024}, possibly unusable~\citep{gonzalez-barahona_revisiting_2023} or outdated, set of scripts, libraries and tools. In deep learning, complex training and testing pipelines add to the difficulty of replication~\citep{liu_reproducibility_2021, abualhaija_replication_2024}.

Besides evaluating threats, replication studies offer the potential to increase confidence in findings of previous studies~\citep{shull_role_2008}, extend~\citep{dinh-trong_freebsd_2005}, complement ~\citep{gonzalez-barahona_reproducibility_2012} or refine~\citep{bock_synchronous_2021} their results and increase the impact of the field~\citep{liang_can_2024}. 
While replications are more common in areas such as defect detection~\citep{mahmood_reproducibility_2018, di_penta_relationship_2020, niu_ablots_2023}, their availability is still limited in the context of developer networks~\citep{herbold_systematic_2021}. 

As part of a mining hackathon at the Mining Software Repositories (MSR) conference, a study from Eng and Sahar~\citeyearpar{eng_replicating_2022} explored the replacement of a traditional data processing pipeline for chat message investigation by the mining tool \textsc{GrimoireLab}. By replicating a prior study, the authors evaluate technical aspects 
such as speed and data consistency, 
and, amongst others, find that the change in pipeline led to different, but more precise results. Exploring the impact of a tool \emph{adoption}, this study is the most similar to ours focusing on tool \emph{interchangeability}.

\section{Methodology}
\label{methodology}

The aim of this study is to understand the \emph{impact} of threats to validity due to technical details such as design decisions and limitations in mining tool pipelines, as motivated by Hoess~\etal~\citeyearpar{hoess_does_2025}, on empirically derived results and conclusions. For this purpose, we quantitatively evaluate the stability of outcomes of four tools with the similar purpose of analysing software evolution in terms of contributor activity, collaboration and artefact changes over time at three levels: baseline and derived data, subsequently calculated metrics and statistical analysis results and, finally, central implications.

\subsection{Overall study design and research questions}
\label{rqs}
The overarching question \emph{How does the technical implementation of tools affect the validity and stability of outcomes in empirical studies?} cannot be answered in general terms for all studies and domains in empirical software engineering. However, both positive and negative examples for data, results and conclusion stability can be important indicators of the validity and generalisability of methods and results in our field and give information on possible areas for improvement. Therefore, to evaluate the agreement of tools in data, empirical results and conclusions, the study conducts three independent replication studies, following the three-step approach depicted in Figure~\ref{fig:method-overall} and guided by the following research questions:

\begin{figure}[htbp]
  \centering
  \includegraphics[width=\linewidth]{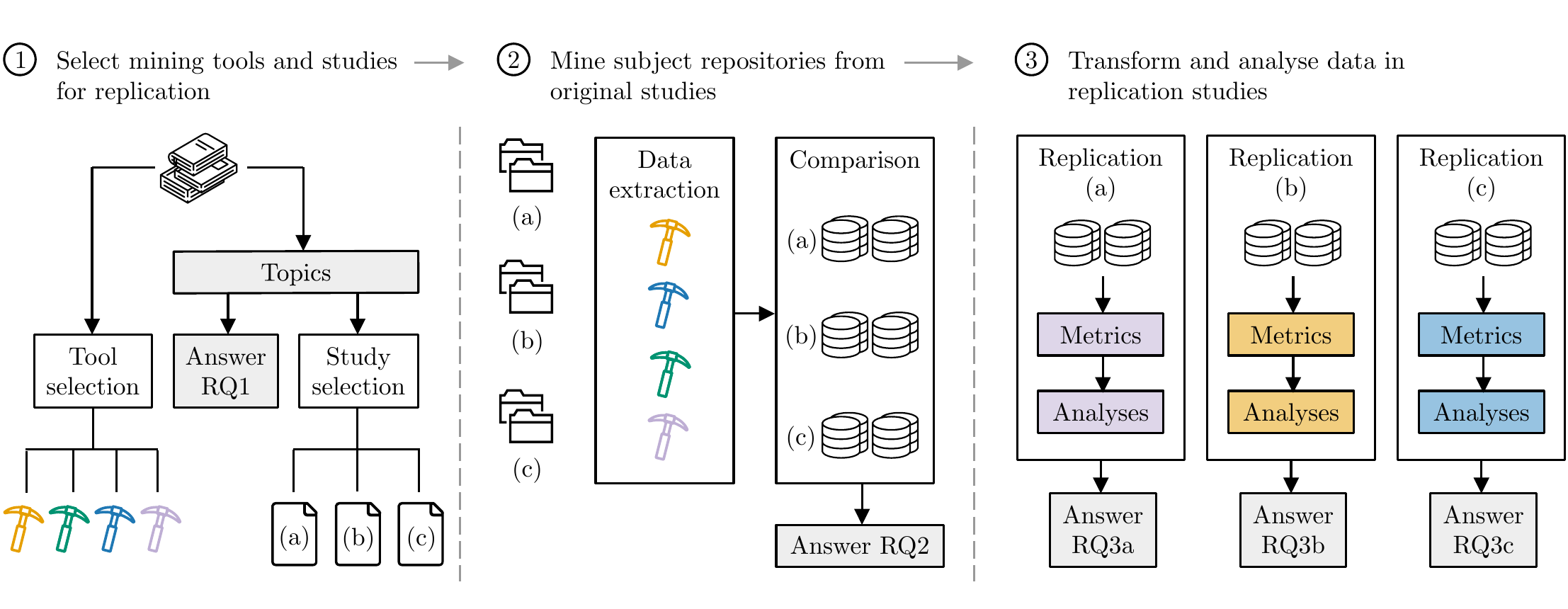}
  \caption{Overall study design.}
  \label{fig:method-overall}
\end{figure}

\paragraph{\textbf{RQ1:} Which topics in empirical software engineering are typically driven by mining software evolution from repositories and could be affected by threats due to differences in tooling?}  We address this question by a lightweight literature review to better understand the relevance and consequences of previous work studying validity in mining software repositories. Knowing the most relevant and potentially vulnerable topics allows for an informed choice of representative tools and original studies for replication.

\paragraph{\textbf{RQ2:} To what extent can we observe discrepancies in the
data obtained from independent software repository mining tools?} 
Each tool makes assumptions on how to extract and process data, which can be
influenced by configuration parameters. Implicit assumptions
and implementation decisions often remain hidden from tool
users. Building on our previous study,
we are interested in evaluating the existence of such
uncertainties for additional mining tools.
In this study, this step illustrates the underlying issue and outlines which tool-specific preparations are required 
prior to conducting the replications as part of RQ3. For instance,
data extracted by each tool follows a different schema, requiring
tailored data processing.
Knowing the technical
reasons for discrepancies in the extracted data also helps to locate where possible 
deviations in subsequent analysis results and conclusions of the replications
originate. This allows for later summarising relevant technical factors of the 
analysis pipeline that researchers and practitioners should implement and
examine particularly carefully to reduce uncertainties.

\paragraph{\textbf{RQ3:} To what extent are results and conclusions derived in empirical software engineering research influenced by the observed discrepancies in the mining tools and their outputs?} This question concerns the actual replication of the empirical studies selected as part of RQ1. After the tool-specific data processing, we apply the same algorithms of the respective studies to all tool datasets to ensure replication conformance. We quantify differences in metrics relevant to each study's outcome and evaluate whether the interpretation and overall conclusion from these results are stable across tools. We define a sub-research question for each individual study to emphasise their relevance as case studies rather than a means to derive a universal answer to this question. 
Together with the technical aspects from RQ2, the gained insights can help others to more accurately assess the scope of validity and generalisability of methods and results.

\subsection{Lightweight literature review and study selection}\label{sec:study-selection} 

Mining software repositories is a broad field, making it impossible to compare all sub areas and tools. We must therefore make restrictions in both studies and tools, but at the same time select enough of both to cover a sufficiently broad range of topics. 
To identify suitable replication studies and tools for comparison as part of RQ1, we take inspiration from the guidelines on systematic literature reviews and mapping studies proposed by  Kitchenham~\etal~\citeyearpar{kitchenham_procedures_2004, kitchenham_systematic_2013}, Petersen~\etal~\citeyearpar{petersen_guidelines_2015} and 
the lightweight literature review conducted by Berger~\etal~\citeyearpar{berger_state_2020}. We collect primary studies from the proceedings and articles of seven high-ranked conferences and journals, and investigate which research topics are frequently driven by mining tools.

In the first phase, we compose a dataset of candidate studies published between 2015 and 2024 in one of these venues using the query (``empirical study'' OR ``data analysis'' OR ``evolution'') AND (``software repositories'' OR ``version control system''). The keywords were selected in line with the previous study to favour studies that primarily deal with the long-term and comprehensive analysis of data from software repositories. 

To identify domains of mining software repositories in which tools play a major role, we read each study's abstract and research questions in the second phase. If required, we read the entire methodology and results sections. We then assign up to three categories to each study, following a three-level taxonomy of primary, secondary and tertiary research fields, where each tertiary field is the most fine-grained subfield within its secondary field, and each secondary field a subfield of its primary field. The research fields are iteratively grouped and updated in the course of the literature review.

In the third phase, we quantitatively evaluate the popularity of each primary, secondary and tertiary field by the number of annotations assigned to it across all studies. As each study receives up to three category annotations following the taxonomy, a single study can contribute to several fields.

In the fourth and final phase, we choose three diverse studies from the most relevant primary fields for replication. We target three as the minimum number required to identify an initial pattern, as a single replication study cannot show whether its result is typical or anomalous, and two studies can contrast without showing whether differences are systematic. Since each study is replicated across all tools, every additional study multiplies the replication effort and reduces the depth we can achieve in each comparison. By drawing the three studies from distinct yet highly prevalent fields, we ensure the selection is both diverse and representative.

As study pipelines vary and may involve steps such as manual processing, we expect that not all studies are replicable with the tools under study. Therefore, we define the following selection criteria: First, tool, script and data availability are necessary for replication, as the absence of these assets substantially increases the risk of accidentally introducing new threats due to incorrect re-implementation. Next, we consider studies particularly suitable to assess the impact of mining threats if they derive conclusions on best practices in software engineering based on statistical relationships in tool data, as these studies inherently present metrics suited to statistical comparison. Conversely, we exclude studies with a high proportion of qualitative manual and subjective data analyses. Finally, we focus on studies that can be reasonably replicated with VCS data only, as support for mining other sources such as mailing lists and issue trackers is often less common and less robust in tools, for instance due to external API changes. Filtering all of these methodological criteria automatically is not reliably feasible, which is why we chose the manual literature review and selection process.

\subsection{Tool selection}
\label{sec:tool-selection}

Relevant studies and suitable tools for replication are closely linked: Studies require specific tool capabilities and tools are suitable only for a subset of studies. To solve this causality dilemma, we take our previous study as a reference, which analysed the technical causes of discrepancies in the data obtained by the mining tools \textsc{Codeface} and \textsc{Kaiaulu}, and follow the tool selection criteria and process a typical user would take when searching for a new or replacement tool. 
Consequently, we search for tools providing the same or a subset of these tools' data extraction and processing capabilities alongside our literature review. These capabilities include 
(1) the extraction of commits from version-control systems over the entire project history, allowing for evolutionary analyses,
(2) the extraction of finer-grained code entities, such as files, functions and interfaces, and 
(3) the possibility to study socio-technical aspects by the construction of developer networks. 
As a further selection criterion (4), we consider the tools' active maintenance, as our prior study showed that extensive discussions with tool authors may be required to clarify technical details. We select every identified tool satisfying all four criteria.

As we demonstrated in our previous study for the tool \textsc{Kaiaulu}, tools can rely on very specific methods that may not be available in others without substantial extensions and adjustments. While not adjusting other tools may itself introduce a threat, we deliberately \emph{preserve} exactly these uncertainties in this study to evaluate their impact on conclusions without introducing additional bias.

\subsection{Baseline data comparison}\label{sec:baseline-comparison}

In our previous study, we conducted an extensive comparison of output data from the tools \textsc{Codeface} and \textsc{Kaiaulu}. In this study, we verify that the identified discrepancies generalise to all tools under study, and give an overview of uncertainties and underlying causes potentially influencing subsequent study results and conclusions. However, to focus on the novel aspects of result and conclusion stability in this study, we only briefly illustrate similarities and differences in the baseline data to answer RQ2. In particular, we consider the in-depth baseline comparison part of our prior work, eliminating the need for a full comparison across all projects in this study. Instead, a representative sample of projects suffices to show that discrepancies are also present in the additional tools. We therefore draw a subset of projects analysed in the replications, so that RQ2 and RQ3 concern the same subjects and configurations. The subset spans diverse application domains, programming languages, project ages and team sizes, allowing us to evaluate whether discrepancies persist across project contexts.

After data extraction, we calculate a set of fundamental measures which are often the basis for more complex metrics. These measures include the number of commits, files, code entities and developers. We compare the metrics quantitatively and visualise the results as time series.

\subsection{Replication methodology and stability assessment}\label{sec:replication-stability}

To measure the impact of discrepancies in tools and their extracted data on empirical results and conclusions, we replicate the \emph{central analyses} of the three representative studies identified from the literature review. 
We trade off depth and breadth by considering the relevance of individual analyses for the central conclusion of each study and the effort for replication. 
Although we aim to minimise operational and conceptual changes to the original studies, some can involve high computational effort or anonymisation, which requires us to limit the replication scope.

In the first step, we evaluate reproducibility of the original study with the supplementary scripts provided. If the study is reproducible, we apply the exact same code to all tool datasets, while fixing any random seeds to improve comparability. In case the study is not fully reproducible, we reconstruct the missing part with an implementation that follows the description in the original paper as closely as possible. Crucially, any such reconstruction is applied identically to all tool datasets and therefore remains constant across the comparison. Although this may cause the replication to deviate from the original study and its results, it does not affect the cross-tool stability comparison we are interested in, as a component applied identically to all tools cannot be the source of any difference between them. We summarise the known technical differences relevant to each study in Appendix~\ref{sec:appendix}, and restate those most relevant to interpreting a specific result directly in the corresponding part of Section~\ref{results}.

In the second step, we configure the replication tools as closely as possible to the documented decisions of the original study to extract the baseline data. 

In the third step, we build a single, shared analysis pipeline from the original study's code. We deviate as little as possible from the original to apply the exact same analysis steps to every tool's dataset. Keeping the common pipeline ensures that the comparison of tools reflects tool-induced variability and excludes variability from the analysis itself. 

In the fourth and most challenging step, we interpret results. Because reproducibility depends on many interacting elements~\citep{gonzalez-barahona_reproducibility_2012}, deciding whether a study is \emph{reproducible} with the same setup is not trivial. Assessing \emph{replicability} with a different setup, such as a different mining tool, is even more challenging. As the~\citet{national_academies_of_sciences_engineering_and_medicine_reproducibility_2019} report summarises, there is no single, universal approach to determine whether results are \emph{sufficiently similar}. However, giving an appropriate answer for RQ2 and RQ3 requires us to adopt some principled approach.

We formalise \emph{result stability}, referring to analysis outputs, and \emph{conclusion stability}, referring to the original study's high-level verdict, mathematically to guide the reader through the process of comparing and interpreting quantitative and complementary visual results~\citep{anscombe_graphs_1973}. Following~\citet{bangash_time-based_2020}, we define study-specific hypotheses and apply thresholds that reflect the original authors' inferential logic as closely as possible. Additionally, we apply confidence intervals and related statistical means tailored to the specific analysis steps and data types of each replication study. 

We emphasise that this formalisation only illustrates our interpretative logic. The hypotheses and thresholds structure our reasoning and the confidence intervals characterise sampling variability, but we draw no claims of statistical significance or generalisations beyond our replication setting.

\section{Results}
\label{results}

In the following, we describe the main results from the study phases shown in Figure~\ref{fig:method-overall}: the literature review, the baseline data comparison and, finally, the three replication studies exploring the actual impact of differences between tools on empirical study results and conclusions. 
For better understandability, we limit metrics, tables and visualisations to the most meaningful examples. The full data are available on our supplementary website\footnotemark[16]. 

\footnotetext[16]{Supplementary website: \href{https://lfd.github.io/emse2025.github.io/}{https://lfd.github.io/emse2025.github.io/}}

\subsection{Lightweight literature review} \label{sec:literature-results} 

The literature review shows that many of the central issues in software engineering are driven by mining software repositories with appropriate tools. From the venues in Table~\ref{tab:venues}, our keyword search resulted in 841 studies to annotate. Figure~\ref{fig:treemap} illustrates the popularity of each primary, secondary and tertiary field according to these annotations. The corresponding datasets are available in our supplementary material. Out of all considered studies, 148 (\(18\%\)) introduce new tools, which emphasises their essential role in the field. 

\begin{table*}[htb]
   \caption{Investigated venues and studies matching our keyword search in the literature review. \label{tab:venues}}
   \begin{tblr}{colspec={Xcr},
                row{odd}={bg=gray!15},  
                row{1}={bg=white,fg=black}}
        \toprule
         Conference & Acronym & Studies \\ \midrule
         Int.\ Conf.\ on Software Engineering & ICSE & 31 \\
         Mining Software Repositories & MSR & 229 \\
         Int.\ Conf.\ on the Foundations of Software Engineering & FSE & 108 \\
         Int.\ Conf.\ on Software Analysis, Evolution and Reengineering & SANER & 95 \\
         IEEE Transactions on Software Engineering & TSE & 41 \\
         ACM Transactions on Software Engineering and Methodology & TOSEM & 73 \\
         Empirical Software Engineering & EMSE & 264 \\ 
         \midrule
         Total & & 841 \\
         \bottomrule
    \end{tblr}
\end{table*}

To select studies for replication, we focus on the most active primary research areas, which include software maintenance, software quality, MSR techniques, development support and automation and collaboration and coordination in descending order. MSR techniques mainly comprises studies proposing tools for specific purposes, such as software package analysis in containers~\citep{Zerouali_conpan_2019}, the creation of datasets~\citep{montgomery_alternative_2022}, or the analysis of threats to validity~\citep{zhu_empirical_2019}. As these studies do not derive conclusions or best practices, we exclude them from our candidates. Similarly, development support and automation targets the implementation of novel, often machine-learning-based algorithms to help developers in their daily work, for instance by generating code~\citep{ciniselli_empirical_2021} and documentation ~\citep{zhang_automatic_2024, gao_evaluating_2023} or recommending libraries~\citep{he_multi-metric_2021}, which requires different capabilities than provided by the tools under study. Therefore, we focus on the remaining most popular topics when selecting the studies for replication:

\begin{figure*}[htbp]
    \includegraphics[width=\textwidth]{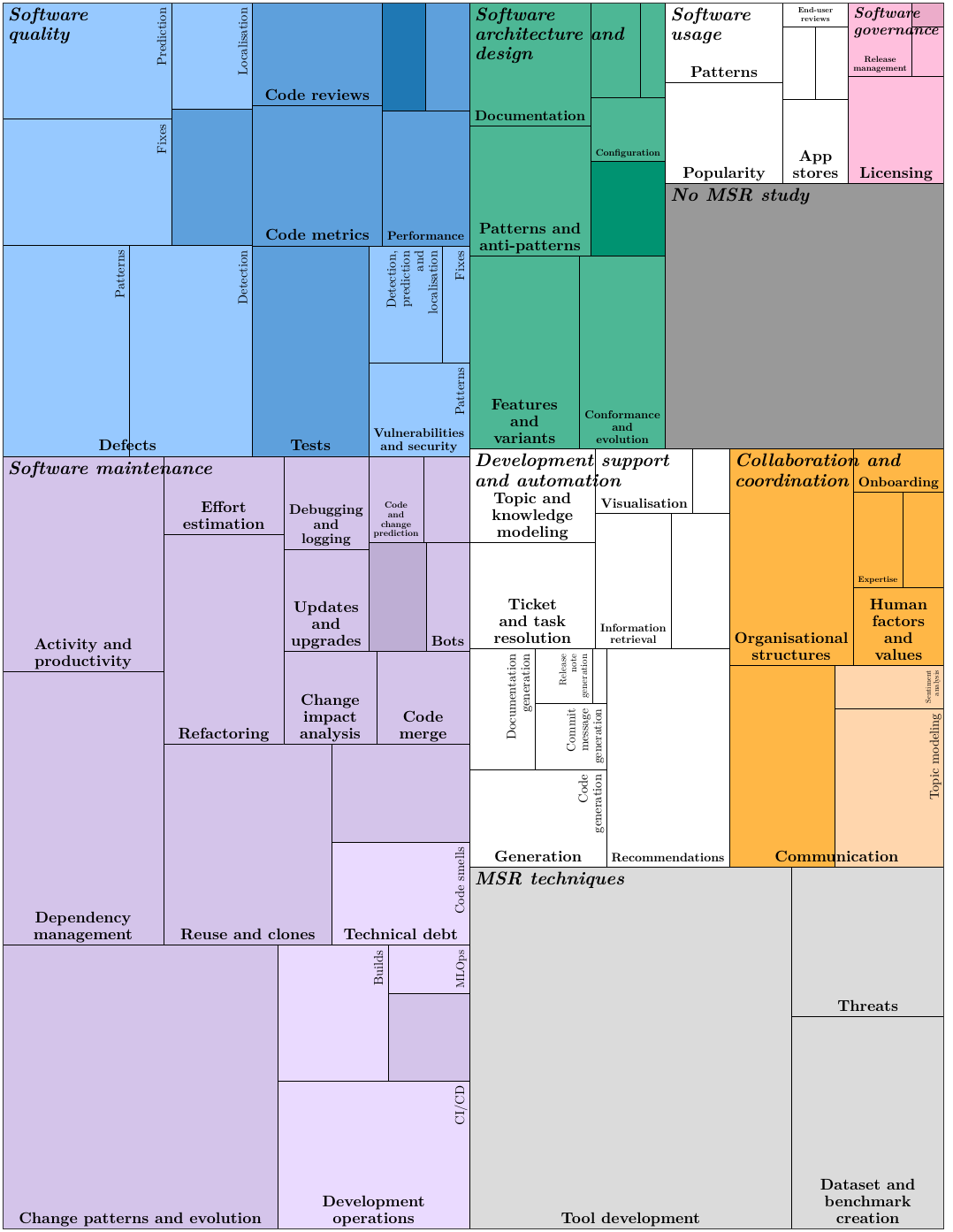}\vspace*{-1em}
    \caption{Tree map visualising the popularity of each field identified in the literature review. The size of a square indicates the popularity of the field, measured by the number of studies addressing it. Large italic headings in the upper left corners represent primary fields; bold headings at the bottom of a group squares indicate secondary categories; vertical texts and smallest, lighter coloured squares represent tertiary categories.}
    \label{fig:treemap}
    \vspace{-0.75em}
\end{figure*}

\paragraph{Software maintenance} The largest field, software maintenance, encompasses diverse subdomains, such as change patterns and evolution, spanning the representation~\citep{pravilov_unsupervised_2021, lin_cct5_2023}, grouping~\citep{jiang_summarizing_2015} and classification of changes~\citep{kiehn_empirical_2019, zeng_colare_2024}, the analysis of maintenance efforts of programming constructs~\citep{zampetti_downside_2024}, code clones~\citep{mondal_is_2018}, components and projects~\citep{arabat_empirical_2024} and guidance on tasks such as commit untangling~\citep{li_utango_2022}.

Studies on estimating maintenance and integration efforts analyse patterns in open-source projects~\citep{jiang_improving_2015, robles_development_2022} and company settings to build models supporting this task~\citep{dehghan_predicting_2017}. Some studies look at activity patterns of developers~\citep{calefato_will_2022} and companies~\citep{zhang_companies_2021}, measures~\citep{oliveira_code_2020} and models for team~\citep{wang_quantifying_2022} and individual productivity~\citep{kuutila_individual_2021}, collaboration~\citep{zhou_scalability_2017, arabat_empirical_2024} and other promoting and hindering factors~\citep{scholtes_aristotle_2016, wessel_github_2023} to optimise development efforts.

Software change impact analysis allows for identifying ripple effects during maintenance. Related studies investigated the relationship between logical and semantic coupling with co-change~\citep{rolfsnes_generalizing_2016, ajienka_empirical_2018} to identify artefacts affected~\citep{nejati_automated_2016, borg_supporting_2017}. Evolutionary couplings were further used to recommend changes~\citep{rolfsnes_aggregating_2018}. Similarly, studies predicted future changes, such as whether a commit is prone to reversal~\citep{yan_characterizing_2019} or conflicts~\citep{accioly_analyzing_2018}. Merge conflicts and their resolution were explored in regards to frequency~\citep{liu_customized_2022}, advanced merge strategies~\citep{seibt_leveraging_2022} and interactions with refactorings~\citep{mahmoudi_are_2019, ellis_operation-based_2023}.

Sequences of code changes are frequently represented by abstract syntax trees (ASTs)~\citep{stevens_querying_2019, tsantalis_refactoringminer_2022} and more coarse-grained code structure trees (CSTs)~\citep{silva_refdiff_2021}, for instance to detect and analyse refactoring practices~\citep{muse_refactoring_2023} and motivations~\citep{silva_why_2016}. Others proposed tools to support and automate refactoring-related operations~\citep{nyamawe_feature_2020, gao_automating_2021} or studied the frequency and tactics of refactoring at the process level~\citep{noei_empirical_2023}.

While refactoring reduces technical debt, its emergence has also been extensively studied, for instance through the investigation of code smells~\citep{tufano_when_2015, wu_cloneripples_2024} and their relationship with software architecture, design~\citep{oliveira_dont_2023} and quality~\citep{wang_empirical_2020, oishwee_exploratory_2022}. Studies also introduced novel types of smells~\citep{rosa_fixing_2024, peruma_tsdetect_2020} and analysed self-admitted technical debt through code comments~\citep{huang_identifying_2018, obrien_23_2022}.

Code reuse has also attracted research attention as a beneficial practice, particularly in software families with divergent forks of the same code base~\citep{businge_reuse_2022, michelon_evolving_2022} and shared commits across repositories~\citep{mockus_complete_2020}. 
Accordingly, the challenges and strategies of keeping reused code~\citep{hata_same_2021}, documentation~\citep{tan_detecting_2023} or apps~\citep{li_cda_2020, wen_keep_2024} up-to-date represent further research directions. Related studies on dependency management explored best practices to migrate libraries~\citep{sawant_reaction_2018, he_large-scale_2021}, frequencies of upgrades and downgrades~\citep{dilhara_understanding_2021}, and proposed methods to support developers in detecting API breaking changes~\citep{brito_apidiff_2018} and incompatibilities~\citep{claes_historical_2015}, and recommending suitable library alternatives~\citep{nafi_mining_2022}. 

For software builds and development operations, methods were proposed to detect unspecified dependencies~\citep{bezemer_empirical_2017}, optimise build times~\citep{ghaleb_empirical_2019, gallaba_lessons_2022}, detect build changes~\citep{macho_extracting_2017} as well as to understand~\citep{zolfagharinia_study_2019}, predict~\citep{santolucito_learning_2022} and resolve build failures~\citep{vassallo_every_2020}. To improve continuous integration (CI) and deployment activities, studies explored patterns in CI specifications~\citep{sidhu_reuse_2019, gallaba_use_2020} and the evolution~\citep{golzadeh_rise_2022} and impact of CI practices~\citep{rahman_characterizing_2018, bernardo_impact_2023}. With increasing popularity of artificial intelligence (AI) components, machine learning operations became another topic of interest~\citep{bernardo_how_2024} with the goal of improving current practices~\citep{aghili_studying_2023, bendimerad_-premise_2023}. Other investigated automation practices include tools~\citep{guo_smartdebug_2016, hashimoto_automated_2018} for debugging and logging~\citep{chen_characterizing_2017, li_studying_2018}, as well as bots, both as mechanisms for automating tasks~\citep{erlenhov_empirical_2020, rombaut_theres_2023} and as entities requiring detection in mining pipelines~\citep{ma_world_2021}.

From the field of software maintenance, we select the study ``Big Data = Big Insights? Operationalising Brooks’ Law in a Massive GitHub dataset'' by Gote~\etal~\citeyearpar{gote_big_2022} for replication. By analysing developer productivity and its relation to team dynamics, the study addresses a central question in this area while leveraging all the data extraction and analysis capabilities required by our tool selection criteria.

\paragraph{Software quality} Ensuring and improving quality is a central concern in software engineering. A major interest in mining studies is in novel methods for identifying and localising defects, ideally just-in-time when a change is performed~\citep{yan_just--time_2022} and generalisable across projects~\citep{zhou_how_2018}. Studies incorporate historical commit information to improve localisation performance~\citep{wen_historical_2021}, and propose and evaluate variants and extensions~\citep{bludau_pr-szz_2022} of the SZZ algorithm~\citep{sliwerski_when_2005} for the identification of bug-introducing commits~\citep{fan_impact_2021} and bug-fixing commits~\citep{lyu_evaluating_2024}. 
Predictive models often rely on supervised machine learning~\citep{pornprasit_jitline_2021, ni_just--time_2022} and evaluate different sets of features, including complexity~\citep{alqadi_slice-based_2020} and socio-technical metrics~\citep{tourani_impact_2016, di_nucci_developer_2018}. As defect prediction remains challenging, studies investigate contributing factors~\citep{foucault_impact_2015, wan_data_2024} and reduce complexity by focusing on specific application domains~\citep{wan_bug_2017, rahman_gang_2020} or bug types~\citep{sellik_learning_2021}. 
Automatically ensuring good coding practices and fixing defects is another goal of research, ranging from the correction of violations~\citep{oumarou_identifying_2015} to automatic program repair~\citep{durieux_empirical_2019, huang_execution-free_2024}. 

As a critical class of software defects, vulnerabilities and their detection~\citep{ponta_detection_2020}, categorisation~\citep{mazuera-rozo_android_2019, rahman_security_2023}, and prediction receive considerable attention, particularly with the growing importance of timely updates and software supply chain security~\citep{sawadogo_sspcatcher_2022, hommersom_automated_2024}.  
Research further examines how vulnerabilities are introduced and evolve, focusing on development practices~\citep{rahman_why_2022, iannone_rubbing_2023}, patterns~\citep{verdi_empirical_2022, almanee_too_2021}, life cycles and propagation~\citep{alfadel_empirical_2023}. Other studies focus on specific cyber attacks~\citep{davis_impact_2018} and best practices to counterfeit them~\citep{santos_counterfeit_2022}. 

Code reviews are another field of quality assurance~\citep{thongtanunam_investigating_2015}, with studies investigating common practices, for instance regarding review coverage, participation and reviewer expertise~\citep{mcintosh_empirical_2016}. To provide guidance and optimise processes, researchers explored factors leading to patch acceptance~\citep{baysal_investigating_2016}, reviewer participation~\citep{ruangwan_impact_2019} and best practices to write useful reviews~\citep{bosu_characteristics_2015}. Automated methods have been proposed for review comment generation~\citep{chatley_diggit_2018}, reviewer recommendation~\citep{zanjani_effective_2016}, linking interdependent reviews~\citep{hirao_review_2019} and tracking the evolution of changes alongside their review comments~\citep{ramsauer_list_2019}.

Code metrics are a prerequisite to measure, compare and monitor various aspects of code quality~\citep{nguyen_empirical_2022}. Studies proposed metrics for readability~\citep{piantadosi_how_2020}, regularity~\citep{gil_correlation_2017}, complexity~\citep{meijer_maintenance_2022, alqadi_slice-based_2020}, changes~\citep{reck_multidimensional_2023}, function usage~\citep{grotov_large-scale_2022}, smells~\citep{jebnoun_scent_2020}, and structure~\citep{nikolaidis_metrics-based_2023}. Others explored the validity of code metrics, for instance regarding their agreement~\citep{o_cinneide_experimental_2017} or correlation with size~\citep{gil_correlation_2017, chowdhury_empirical_2022}, social factors~\citep{youssef_impact_2015} and bug density~\citep{reck_multidimensional_2023}.

Complementing static analysis, testing is the primary dynamic approach for quality assurance. To support testing practices, researchers explored the benefits of integrating historic code changes into test selection~\citep{soetens_change-based_2016, kauhanen_regression_2021}, proposed novel methods to detect flaky tests~\citep{parry_empirically_2023} and explored patterns in test failures~\citep{chen_t-evos_2023}. 

Other studies aim to improve software performance by automatically identifying code changes responsible for performance regressions~\citep{luo_mining_2016}, predicting performance~\citep{gong_does_2022} and exploring (in-)efficient programming patterns to optimise energy efficiency~\citep{rua_large-scale_2023}.

From the field of software quality, we select the study ``Impact of Developer Turnover on Quality in Open-Source Software'' by Foucault~\etal~\citeyearpar{foucault_impact_2015} for replication. Addressing software defects, the largest sub-area within this field according to Figure~\ref{fig:treemap}, and their relationship with developer turnover, it relies only on tool capabilities covered by our selection criteria.

\paragraph{Collaboration and coordination} Coordination is essential in global open-source projects, representing a primary subject in empirical studies. Popular topics include detecting organisational structures such as developer roles~\citep{pinto_more_2016, milewicz_characterizing_2019, jiang_bringing_2024}, hierarchy~\citep{joblin_hierarchical_2023}, reputation~\citep{rahman_snakes_2019}, communities~\citep{kannee_intertwining_2023}, as well as their stability~\citep{sharif_studying_2016} and future evolution~\citep{wang_quantifying_2022, zhang_measuring_2025}. Often, analyses are driven by the construction of developer networks at different levels of granularity, such as for individual software projects~\citep{ashraf_mixed_2020, maddila_nalanda_2022, bock_measuring_2021} or entire ecosystems~\citep{lamba_heard_2020, zhang_how_2020}. 

Communication between developers provides insights into technical issues~\citep{croft_empirical_2021} or sentiments~\citep{calefato_sentiment_2018}, allowing for identifying areas for improvement. The analysed means of communication range from mailing lists to chats~\citep{alkadhi_rationale_2017, mezouar_exploring_2022}, GitHub issues and pull requests~\citep{brisson_we_2020}, including reactions~\citep{batoun_empirical_2023} and tags for social coding~\citep{foundjem_mixed-methods_2022}. Particular interest also lies in mining topics from discussions on Q\&A websites~\citep{kamienski_empirical_2021} such as Stack Overflow~\citep{beyer_grouping_2016, uddin_empirical_2021}.

Other studies analysed and modelled developer expertise, for instance to support task assignment~\citep{da_silva_niche_2015, milano_navigating_2024} or matching competencies and projects~\citep{fang_matching_2023}. Similarly, studies derived best practices for onboarding~\citep{gautam_empirical_2017}, newcomer tasks~\citep{santos_tag_2023}, mentoring and long-term motivation~\citep{norikane_which_2017}.

Studies also investigated human factors~\citep{avgustinov_tracking_2015} and compliance with human values, for instance regarding privacy and inclusiveness~\citep{nurwidyantoro_integrating_2023, khalajzadeh_supporting_2023}, team diversity~\citep{rossi_geographic_2022}, the promotion of ethical developer behaviour~\citep{tourani_code_2017, win_towards_2023} and bias~\citep{treude_she_2023}.

From the field of collaboration and coordination, we select the study ``Classifying Developers into Core and Peripheral: An Empirical Study on Count and Network Metrics'' by Joblin~\etal~\citeyearpar{joblin_classifying_2017} for replication. It addresses organisational structures, the largest sub-area within this field according to Figure~\ref{fig:treemap}, and relies on the capabilities of \textsc{Codeface}, a tool from our previous study that informed the tool selection criteria in this work.

\begin{answerbox}
\textbf{Answer to RQ1} \emph{Which topics in empirical software engineering are typically driven by mining software evolution from repositories and could be affected by threats due to differences in tooling?}: Besides the development of automation and mining techniques, most tool-driven insights are gained in the fields of software maintenance, software quality and collaboration and coordination. Other, slightly less popular fields include software architecture and design, software usage and software governance, respectively. All of these fields can potentially be affected by threats in mining tools, as they rely on complex, very specific analytical capabilities, which inherently require numerous implementation decisions.
\end{answerbox}

\subsection{Mining tools}

Alongside the literature review, we collected mining tools fulfilling a subset of the selection criteria from Section~\ref{sec:tool-selection}. Table \ref{tab:tools} summarises the resulting candidate open-source tools. All tools serve the purpose of evolutionary software analyses, but pursue different goals and provide different functionalities, which we determined based on papers and repositories. To support future users in tool selection, we also listed some of the capabilities beyond our criteria. The full list of 111
open-source tools, their primary purpose, repository and, if applicable, introducing paper can be found in the supplementary material. 

From Table~\ref{tab:tools}, four tools fulfil all of the selection criteria, namely \textsc{Codeface}, \textsc{git2net}, \textsc{GrimoireLab} and \textsc{Kaiaulu}. 
In the following, we outline their pipelines to illustrate similarities and differences:

\begin{table}[htb]
    \caption{Tools for mining software repository evolution with similar capabilities: parsing of commits (VCS), finer-grained code entities (Entity), mailing lists (Mail), issues from bug trackers and advanced analyses including developer network construction (Net.), code complexity analysis (Comp.) and construction of abstract syntax trees (AST). \label{tab:tools}}
   \begin{tblr}{colspec={Xcccccccc},
                row{odd}={bg=gray!15},  
                row{1}={bg=white,fg=black},
                row{2}={bg=lfd-yellow!50,fg=black},
                row{6}={bg=lfd-green!50,fg=black},
                row{8}={bg=lfd-lilac!50,fg=black},
                row{9}={bg=lfd-blue!50,fg=black}}
        \toprule
        Tool & VCS & Entity & Mail & Issue & Net. & Comp. & AST & Active\\ \midrule
        \textsc{Codeface}\footnotemark[1] & x & x & x & x & x & x & -- & 2010--now \\
        \textsc{Crossminer}\footnotemark[2] & x & x & x & x & x & x & -- & 2013--2019 \\
        \textsc{Diggit}\footnotemark[3] & x & x & -- & -- & -- & x & -- & 2014--2021 \\
        \textsc{Dominoes}\footnotemark[4] & x & x & x & x & x & -- & x & 2014-2020 \\
        \textsc{git2net}\footnotemark[5] & x & x & -- & -- & x & x & -- & 2019--now \\
        \textsc{GHTorrent}\footnotemark[6] & x & -- & -- & -- & -- & -- & -- & 2013--2019 \\
        \textsc{GrimoireLab}\footnotemark[7] & x & x & x & x & x & x & -- & 2015--now \\
        \textsc{Kaiaulu}\footnotemark[8] & x & x & x & x & x & -- & -- & 2020--now \\
        \textsc{Lagoon}\footnotemark[9] & x & x & x & -- & x & -- & -- & 2021--2022 \\
        \textsc{LibVCS4j}\footnotemark[10] & x & -- & -- & x & -- & x & -- & 2018--2024 \\
        \textsc{Lisa}\footnotemark[11] & x & -- & -- & -- & -- & -- & x & 2016--2019\\
        \textsc{Pandora}\footnotemark[12] & x & -- & -- & x & -- & -- & -- & 2020--2021 \\
        \textsc{PyDriller}\footnotemark[13] & x & x & -- & -- & -- & x & -- & 2018--now \\
        \textsc{SmartSHARK}\footnotemark[14] & x & x & x & x & -- & x & x & 2015--now \\
        \textsc{Topleet}\footnotemark[15] & x & -- & -- & x & -- & x & -- & 2019--2021 \\
        \bottomrule
    \end{tblr}
\end{table}

\footnotetext[1]{\href{https://github.com/lfd/codeface}{https://github.com/lfd/codeface},~\cite{joblin_developer_2015}}
\footnotetext[2]{\href{https://github.com/crossminer/crossminer}{https://github.com/crossminer/crossminer},~\cite{di_rocco_development_2021}}
\footnotetext[3]{\href{https://github.com/lawrencejones/diggit}{https://github.com/lawrencejones/diggit},~\cite{chatley_diggit_2018}}
\footnotetext[4]{\href{https://github.com/gems-uff/dominoes}{https://github.com/gems-uff/dominoes},~\cite{da_silva_junior_dominoes_2022}}
\footnotetext[5]{\href{https://github.com/gotec/git2net}{https://github.com/gotec/git2net},~\cite{gote_git2net_2019}}
\footnotetext[6]{\href{https://github.com/ghtorrent/ghtorrent.org}{https://github.com/ghtorrent/ghtorrent.org},~\cite{gousios_ghtorent_2013}}
\footnotetext[7]{\href{https://github.com/chaoss/grimoirelab}{https://github.com/chaoss/grimoirelab},~\cite{duenas_grimoirelab_2021}}
\footnotetext[8]{\href{https://github.com/sailuh/kaiaulu}{https://github.com/sailuh/kaiaulu},~\cite{paradis_building_2022}}
\footnotetext[9]{\href{https://github.com/GaloisInc/SocialCyberLAGOON}{https://github.com/GaloisInc/SocialCyberLAGOON},~\cite{dey_lagoon_2022}}
\footnotetext[10]{\href{https://github.com/uni-bremen-agst/libvcs4j}{https://github.com/uni-bremen-agst/libvcs4j},~\cite{steinbeck_mining_2020}}
\footnotetext[11]{\href{https://bitbucket.org/sealuzh/lisa/src/master/}{https://bitbucket.org/sealuzh/lisa/src/master/},~\cite{alexandru_rapid_2015}}
\footnotetext[12]{\href{https://github.com/clowee/PANDORA}{https://github.com/clowee/PANDORA},~\cite{nguyen_pandora_2022}}
\footnotetext[13]{\href{https://github.com/ishepard/pydriller/}{https://github.com/ishepard/pydriller/},~\cite{spadini_pydriller_2018}}
\footnotetext[14]{\href{https://github.com/smartshark}{https://github.com/smartshark},~\cite{trautsch_addressing_2018}}
\footnotetext[15]{\href{https://github.com/topleet/topleet/}{https://github.com/topleet/topleet/},~\cite{hartel_incremental_2020}}

\begin{figure}[htbp]
  \centering
  \includegraphics[width=\linewidth]{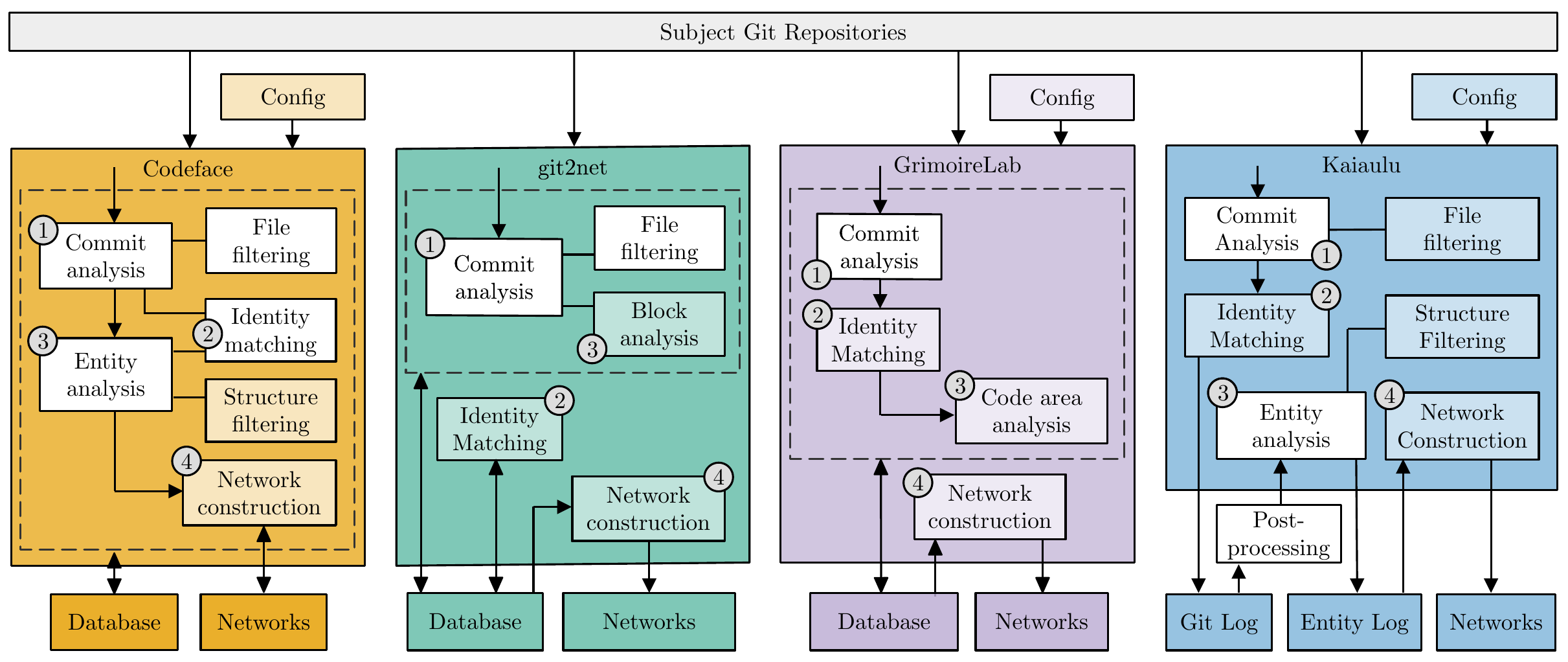}
  \caption{Informal overview of structural components of the mining tools
\textsc{Codeface}, \textsc{git2net}, \textsc{GrimoireLab} and \textsc{Kaiaulu}. White boxes indicate fixed parts of the data extraction pipeline, while coloured boxes visualise configurable steps. Although all tools perform similar analysis steps, their interaction and data structure differ.}
  \label{fig:pipeline}
\end{figure}

\paragraph{Codeface:} Starting as an industrial software analysis tool from Siemens AG, \textsc{Codeface} evolved to a research tool for socio-technical aspects in software development. One of its key features is the efficient mining of Git repositories for evolutionary analyses. Figure \ref{fig:pipeline} shows the informal components and steps involved in this process: Codeface supports different analysis modes to build a MySQL database of commits, persons, mails, and fine-grained changes affecting entities such as files, overarching features or functions. During analysis, the tool automatically constructs temporal, directed and weighted developer networks from jointly edited entities and detects communities~\citep{joblin_developer_2015}. \textsc{Codeface} also offers a code complexity model for effort estimation. Built for industrial scales, the tool implements parallelisation at multiple levels and provides a dashboard for managers. The tool sets meaningful defaults for aspects such as file filtering, entity parsing, identity matching and network construction and hides its complexity under a command-line interface (CLI). Configuration files enable users to modify few parts of the pipeline, such as the time interval length or entity granularity for analysis. In the database, all information is related to a release range resulting from the time intervals. For reproducibility, \textsc{Codeface} provides a Docker image with all dependencies.

Codeface has been used to explore organisational structures~\citep{joblin_evolutionary_2017} such as hierarchy~\citep{joblin_hierarchical_2023}, developer roles~\citep{joblin_classifying_2017, bock_automatic_2023, picha_towards_2017} and communities~\citep{bock_measuring_2021}. \textsc{Codeface4Smells}~\citep{tamburri_exploring_2021} is an expansion tool to detect community smells as sub-optimal patterns of social organisation. The prevalence, prediction~\citep{palomba_predicting_2021, almarimi_learning_2020} and causes~\citep{catolino_gender_2019, lambiase_good_2022} of these smells have been studied along with their impact~\citep{tamburri_exploring_2021, de_stefano_splicing_2020, palomba_beyond_2021, eken_empirical_2021, stefano_impacts_2022}. Other studies used \textsc{Codeface} to track development process conformance~\citep{hunsen_fulfillment_2020, bock_synchronous_2021}, design community-aware software forges~\citep{tamburri_canary_2020, tamburri_evolving_2021} and analyse factors influencing project success~\citep{joblin_how_2022} and software quality~\citep{mauerer_search_2022}.

\paragraph{git2net:} \textsc{git2net} was specifically designed for developer network construction at line granularity. The tool detects changes in exact line ownerships and uses text mining to extract textual information from files. To mine the baseline data for network construction from Git repositories, \textsc{git2net} relies on the mining tool \textsc{PyDriller} and stores all information related to commits, code editing operations and persons in its SQLite database~\citep{gote_git2net_2019, gote_analysing_2021}, as shown in figure \ref{fig:pipeline}. Besides the construction of directed co-editing networks for user-defined or automatically determined time windows, \textsc{git2net} also supports bipartite graphs and projections connecting developers who contributed to the same files~\citep{gote_git2net_2019, gote_analysing_2021}. For identity matching, \textsc{git2net} integrates \textsc{Gambit}~\citep{gote_gambit_2021}, a tool disambiguating developers based on their name and e-mail address similarity. To measure code complexity, \textsc{git2net} relies on the external tool \textsc{lizard}~\citep{gote_big_2022}. Built for large-scale mining, \textsc{git2net} internally implements parallelisation. The tool provides a CLI for standard analyses and an API for more advanced, user-defined pipelines~\citep{gote_git2net_2019, gote_analysing_2021, gote_asonam_2022}. In this study, we use the API for increased flexibility and replication conformance.

In previous work, \textsc{git2net} was used to detect and optimise interactions of team member roles~\citep{zingg_detecting_2023} and to analyse the impact of overhead to coordinate with other developers on individual developers' productivity~\citep{gote_analysing_2021}, illustrating the validity of Brooks' law~\citep{gote_big_2022}. In the context of development operations, future work will employ \textsc{git2net} to explore the effect of specific GitHub actions such as code review bots on team collaboration structures~\citep{roseler_network_2023}.

\paragraph{GrimoireLab:} Introduced by the company Bitergia, \textsc{GrimoireLab} was designed for the free, open source software (FOSS) community to meet industrial requirements in aspects such as automation, configurability and diversity of metrics~\citep{duenas_grimoirelab_2021, gonzalez-barahona_software_2022}. \textsc{Perceval}~\citep{duenas_perceval_2018} is \textsc{GrimoireLab}'s unified API which extracts data from diverse sources such as git repositories, code reviews, mailing lists, issue trackers, project wikis and chats with non-uniform access to a standard format. \textsc{Graal}~\citep{cosentino_engineering_2018} provides additional source code analyses such as evaluating complexity with the help of external tools. In \textsc{GrimoireLab}, output of all analyses is stored in an ElasticSearch database. Analyses focusing on specific information such as commit activity or file changes are isolated in different indexes for efficient querying. Users can specify the desired analyses in configuration files. After analysis, results can be inspected on a Kibana dashboard, where users can visualise bipartite networks and projections from diverse information in the ElasticSearch backend, such as developers, files or entire repositories, as illustrated in Figure~\ref{fig:pipeline}. While \textsc{GrimoireLab}, contrary to other tools, is able to merge information across projects~\citep{duenas_grimoirelab_2021, gonzalez-barahona_software_2022}, for instance by matching developer identities with the \textsc{SortingHat} component~\citep{moreno_sortinghat_2019}, users can also apply filters in the frontend for diverse interests. \textsc{GrimoireLab} optimises computational efforts by means of parallelisation with components such as \textsc{Mordred} and minimal interactions with data sources, for instance in case of processing only new items when refreshing data. To promote reproducibility, \textsc{GrimoireLab} provides multiple docker images for its components~\citep{duenas_grimoirelab_2021, gonzalez-barahona_software_2022}.

Research based on \textsc{GrimoireLab} designed and developed tool extensions, for instance a community dashboard to overview team diversity and developer turnover~\citep{guizani_unveiling_2023}, commit activity forecasting~\citep{decan_gap_2020} and bot detection~\citep{chidambaram_bot_2022}. Besides tailored analyses implemented with the Bitergia analytics platform for customers including the Apache Software Foundation, GitLab, Google, WikiMedia, and others, \textsc{GrimoireLab} has been widely adopted to optimise software development processes. For instance, data and visualisations gathered by \textsc{GrimoireLab} and its predecessors were used to monitor project health~\citep{goggins_making_2021}, improve code review processes~\citep{izquierdo_software_2019, tecimer_detection_2021}, automate security analyses for sensitive software applications~\citep{sonnekalb_towards_2020}, facilitate effort estimation in open-source projects~\citep{robles_estimating_2014} and study the impact of and best practices for continuous integration (CI)~\citep{zhao_impact_2017}. The tool was further used to gain insights into challenges of reusing pre-trained deep learning models~\citep{taraghi_deep_2024}. 

In socio-technical research, \textsc{GrimoireLab} was used for analyses of developer emotions and affective states~\citep{claes_use_2018, kuutila_using_2018}, to study contributors' behavioural patterns outside regular working hours~\citep{claes_abnormal_2017}, identify paid developers~\citep{claes_towards_2018}, explore the effect of stronger formality on development-related risks~\citep{gaughan_engineering_2024} and study engagement of and collaboration across different teams and organisations in open-~\citep{newton_leveraging_2023, robles_industrial_2024} and inner-source projects~\citep{izquierdo-cortazar_starting_2022}. Other studies analysed developer onboarding~\citep{foundjem_onboarding_2021, foundjem_open_2021} and factors influencing community sustainability together with its effect on other aspects such as productivity and quality~\citep{alami_free_2024, alami_incubation_2025}. \textsc{GrimoireLab} was also used in studies targeting code contributors and users, for example ranking open-source repositories based on quality, popularity and maintainability~\citep{hasabnis_gitrank_2022}.

\paragraph{Kaiaulu:}~\cite{paradis_building_2022} built \textsc{Kaiaulu} as a tool for empirical software engineering research, following capabilities and design principles observed in other, often retired, mining tools and focusing on understandability and ease of use~\citep{paradis_analyzing_2024}. As shown in Figure~\ref{fig:pipeline}, the analysis pipeline starts with commit parsing, where users must specify several configuration options regarding file filtering. Contrary to the other mining tools, \textsc{Kaiaulu} stores data in CSV files instead of a database. Additional analyses such as the detection of functions and classes are performed on demand with user-provided settings. Identity matching in \textsc{Kaiaulu} is optional and by default only performed within a single column of a single table. In our replications, we use the scripts from our previous study to perform identity matching in both author and committer columns across all tables. Subsequent network construction supports multiple modes, for instance connecting developers to commonly edited files or entities by bipartite graphs and to each other by bipartite projections. Alternatively, temporal networks connect developers in the order of contributions to the respective artefacts. Edge weights in \textsc{Kaiaulu}'s graphs are aggregated by weight schemes chosen by the user. To support different needs, \textsc{Kaiaulu} provides both a CLI and an API~\citep{paradis_building_2022}. In our work, we rely on the API to leverage and enhance \textsc{Kaiaulu}'s flexibility for higher replication conformance. 

In practice, \textsc{Kaiaulu} has been used to measure process compliance with requirements at NASA~\citep{paradis_towards_2023}. Research extended \textsc{Kaiaulu} to detect social smells~\citep{paradis_analyzing_2024} and examined their relationship with design smells~\citep{mumtaz_analyzing_2022}, GitHub features~\citep{mumtaz_preliminary_2022} and software vulnerabilities~\citep{paradis_socio-technical_2024}.

As illustrated in Figure~\ref{fig:pipeline}, all tools perform similar analysis steps, but the order and configurability differ. From a higher level of abstraction, the implementations of individual steps are also similar. For example, \textsc{Codeface}, \textsc{git2net} and \textsc{Kaiaulu} all use the git blame to identify code entities. Tool users may therefore expect very similar results across tools and assume that switching tools is possible without negative consequences.

\subsection{Baseline data comparison} \label{sec:data-comparison}

In previous work, we compared the baseline data of two mining tools -- \textsc{Codeface} and \textsc{Kaiaulu} -- and examined some of the factors responsible for discrepancies in technical detail. 
This section summarises the most important findings from this initial study and demonstrates the generalisability of the observed threats to validity for two additional mining tools -- \textsc{git2net} and \textsc{GrimoireLab}. 
For this purpose, as explained in Section~\ref{sec:baseline-comparison}, we examine a representative subset of all repositories analysed in the three replication studies. Table~\ref{tab:subject-repositories} provides an overview of the projects' characteristics.

\begin{table}[htb]
   \caption{Descriptive statistics of subject projects sampled from the original studies for the baseline data comparison across tools. Primary language and lines of code (LOC) are determined via \textsc{cloc}~\citep{danial_cloc_2025}. Commits and developers are counted directly via git, without further processing such as identity matching. Depending on the study a project was selected from, configuration parameters such as the time window size \(t\) in months differ. The statistics refer to the repository states checked out in the original studies. \label{tab:subject-repositories}}
   \begin{tblr}{colspec={Xcccccc},
                row{odd}={bg=gray!15},  
                row{1}={bg=white,fg=black}}
        \toprule
        Project & Domain & Language & Commits & Team & LOC[k] & \(t\)[m] \\ \midrule
        \textsc{Birt} & Data visualisation & Java & 32,303 & 236 & 2,538 & 9\\
        \textsc{Conductor} & Orchestration & Java & 2,141 & 344 & 149 & 9 \\
        \textsc{Django} & Web framework & Python & 21,786 & 3,121 & 513 & 3 \\
        \textsc{Flink} & Stream processing & Java, Scala & 22,567 & 1,958 & 1,597 & 9 \\
        \textsc{PostgreSQL} & DBMS & C & 39,375 & 60 & 1,111 & 3 \\
        \textsc{Qemu} & Hardware virtualiser & C & 41,947 & 2,707 & 1,000 & 3 \\ 
        \textsc{U-Boot} & Boot loader & C & 33,496 & 3,089 & 1,261 & 3 \\
        \textsc{Wine} & Compatibility layer & C & 108,690 & 1,854 & 3,334 & 3 \\
        \bottomrule
    \end{tblr}
\end{table}

\begin{figure}[htbp]
  \centering
  \includegraphics[width=\linewidth]{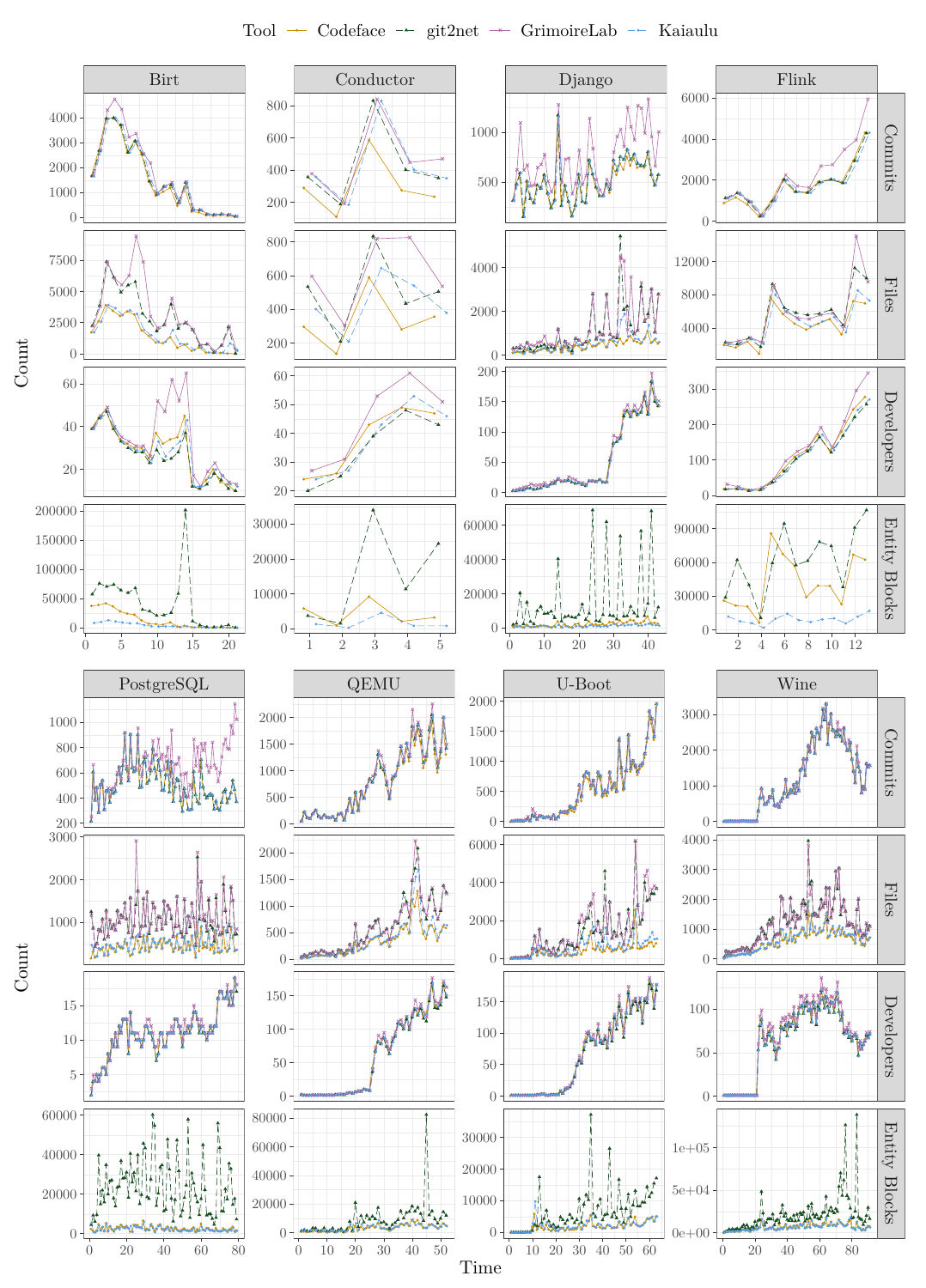}
  \caption{Time series of simple count-based metrics (number of commits, files, developers and finer-grained code entity blocks) calculated based on the git log extracted by \textsc{Codeface}, \textsc{git2net}, \textsc{GrimoireLab} and \textsc{Kaiaulu} with the most similar configurations for the respective replication. Lines are plotted with a small offset to visualise overlapping lines.}
  \label{fig:baseline-ts}
\end{figure}

A fundamental task of mining tools that capture evolutionary software development processes is the consistent extraction of commits and related information such as edited files and developers. The simple counts of these units often form the basis for more complex metrics in the course of a study and should therefore not differ between tools. In figure \ref{fig:baseline-ts}, however, we illustrate that, even with similar tool configurations, the time series for basic metrics such as the number of commits, developers, edited files and edited entity blocks can vary notably across four well-known mining tools.

As in our previous study, we find that the differences between tools vary depending on project characteristics. In addition, the plot reveals differences due to the analysis setup. Since we adopt each original study's configuration, the projects Birt, Conductor and Flink from Gote~\etal~\citeyear{gote_big_2022} are analysed in nine-month windows, whereas those from Joblin~\etal~\citeyearpar{joblin_classifying_2017} use three-month windows. Comparing these two window sizes shows that larger aggregation windows make differences in the baseline data appear more pronounced. 

Commit counts, which form the basis for all other metrics, are often substantially higher for \textsc{GrimoireLab} than for the other tools. \textsc{GrimoireLab} collects commits from all branches in a GitHub repository, whereas the other tools analyse a single, locally checked-out branch. This has notable impact in subject project PostgreSQL: According to \textsc{GrimoireLab}, commit activity reaches its maximum in most recent time windows, while the other tools indicate an overall activity decrease. In our previous study, we identified the order of filtering operations, commit parsing and commit storage as further sources of variation in the commit time series. 

The file counts show the effect of file filtering and data scheme more directly. For instance, \textsc{Codeface} ignores changes in documentation or lock files and only stores file information if one of its code entities changed, therefore identifying less than half as many files as \textsc{GrimoireLab} in subject project Conductor. \textsc{Kaiaulu} provides configuration options to filter only file endings relevant to the user, while \textsc{git2net} internally excludes binary file changes.

Developer identities, by contrast, are largely consistent across tools. One exception is subject project Birt, where \textsc{GrimoireLab}'s identity matching assigns multiple identities to developers sharing the same full name, whereas the other tools merge them into a single identity.

Although all tools support finer-grained code analysis, \textsc{GrimoireLab} does not store function- or block-level entities in its ElasticSearch backend, which is why Figure~\ref{fig:baseline-ts} shows the number of identified entity blocks only for \textsc{Codeface}, \textsc{git2net} and \textsc{Kaiaulu}. The time series indicate that in almost all cases, the number of blocks identified by \textsc{git2net} is an order of magnitude higher than for the other tools. 
This is expected due to differing data schemes: \textsc{Codeface} summarises changed blocks per named entity and commit; \textsc{Kaiaulu} additionally distinguishes \emph{multiple} blocks for the same entity and \textsc{git2net} does not identify entities by their name, but only captures separate code line blocks in local proximity~\citep{gote_analysing_2021}. Despite \textsc{Kaiaulu}'s data scheme being closer to \textsc{git2net}'s, \textsc{Codeface}'s absolute counts are more similar to those of \textsc{git2net}. This indicates that beyond schema differences, other technical details, such as the entity parsing, interfere. Although \textsc{Codeface} and \textsc{Kaiaulu} follow an overall similar approach, \textsc{Codeface} extracts an internally defined set of entities via Doxygen, ctags and a custom SQL parser for predefined programming languages, while \textsc{Kaiaulu} relies solely on user-configured ctags. To date, \textsc{Kaiaulu} does not yet support ctags for all languages, limiting entity analysis to C, C++, Java, Python and R.\\

\begin{figure}[htbp]
  \centering
  \includegraphics[width=\linewidth]{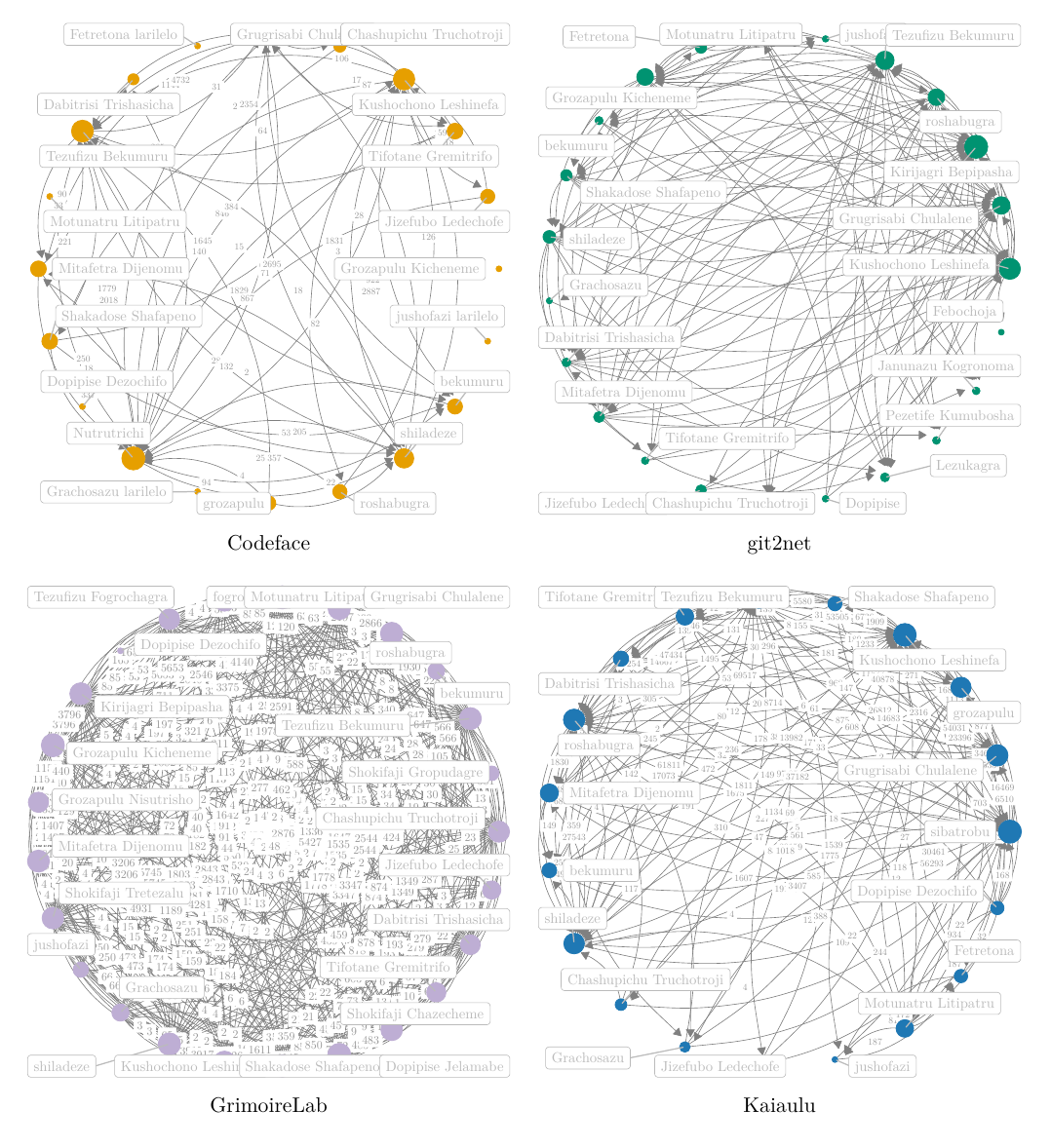}
  \caption{Developer networks constructed by \textsc{Codeface}, \textsc{git2net}, \textsc{GrimoireLab} and \textsc{Kaiaulu} for the same time interval of subject project Flink with the most similar configurations to replicate~\cite{gote_big_2022}. Names have been pseudonymised.}
  \label{fig:baseline-networks}
\end{figure}

In subsequent, more complex data processing steps, results can diverge even further. Figure~\ref{fig:baseline-networks} demonstrates the effect of different network construction methods on the resulting developer collaboration graph. Each tool constructs a network for the same time interval in the history of subject project Flink, with nodes representing developers and edges indicating collaborations.

The overall network structure shows that there are substantial differences between tools. The number of nodes is much higher in the network constructed by \textsc{git2net} compared to all other tools. In addition, the edge density and weights differ across tools. 

These differences largely stem from how each tool interprets collaboration: \textsc{Codeface} and \textsc{Kaiaulu} construct the network based on code entities jointly edited by developers. Here, a directed edge from developer \(d_1\) to \(d_2\) represents \(d_1\) contributing to code written previously by developer \(d_2\). To further indicate the strength of this collaboration, both tools assign weights to the edges, which aggregate the lines of code contributed to the other developer. However, as we showed in previous work, this aggregation differs as well. For instance, the number of lines can count lines contributed by \(d_1\) once (\textsc{Kaiaulu}) or multiple times (\textsc{Codeface}). 

\textsc{git2net} pursues a different approach. Here, results are based on the line-granularity analysis mode, as the block-granularity mode used above is not yet supported for network construction. Contrary to \textsc{Codeface} and \textsc{Kaiaulu}, edges in \textsc{git2net}'s network graphs indicate a change in code line ownership from \(d_1\) to \(d_2\). The changes in code ownership are not aggregated by default, but the graph retains separate edges in a multigraph~\citep{gote_big_2022}. 

\textsc{GrimoireLab} supports the construction of bipartite graphs for any two fields in its database. We chose developer identities as the first and file paths, as the closest code-related option available, for the second node type. 
When calculating the bipartite projection of this graph, \textsc{GrimoireLab} supports a custom weight function, which we define as the number of jointly edited files, weighted by the respective developer's number of changes. As explained by Joblin~\etal~\citeyearpar{joblin_developer_2015} and Gote~\etal~\citeyearpar{gote_git2net_2019}, networks constructed at file-granularity detect more edges in general, but many of them carry little meaning.

Besides network construction, the differences in the baseline data propagate to the networks. 
For instance, by default, \textsc{git2net} uses a different identity matching technique, Gambit~\citep{gote_gambit_2021}, which rates similarity of developer names and e-mails using Levenshtein distance, compared to the exact partial string matching supported by \textsc{Codeface}, \textsc{GrimoireLab} and \textsc{Kaiaulu}. This technique tolerates greater differences between identities, which can produce more false positives. For instance, \textsc{git2net} merges the developers Kushochono
Leshinefa with Febochoj (pseudonymised), causing the latter to appear in the network graph despite contributing no commits in the considered time window. Conversely, the string matching performed by \textsc{Codeface}, \textsc{GrimoireLab} and \textsc{Kaiaulu} leaves several duplicate nodes where identity matching was too conservative. 

\begin{answerbox}
\textbf{Answer to RQ2} \emph{(To what extent can we observe discrepancies in the data obtained
from independent software repository mining tools?)}: In a sample of eight subject projects with different characteristics and analysis parameters, we find mostly similar evolutionary trends in high-level metrics such as the number of commits, developers and files. However, the more fine-grained and downstream in the pipeline a data preparation step is, the higher seems the inherent uncertainty. For instance, notable differences exist in code entity parsing and developer network construction. This indicates that despite offering similar analyses, the tools under study cannot be considered readily interchangeable.
\end{answerbox}

\subsection{First replication: software maintenance}

The study ``Big Data = Big Insights? Operationalising Brooks’ Law in a Massive GitHub dataset'' by Gote~\etal~\citeyearpar{gote_big_2022} represents the field of software maintenance, addressing developer productivity. Brooks' law states that adding manpower to a late software project makes it later, which is similar to the Ringelmann effect in psychology describing that productivity linearly decreases with team size. The authors explore the relationship between numerous productivity and collaboration metrics through a large-scale study using correlation analysis and regression modeling. The results confirm a negative relationship in all cases, leading the authors to conclude that the Ringelmann effect and Brooks' law hold in software engineering. We adopt the central analyses to answer \emph{\textbf{RQ3a:} Do all tools agree that the relationship between team size
and productivity is negative, as claimed by Brooks' law?}

\paragraph{Reproducibility:} The original study references the \textsc{git2net} repository and a reproduction package with a dump of the extracted and aggregated data together with analysis scripts. The code aggregating data and calculating metrics, however, is partially missing. The original author confirmed that, due to pipeline complexity, it cannot be \emph{fully} recovered. With the author's help, we implemented a reconstructed analysis component and, as explained in Section~\ref{sec:replication-stability}, applied it identically to \emph{all} tools. Due to compute constraints, we limit the set of over 200 subject projects analysed in the original work to ten that preserve a similar distribution of project characteristics.

\paragraph{Result stability:} As reasoned in Section~\ref{sec:replication-stability}, the following criteria illustrate our interpretative logic for assessing stability: We consider results as \emph{stable} if the maximum deviation across all tools $t \in T$ remains below the absolute thresholds $\delta_\rho=\delta_\beta=0.1$ for both (i) Pearson correlations $\rho_{m_p,m_c,t}$ between productivity metrics $m_p \in M_{\text{prod}}$ and collaboration metrics $m_c \in M_{\text{collab}}$ and (ii) linear regression coefficients $\beta_{m_p,\text{TS},t}$ for productivity metrics regressing team size $\text{TS}$, each pooled across all project and time window observations. While $0.05$ is a popular threshold in empirical evaluation, we deliberately choose $0.1$ as a more relaxed threshold based on a priori scientific assessment, as suggested by McShane~\etal\citeyearpar{mcshane_abandon_2019}:

\begin{equation*}
H_{\text{aR}}:\quad
\begin{aligned}[t]
& \forall m_p \in M_\text{prod},\; \forall m_c \in M_\text{collab}:\quad
\max_{t \in T} \rho_{m_p, m_c, t} - \min_{t \in T} \rho_{m_p, m_c, t} \leq \delta_\rho \\
& \forall m_p \in M_\text{prod}:\quad
\max_{t \in T} \beta_{m_p, \text{TS}, t} - \min_{t \in T} \beta_{m_p, \text{TS}, t} \leq \delta_\beta
\end{aligned}
\end{equation*}

To evaluate $H_{\text{aR}}$, we construct 95\% bootstrap CIs for each $\rho_{m_p, m_c, t}$ and $\beta_{m_p,\text{TS},t}$ using $B = 10{,}000$ resamples of project and time window observations. We consider results as \emph{not stable} if the deviation across tools, either in the upper bounds or in the lower bounds of the CIs, exceeds the respective threshold $\delta_\rho$ or $\delta_\beta$ for at least one metric combination.

\paragraph{Conclusion stability:} Following Gote~\etal, we consider the verdict of Brooks' law supported by an individual tool if the replicated correlation coefficients $\rho_{m_p,\text{TS},t}$ and linear regression coefficients $\beta_{m_p,\text{TS},t}$ between each productivity metric and team size are consistently negative:

\begin{equation*}
V_t = 
\begin{cases}
1 & \text{if } 
\begin{aligned}[t]
& 
\forall m_p \in M_\text{prod},\;
\rho_{m_p,\text{TS},t} < 0
\quad \text{and}\quad
\beta_{m_p,\text{TS},t} < 0
\end{aligned}
\\
0 & \text{otherwise}
\end{cases}
\end{equation*}

Conclusions are therefore considered \emph{stable} across tools if all of their verdicts agree unanimously, and \emph{not stable} if any tool yields $V_t = 0$:
\begin{equation*}
H_{\text{aC}}: \quad \forall\, t \in T :\quad V_t = 1
\end{equation*}

\begin{figure}[htbp]
  \centering
  \begin{subfigure}[b]{0.8\textwidth}
    \includegraphics[width=\linewidth, trim=0cm 0.8cm 0cm 0.5cm, clip]{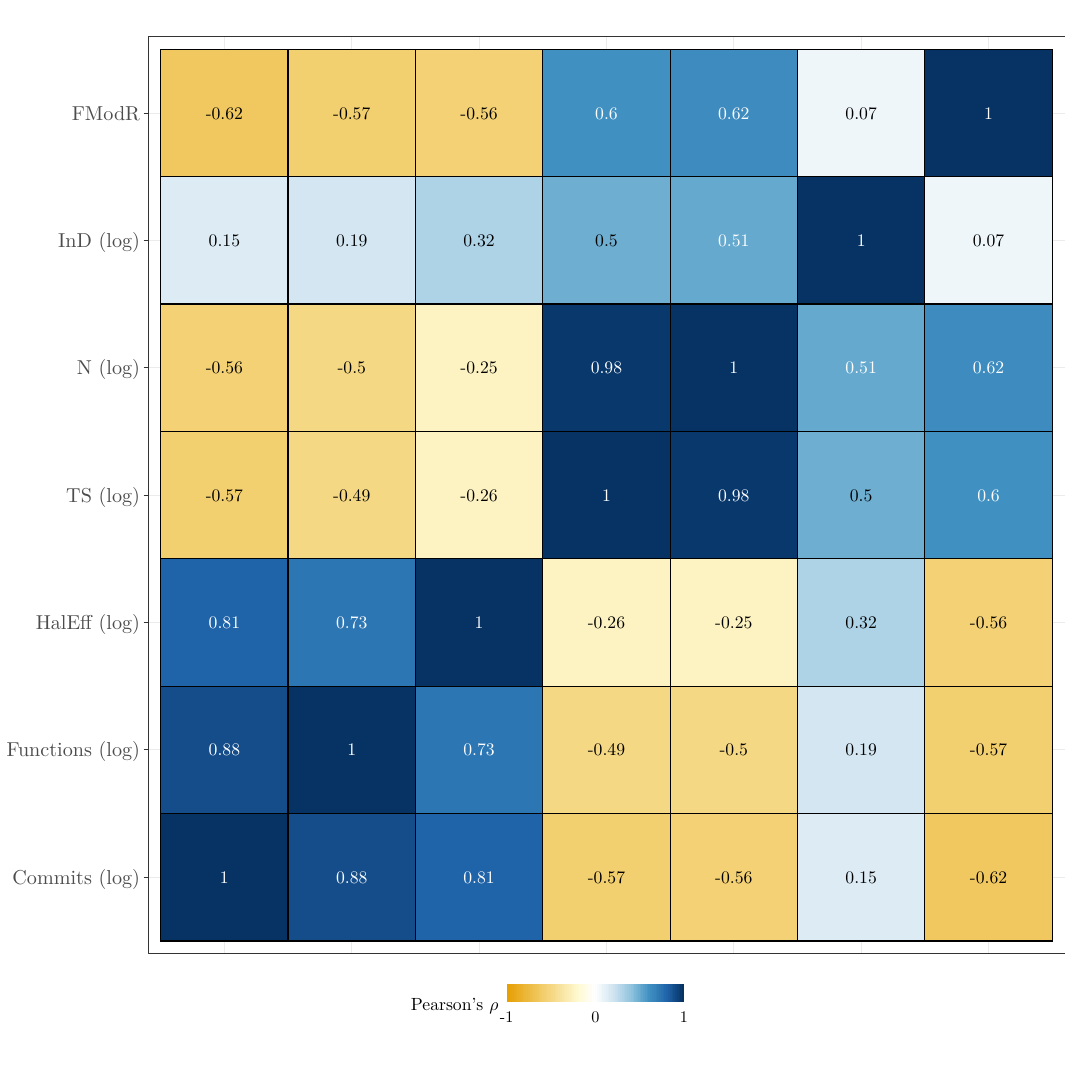}
    \includegraphics[width=\linewidth]{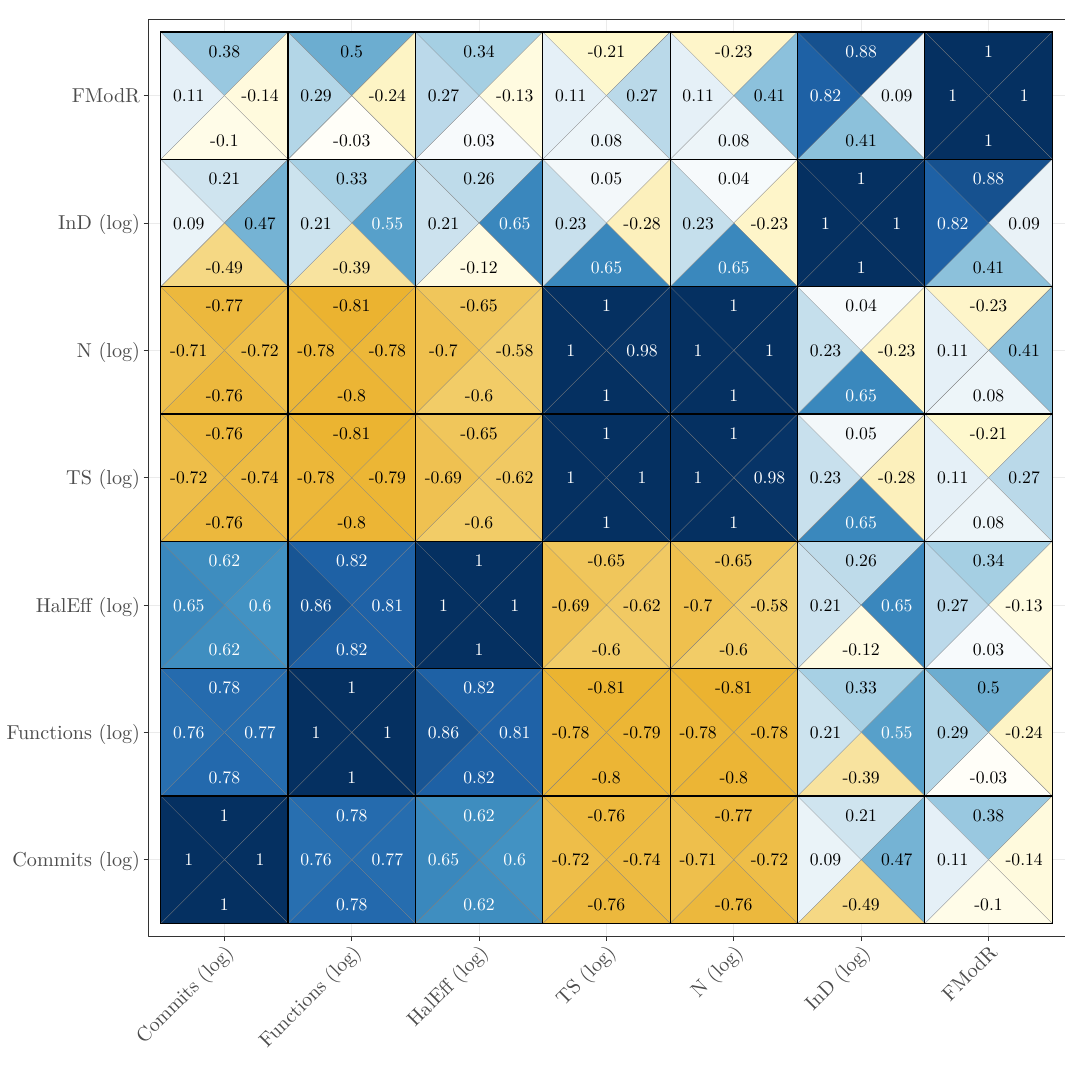}
  \end{subfigure}
  \caption{Pearson correlation between the transformed productivity and network metrics, calculated by the original study (upper plot) and by the replication (lower plot) using data extracted by \textsc{Codeface} (top tile), \textsc{git2net} (right tile), \textsc{GrimoireLab} (bottom tile) and \textsc{Kaiaulu} (left tile).}
  \label{fig:corr-combined}
\end{figure}

\paragraph{Correlation between productivity and collaboration metrics:} Figure \ref{fig:corr-combined} shows the Pearson correlations $\rho_{m_p,m_c,t}$ between the transformed productivity metrics $M_{\text{prod}}$, consisting of the number of commits, functions and Halstead effort, and the collaboration metrics $M_{\text{collab}}$, including team size, number of nodes in the developer network graph, mean in-degree of developers and mean foreign modification ratio (FModR).

The upper plot is computed directly from the aggregated data in the original reproduction package, restricted to the reduced set of projects described above. Despite this reduction, the observed values are consistent with the results obtained for the full original data. Most importantly, the negative correlation between team size and each of the productivity metrics is preserved.

The lower plot shows the correlations calculated on data extracted by the tools \textsc{Codeface}, \textsc{git2net}, \textsc{GrimoireLab} and \textsc{Kaiaulu}. Across all four tools, we observe the central negative correlation between the team size and the productivity metrics. The magnitudes are also closely aligned. For instance, the correlation between team size and number of commits varies by at most 0.04 across tools, corresponding to $2\%$ of the coefficient's possible range.

However, the network metrics used as control variables in the original study show differences. For instance, the correlation between productivity and mean in-degree of developers is positive for \textsc{Codeface}, \textsc{git2net} and \textsc{Kaiaulu}, but negative for \textsc{GrimoireLab}. This stems from how each tool constructs developer networks. \textsc{Codeface} and \textsc{Kaiaulu} aggregate directed edges between developers according to a weight scheme, thus yielding similar results. \textsc{git2net} also constructs directed networks, but, as shown in Section~\ref{sec:data-comparison}, retains multiple edges, causing stronger correlation. \textsc{GrimoireLab} constructs undirected networks, where in-degree cannot be computed. As a substitute, we calculate the \emph{total} node degree, which merges in- and out-degree into a single measure, likely explaining the inverted sign. 

\begin{figure}[htbp]
  \centering
  \includegraphics[width=\textwidth]{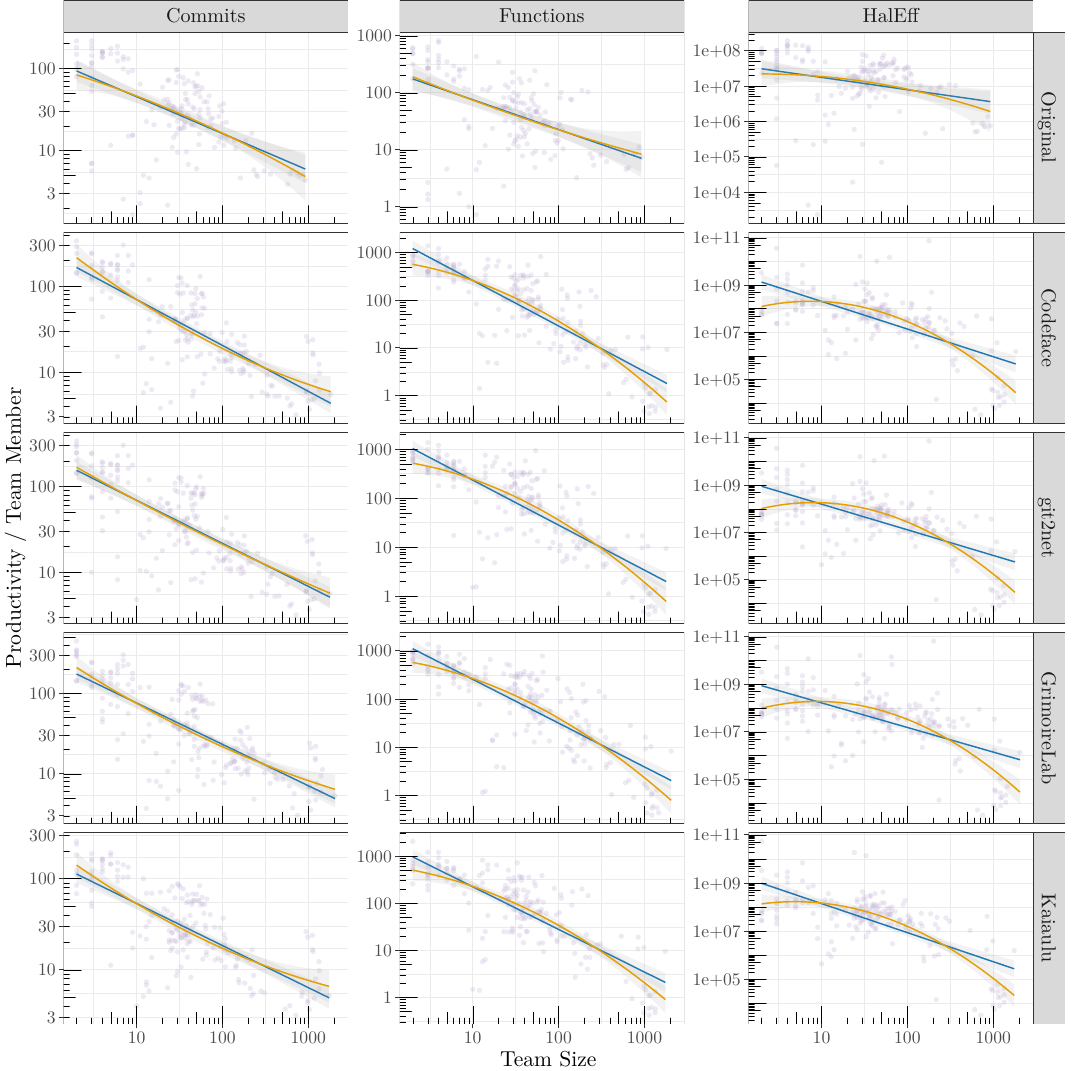}
  \caption{Productivity per team member as a function of team size estimated by linear (blue) and quadratic (yellow) regression models. The models are fit on log-transformed metrics and the axes are log-scaled, as in the original study. Each panel shows the models for the original dataset or one of the four replication tools \textsc{Codeface}, \textsc{git2net}, \textsc{GrimoireLab} and \textsc{Kaiaulu}.}
  \label{fig:regression-plot}
\end{figure}

\begin{table}
\centering
\caption{Regression models fitted to data from the original study and with the
         replicated data from the four mining tools. Each table column shows a separate model with the productivity metric as target variable and the team size
         (TS) as covariate. Yellow indicates a negative regression coefficient, blue indicates a positive coefficient. The intensity shows the coefficient magnitude. Standard errors are shown in parentheses. 
               \label{tab:regression-models}}
\centering
\resizebox{\ifdim\width>\linewidth\linewidth\else\width\fi}{!}{
\begin{tabular}[t]{c|cccc|cccc}
\toprule
& \multicolumn{4}{c|}{\textbf{Linear relationship}} & \multicolumn{4}{c}{\textbf{Quadratic relationship}}\\
\cmidrule(){2-9}
Tool & Term & Commits & Functions & HalEff & Term & Commits & Functions & HalEff \\
\midrule
\multirow{8}{*}{\rotatebox[origin=c]{90}{Original}} 
& (IC) & $4.84^{}$ & $5.53^{}$ & $17.48^{}$ & (IC) & $4.64^{}$ & $5.69^{}$ & $16.87^{}$\\
& & $(0.19)$ & $(0.27)$ & $(0.38)$ & & $(0.32)$ & $(0.45)$ & $(0.64)$\\
& TS &  \colorcell{-0.45}{-0.45^{}} & \colorcell{-0.52}{-0.52^{}} & \colorcell{-0.35}{-0.35^{}} & TS & \colorcell{-0.30}{-0.30^{}} & \colorcell{-0.64}{-0.64^{}} & \colorcell{0.09}{0.09^{}}\\
& & $(0.05)$ & $(0.07)$ & $(0.11)$ & & $(0.20)$ & $(0.28)$ & $(0.39)$\\
& & & & & TS$^{2}$ & \colorcell{-0.02}{-0.02^{}} & \colorcell{0.02}{0.02^{}} & \colorcell{-0.07}{-0.07^{}}\\
& & & & & & $(0.03)$ & $(0.04)$ & $(0.06)$\\
\cmidrule(){2-9}
& R$^{2}$ & 0.32 & 0.24 & 0.07 & R$^{2}$ & 0.32 & 0.24 & 0.07\\
& Adj. R$^{2}$ & 0.32 & 0.24 & 0.06 & Adj. R$^{2}$ & 0.31 & 0.23 & 0.06\\
\midrule
\multirow{8}{*}{\rotatebox[origin=c]{90}{Codeface}} 
& (IC) & $5.48^{}$ & $7.75^{}$ & $21.82^{}$ & (IC) & $5.92^{}$ & $6.51^{}$ & $17.95^{}$\\
& & $(0.14)$ & $(0.22)$ & $(0.44)$ & & $(0.26)$ & $(0.39)$ & $(0.73)$\\
& TS & \colorcell{-0.54}{-0.54^{}} & \colorcell{-0.95}{-0.95^{}} & \colorcell{-1.17}{-1.17^{}} & TS & \colorcell{-0.80}{-0.80^{}} & \colorcell{-0.20}{-0.20^{}} & \colorcell{1.19}{1.19^{}}\\
& & $(0.03)$ & $(0.05)$ & $(0.10)$ &  & $(0.14)$ & $(0.21)$ & $(0.38)$\\
& & & & & TS$^{2}$ & \colorcell{0.03}{0.03^{}} & \colorcell{-0.09}{-0.09^{}} & \colorcell{-0.29}{-0.29^{}}\\
& & & & & & $(0.02)$ & $(0.02)$ & $(0.05)$\\
\cmidrule(){2-9}
& R$^{2}$ & 0.58 & 0.65 & 0.42 & R$^{2}$ & 0.59 & 0.68 & 0.53\\
& Adj. R$^{2}$ & 0.58 & 0.65 & 0.42 & Adj. R$^{2}$ & 0.59 & 0.68 & 0.52\\
\midrule
\multirow{8}{*}{\rotatebox[origin=c]{90}{git2net}}
& (IC) & $5.39^{}$ & $7.57^{}$ & $21.37^{}$ & (IC) & $5.52^{}$ & $6.42^{}$ & $17.72^{}$\\
& & $(0.14)$ & $(0.21)$ & $(0.42)$ &  & $(0.24)$ & $(0.37)$ & $(0.67)$\\
& TS & \colorcell{-0.50}{-0.50^{}} & \colorcell{-0.92}{-0.92^{}} & \colorcell{-1.08}{-1.08^{}} & TS & \colorcell{-0.59}{-0.59^{}} & \colorcell{-0.18}{-0.18^{}} & \colorcell{1.25}{1.25^{}}\\
& & $(0.03)$ & $(0.05)$ & $(0.10)$ & & $(0.13)$ & $(0.20)$ & $(0.37)$\\
& & & & & TS$^{2}$ & \colorcell{0.01}{0.01^{}} & \colorcell{-0.09}{-0.09^{}} & \colorcell{-0.30}{-0.30^{}}\\
& & & & & & $(0.02)$ & $(0.03)$ & $(0.05)$\\
\cmidrule(){2-9}
& R$^{2}$ & 0.55 & 0.63 & 0.38 & R$^{2}$ & 0.55 & 0.65 & 0.50\\
& Adj. R$^{2}$ & 0.55 & 0.63 & 0.38 & Adj. R$^{2}$ & 0.55 & 0.65 & 0.50\\
\midrule
\multirow{8}{*}{\rotatebox[origin=c]{90}{GrimoireLab}}
& (IC) & $5.52^{}$ & $7.64^{}$ & $21.31^{}$ & (IC) & $5.83^{}$ & $6.54^{}$ & $17.65^{}$\\
& & $(0.13)$ & $(0.21)$ & $(0.43)$ & & $(0.24)$ & $(0.37)$ & $(0.68)$\\
& TS & \colorcell{-0.52}{-0.52^{}} & \colorcell{-0.91}{-0.91^{}} & \colorcell{-1.03}{-1.03^{}} & TS & \colorcell{-0.71}{-0.71^{}} & \colorcell{-0.21}{-0.21^{}} & \colorcell{1.30}{1.30^{}}\\
& $(0.03)$ & $(0.05)$ & $(0.10)$ & & $(0.13)$ & $(0.20)$ & $(0.37)$\\
& & & & & TS$^{2}$ & \colorcell{0.03}{0.03^{}} & \colorcell{-0.09}{-0.09^{}} & \colorcell{-0.30}{-0.30^{}}\\
& & & & & & $(0.02)$ & $(0.02)$ & $(0.05)$\\
\cmidrule(){2-9}
& R$^{2}$ & 0.58 & 0.63 & 0.36 & R$^{2}$ & 0.59 & 0.66 & 0.48\\
& Adj. R$^{2}$ & 0.58 & 0.63 & 0.36 & Adj. R$^{2}$ & 0.59 & 0.65 & 0.47\\
\midrule
\multirow{8}{*}{\rotatebox[origin=c]{90}{Kaiaulu}}
& (IC) & $5.04^{}$ & $7.49^{}$ & $21.54^{}$ & (IC) & $5.42^{}$ & $6.42^{}$ & $18.29^{}$\\
& & $(0.13)$ & $(0.22)$ & $(0.38)$ & & $(0.24)$ & $(0.38)$ & $(0.61)$\\
& TS & \colorcell{-0.46}{-0.46^{}} & \colorcell{-0.90}{-0.90^{}} & \colorcell{-1.21}{-1.21^{}} & TS & \colorcell{-0.69}{-0.69^{}} & \colorcell{-0.24}{-0.24^{}} & \colorcell{0.82}{0.82^{}}\\
& & $(0.03)$ & $(0.05)$ & $(0.10)$ & & $(0.13)$ & $(0.20)$ & $(0.33)$\\
& & & & & TS$^{2}$ & \colorcell{0.03}{0.03^{}} & \colorcell{-0.09}{-0.09^{}} & \colorcell{-0.26}{-0.26^{}}\\
& & & & & & $(0.02)$ & $(0.02)$ & $(0.04)$\\
\cmidrule(){2-9}
& R$^{2}$ & 0.52 & 0.61 & 0.47 & R$^{2}$ & 0.53 & 0.63 & 0.57\\
& Adj. R$^{2}$ & 0.51 & 0.60 & 0.47 & Adj. R$^{2}$ & 0.52 & 0.63 & 0.57\\
\bottomrule
\end{tabular}}
\end{table}

\paragraph{Regression models relating team size to productivity:} Figure~\ref{fig:regression-plot} shows the linear and quadratic regression models fitted to the individual productivity metrics to investigate the effect of team size. Following the original study, the models are constructed separately for each productivity metric, for instance explaining the average difference in number of commits by team size ($\text{Commits}\sim\text{TS}$) in the linear model and $\text{Commits}\sim\text{TS}+\text{TS}^2$ in the quadratic model. Although the overall shape of the quadratic curve appears to differ slightly across tools, the negative relationship between team size and individual productivity holds across all combinations of datasets and productivity metrics.

The corresponding quantitative regression coefficients for the linear and quadratic models fitted to the reduced original dataset and to data extracted by the replication tools are shown in Table~\ref{tab:regression-models}. 
The linear models consistently report negative regression coefficients for team size, indicating that individual productivity in terms of difference in commits, functions or Halstead effort decreases as team size grows. 
Some quadratic model coefficients for our reduced original dataset differ from those reported for the full dataset. This is plausible due to the sensitivity of higher-order terms in our smaller sample. As this study focuses on cross-tool stability rather than conformance to the original study, what matters is that the signs of coefficients are consistent across tools, which holds throughout.

For clarity, the regression models including the network metrics as control variables are provided in the supplementary material. Here, the negative coefficient for \(TS\) remains consistent across all tools in the linear models, while coefficients for control variables vary, as expected due to the previously observed differences in the networks.

\paragraph{Stability assessment:} 
The bootstrap CIs for the Pearson correlations $\rho_{m_p,m_c,t}$ and the linear regression coefficients $\beta_{m_p,\text{TS},t}$ corroborate and quantify our qualitative findings: As shown in Table~\ref{tab:gote-result-stability}, the cross-tool CI bounds differ well beyond the threshold for the correlations involving mean in-degree ($\text{InD}$) and the foreign modification ratio ($\text{FModR}$), and for the regression coefficient $\beta_{\text{HalEff},\text{TS}}$, whereas the remaining coefficients stay within it. We therefore consider results \emph{not stable} across tools. 
Despite the varying magnitude, the sign of every $\rho_{m_p,\text{TS},t}$ and $\beta_{m_p,\text{TS},t}$ remains negative across all tools. Consequently, all tools yield the same verdict and we consider the conclusion \emph{stable}.

\begin{table}
\centering
\begin{minipage}[t]{0.55\textwidth}
  \centering
  \caption{Result stability: cross-tool range of the 95\% bootstrap CI bounds (absolute and percentage, relative to the $[-1,1]$ range for~$\rho$ and to the average of the respective bound across tools for~$\beta$). Deviations exceeding $\delta_\rho = \delta_\beta = 0.1$ are highlighted.\label{tab:gote-result-stability}}
  {\scriptsize
\setlength{\tabcolsep}{4pt}
\begin{tabular}{lrrrr}
\toprule
 & $\text{CI}_{\text{lo}}$ & \% & $\text{CI}_{\text{hi}}$ & \%\\
\midrule
$\rho_{\text{Commits},\text{N}}$ & 0.05 & (2\%) & 0.08 & (4\%)\\
$\rho_{\text{Commits},\text{InD}}$ & \cellcolor{lfd-lilac!60}{0.91} & \cellcolor{lfd-lilac!60}{(45\%)} & \cellcolor{lfd-lilac!60}{0.99} & \cellcolor{lfd-lilac!60}{(49\%)}\\
$\rho_{\text{Commits},\text{FModR}}$ & \cellcolor{lfd-lilac!60}{0.53} & \cellcolor{lfd-lilac!60}{(27\%)} & \cellcolor{lfd-lilac!60}{0.50} & \cellcolor{lfd-lilac!60}{(25\%)}\\
$\rho_{\text{Functions},\text{N}}$ & 0.03 & (2\%) & 0.05 & (3\%)\\
$\rho_{\text{Functions},\text{InD}}$ & \cellcolor{lfd-lilac!60}{0.88} & \cellcolor{lfd-lilac!60}{(44\%)} & \cellcolor{lfd-lilac!60}{0.98} & \cellcolor{lfd-lilac!60}{(49\%)}\\
$\rho_{\text{Functions},\text{FModR}}$ & \cellcolor{lfd-lilac!60}{0.76} & \cellcolor{lfd-lilac!60}{(38\%)} & \cellcolor{lfd-lilac!60}{0.70} & \cellcolor{lfd-lilac!60}{(35\%)}\\
$\rho_{\text{HalEff},\text{N}}$ & \cellcolor{lfd-lilac!60}{0.10} & \cellcolor{lfd-lilac!60}{(5\%)} & \cellcolor{lfd-lilac!60}{0.13} & \cellcolor{lfd-lilac!60}{(6\%)}\\
$\rho_{\text{HalEff},\text{InD}}$ & \cellcolor{lfd-lilac!60}{0.75} & \cellcolor{lfd-lilac!60}{(38\%)} & \cellcolor{lfd-lilac!60}{0.79} & \cellcolor{lfd-lilac!60}{(39\%)}\\
$\rho_{\text{HalEff},\text{FModR}}$ & \cellcolor{lfd-lilac!60}{0.48} & \cellcolor{lfd-lilac!60}{(24\%)} & \cellcolor{lfd-lilac!60}{0.45} & \cellcolor{lfd-lilac!60}{(22\%)}\\
$\beta_{\text{Commits},\text{TS}}$ & 0.07 & (13\%) & 0.07 & (17\%)\\
$\beta_{\text{Functions},\text{TS}}$ & 0.05 & (5\%) & 0.06 & (8\%)\\
$\beta_{\text{HalEff},\text{TS}}$ & \cellcolor{lfd-lilac!60}{0.18} & \cellcolor{lfd-lilac!60}{(13\%)} & \cellcolor{lfd-lilac!60}{0.15} & \cellcolor{lfd-lilac!60}{(18\%)}\\
\bottomrule
\end{tabular}
}

\end{minipage}\hfill
\begin{minipage}[t]{0.42\textwidth}
  \centering
  \caption{Conclusion stability: per-tool Brooks' law verdicts $V_t$. A tool supports the verdict if all correlations $\rho_{m_p,\text{TS},t}$ and regression coefficients $\beta_{m_p,\text{TS},t}$ are negative.
  \label{tab:gote-conclusion-stability}}
  {\scriptsize
\setlength{\tabcolsep}{4pt}
\begin{tabular}{lcccc}
\toprule
 & \rotatebox{90}{Codeface} & \rotatebox{90}{git2net} & \rotatebox{90}{Grimoire} & \rotatebox{90}{Kaiaulu}\\
\midrule
$\rho_{\text{Commits},\text{TS}}$ & $<0$ & $<0$ & $<0$ & $<0$\\
$\beta_{\text{Commits},\text{TS}}$ & $<0$ & $<0$ & $<0$ & $<0$\\
$\rho_{\text{Functions},\text{TS}}$ & $<0$ & $<0$ & $<0$ & $<0$\\
$\beta_{\text{Functions},\text{TS}}$ & $<0$ & $<0$ & $<0$ & $<0$\\
$\rho_{\text{HalEff},\text{TS}}$ & $<0$ & $<0$ & $<0$ & $<0$\\
$\beta_{\text{HalEff},\text{TS}}$ & $<0$ & $<0$ & $<0$ & $<0$\\
\midrule
$V_t$ & 1 & 1 & 1 & 1\\
\bottomrule
\end{tabular}
}

\end{minipage}
\end{table}

\begin{answerbox}
\textbf{Answer to RQ3a:} 
The \emph{conclusion} that team size and productivity are negatively related is \emph{stable} across all tools, consistent with the central finding on Brooks' law reported by Gote~\etal~\citeyearpar{gote_big_2022}.
However, individual \emph{results} such as correlations and regression coefficients for some network metrics are \emph{not stable} across tools, indicating that other conclusions drawn from them would likely differ depending on the tool used.
\end{answerbox}

\subsection{Second replication: software quality}

The study ``Impact of Developer Turnover on Quality in Open-Source Software'' by Foucault~\etal~\citeyearpar{foucault_impact_2015} investigates the relationship between software quality and developer turnover, describing the phenomenon of developers newly joining or leaving a project in the context of open-source software. The authors modularise the source code, calculate multiple turnover rates and finally correlate them to the bug density per module. The results indicate that the role of turnover in open-source software projects differs from the one in industrial settings reported by prior studies. Based on the correlations, the authors conclude that external turnover at project level negatively impacts software quality in open-source contexts, while internal turnover is not problematic. We replicate the analyses required for this central finding to answer \emph{\textbf{RQ3b:} Is the relationship between turnover and software quality consistent across all tools, indicating that external turnover negatively impacts module quality?}

\paragraph{Reproducibility:} The original studies provide a reproduction package in the form of a self-contained VirtualBox image with all required data, scripts and a snapshot of the tool \textsc{Diggit}. Although the code to calculate confidence intervals was missing, we recovered it from an updated version of the authors' Git repository~\citep{foucault_rdeveloperturnover_2016}. We were able to re-run the entire analysis pipeline and reproduce the original results, up to negligible differences in the confidence intervals arising from non-fixed random seeds. In our replication component, we fixed seeds to ensure that differences in the intervals are attributable to the tools, as explained in Section~\ref{sec:replication-stability}. To rule out that our choice affects results, we evaluated multiple seeds and confirmed that the impact remains negligible.

\paragraph{Result stability:} As reasoned in Section~\ref{sec:replication-stability}, the following criteria illustrate our interpretative logic for assessing stability: We consider results across tools $t \in T$ as \emph{stable} if the maximum deviations in the bounds of the bootstrap confidence intervals $\text{CI}_{95}(\rho_{m,\text{BD},p,t})$ of Spearman correlations between bug density ($\text{BD}$) and turnover, for every project $p \in P$ and turnover metric $m \in M$, stay below a threshold $\delta_\rho = 0.1$:

\begin{equation*}
H_{\text{bR}}:
\forall p \in P\,
\forall m \in M: 
\bigwedge_{\xi \in \{\text{lo},\text{hi}\}} \max_{t \in T} \text{CI}_{95}^{\xi}(\rho_{m,\text{BD},p,t}) - \min_{t \in T} \text{CI}_{95}^{\xi}(\rho_{m,\text{BD},p,t}) \leq \delta_\rho
\end{equation*}

To evaluate $H_{\text{bR}}$ for Foucault et al., we directly use the 95\% bootstrap CIs with $B = 1{,}000$ resamples per tool, following the original methodology. We consider results as \emph{not stable} if the deviation across tools exceeds $\delta_\rho$ either in the upper bounds or in the lower bounds of the CIs for at least one combination of project and turnover metric.

\paragraph{Conclusion stability:} Following the interpretative logic of Foucault~\etal, we consider the verdict supported by an individual tool if the replicated bootstrap CIs for external newcomer activity $\text{ENA}$ and bug density $\text{BD}$ include only positive values for the majority of projects ($\lceil |P|/2 \rceil = 3$ of $|P| = 5$):

\begin{equation*}
V_t = 
\begin{cases}
1 & \text{if } \left|\{p \in P :\; \text{CI}_{95}(\rho_{\text{ENA}, \text{BD}, p, t}) \subset \mathbb{R}_{>0}\}\right| \geq \left\lceil |P|/2 \right\rceil \\
0 & \text{otherwise}
\end{cases}
\end{equation*}

We consider conclusions as \emph{stable} if all tools' verdicts agree unanimously, and \emph{not stable} if any tool yields $V_t = 0$:
\begin{equation*}
H_{\text{bC}}: \quad \forall\, t \in T :\quad V_t = 1
\end{equation*}

\begin{figure}[htbp]
  \centering
  \includegraphics[width=\linewidth]{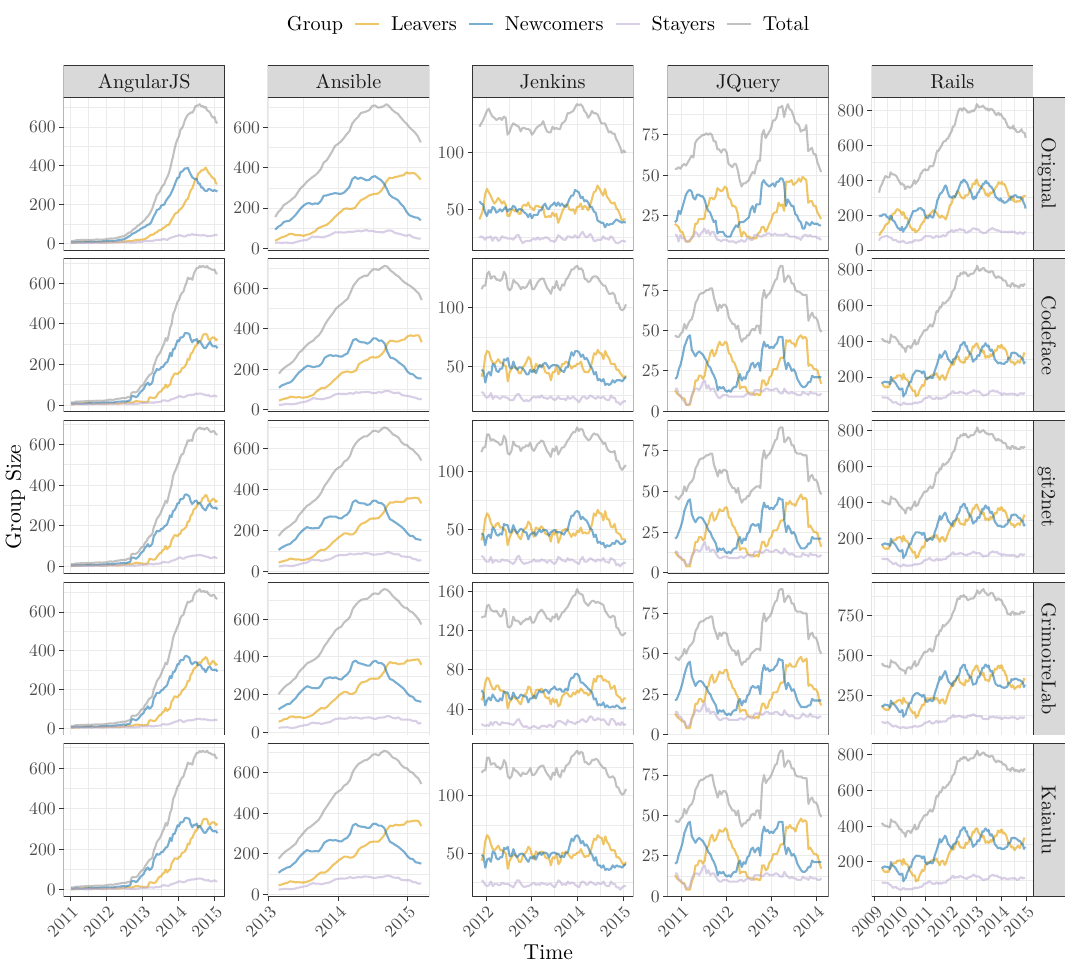}
  \caption{Evolution of developer turnover in the original study and the replications. The curves show the total number of developers (grey) and the numbers of stayers (lilac), external newcomers (blue), and external leavers (yellow).}
  \label{fig:turnover-evol}
\end{figure}

\paragraph{Relevance of developer turnover:} First, we analyse the relevance of turnover in open-source projects through time series evolution. Figure~\ref{fig:turnover-evol} compares the trends of contributions from external newcomers, external leavers, stayers and all developers across the time windows analysed of the original study and our replications. All time series are visually very similar, despite \textsc{Diggit} and \textsc{GrimoireLab} analysing all branches of a repository, while the remaining tools only consider the main branch. Marginal differences exist in the leavers and newcomers activity for Jenkins between 2014 and 2015. As in the original study, we find that  at least 80\% of developers are either newcomers or leavers throughout the project history.

\begin{figure}
  \centering
  \includegraphics[width=\linewidth]{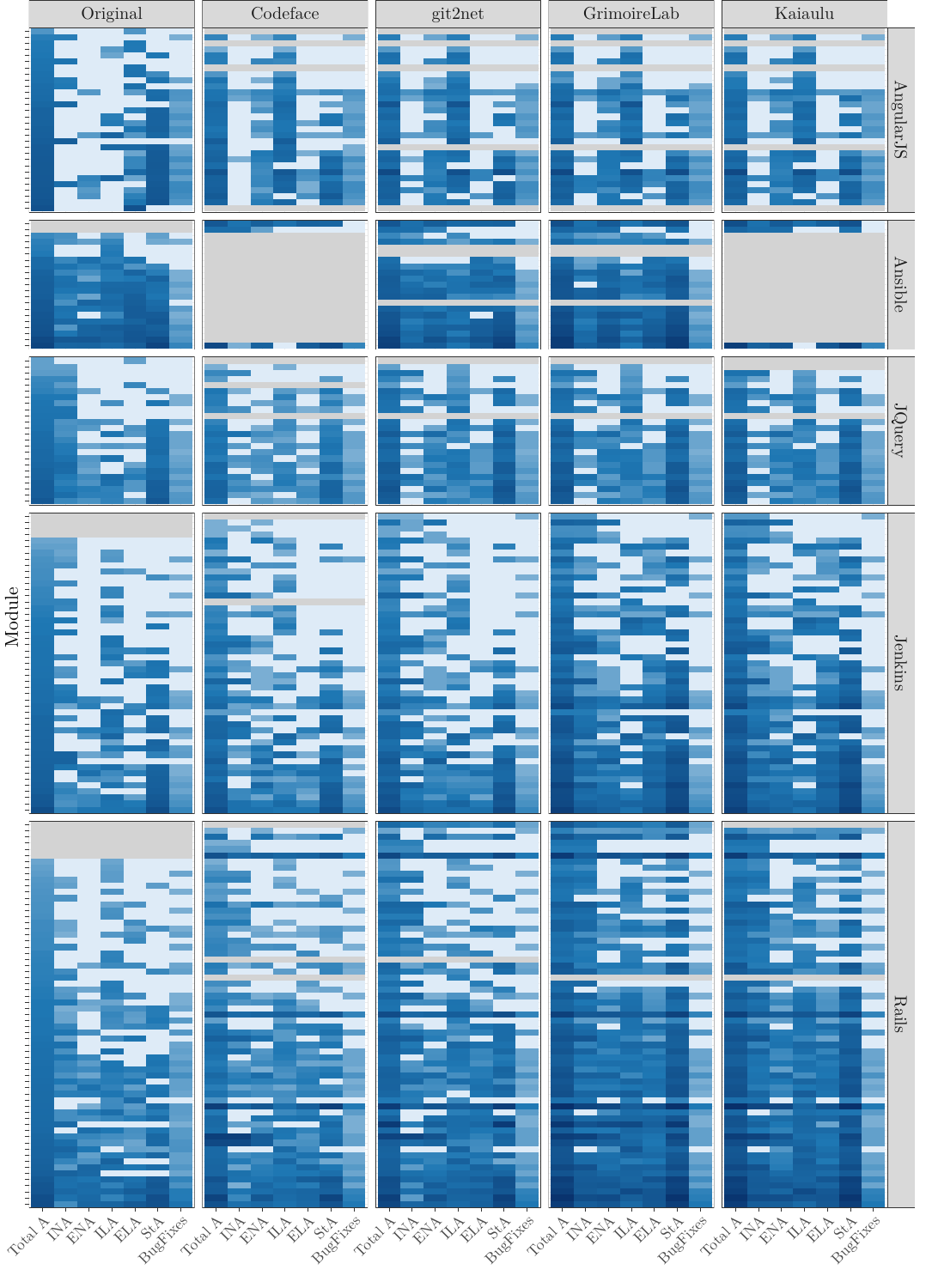}
  \caption{Visualization of developer groups' activity and the quantity of bugfixes for each module identified by the original study and the four replication tools \textsc{Codeface}, \textsc{git2net}, \textsc{GrimoireLab} and \textsc{Kaiaulu}. Each horizontal line of blocks represents a module. The darker the colour, the higher the metric value. Grey lines indicate that the respective module was not detected by a tool.}
  \label{fig:module-bug-patterns}
\end{figure}

\paragraph{Patterns of activity and bug fixes:} Figure~\ref{fig:module-bug-patterns} shows the activity of external and internal newcomers (ENA, INA), leavers (ELA, ILA), stayers (StA) and the number of bugs per module in the six months before and after the release chosen by the authors for each project. Modules are determined by matching files identified by each tool via regular expressions. Grey modules indicate that no file in the module was detected by a certain tool, for instance due to differences in raw data parsing. The most prominent discrepancy can be observed for subject project Ansible. A large majority of modules are missed by the tools \textsc{Codeface} and \textsc{Kaiaulu}, while \textsc{Diggit} (original), \textsc{git2net} and \textsc{GrimoireLab} found them. Manually inspecting the missed modules reveals that the repository contains several Python code files without the appropriate file ending \texttt{.py}. Since \textsc{Codeface} and \textsc{Kaiaulu} both filter files for analysis by their extensions, these could not be detected.

Another surprising aspect is the relatively low number of modules in the Ansible project. When inspecting the data, we note that several modules identified in the manual clustering by Foucault~\etal were not matched, both in the original and our replication results. The code reveals to which module a file belongs. The original implementation traverses a list of regular expressions and groups a file based on the first match. In Ansible, superordinate modules appeared first in the list, so more fine-grained modules were skipped. As this behaviour corresponds to the original implementation of \textsc{Diggit}, we keep it to maintain comparability across tools and results. Other modules are detected by the replication tools while being absent in the original results. This occurs for files that \textsc{Diggit} uses a more complex regular expression, which the other tools cannot replicate given their configuration options.

In general, the developer groups' activity found for each tool corresponds to that observed in the original study. However, comparing specific modules across tools reveals differences. For instance, \textsc{GrimoireLab} and \textsc{Kaiaulu} yield a higher total activity in all projects and notably higher external newcomers and stayers activity in Jenkins and Rails. More strikingly, the original study states that no module was changed exclusively by external newcomers, indicating that supervision by experienced developers always took place. With \textsc{Codeface} and \textsc{git2net}, this does not hold for the second module in subject project Jenkins, where contributions come only from external newcomers.

Figure~\ref{fig:module-bug-patterns} further shows the number of bug fixes identified by each tool for each module. Although we rely on Foucault~\etal's manually identified list of bug-fixing commits to detect affected files and modules, minor differences across tools remain due to the baseline data. In Jenkins, for instance, \textsc{Codeface} failed to identify certain bug-fixing commits or affected files. While minor in magnitude, these differences nonetheless affect the activity and bug density per module, propagating into the subsequent analysis.

\paragraph{Correlation between developer turnover and software module quality:} Table~\ref{tab:corr-table} shows the 95\% confidence intervals for Spearman's correlation between the developer groups' activity and bug density per module for all tools. Bug density is normalised by the number of lines of code measured by \textsc{cloc}~\citep{danial_cloc_2025} across all files of a module. As in the original study, this includes files recognised by \textsc{cloc} but not by the mining tool. As a further check, we recomputed lines of code to only include files actually identified by the mining tool. Despite differences in the line counts, impact on the overall result was minor, so we report it only in the supplementary material. 
The confidence intervals in Table~\ref{tab:corr-table} are wide for all turnover metrics and tools. For \textsc{Codeface} and \textsc{Kaiaulu}, the low number of identified bug fixes
and modules in Ansible prevented meaningful bootstrapping, causing the missing values. Other correlations differ substantially across tools: the confidence intervals for external leaver activity and bug density in Jenkins lie near zero for \textsc{Diggit} (original), \textsc{GrimoireLab} and \textsc{Kaiaulu}, but are clearly positive for \textsc{Codeface} and \textsc{git2net}.

\begin{table}
\centering
\caption{Spearman correlation coefficients between turnover metrics and
         bug density (bug-fixing commits normalised by code size) per module and subject project.
         Confidence intervals are computed per tool data set using bootstrap.
         Turnover metrics include internal newcomers activity (INA),
         internal leavers activity (ILA), external newcomers activity (ENA), 
         external leavers activity (ELA), stayers activity (StA) and total activity of all developers (A). Coloured cells indicate correlations meeting the \emph{significance} criterion defined by Foucault~\etal, for the original study (grey) and in the replications with Codeface (yellow), git2net (green), GrimoireLab (lilac) and Kaiaulu (blue).
         \label{tab:corr-table}}
\centering
\resizebox{\ifdim\width>\linewidth\linewidth\else\width\fi}{!}{
\begin{tabular}[t]{clcccccc}
\toprule
& \textbf{Tool} & \textbf{INA} & \textbf{ILA} & \textbf{ENA} & \textbf{ELA} & \textbf{StA} & \textbf{Total A}\\
\midrule
\multirow{5}{*}{\rotatebox[origin=c]{90}{AngularJS}} & Original & {}[-0.44,  0.38 ] & {}[-0.36,  0.30 ] & \cellcolor{lfd-gray!50}{[ 0.11,  0.68 ]} & {}[-0.56,  0.20 ] & \cellcolor{lfd-gray!50}{[ 0.16,  0.83 ]} & {}[-0.14,  0.69 ]\\
 & Codeface & {}[-0.04,  0.62 ] & {}[-0.27,  0.47 ] & \cellcolor{lfd-yellow!50}{[ 0.01,  0.72 ]} & \cellcolor{lfd-yellow!50}{[ 0.06,  0.69 ]} & \cellcolor{lfd-yellow!50}{[ 0.12,  0.81 ]} & {}[-0.21,  0.69 ]\\
 & git2net & \cellcolor{lfd-green!50}{[ 0.02,  0.65 ]} & {}[-0.41,  0.40 ] & \cellcolor{lfd-green!50}{[ 0.15,  0.82 ]} & {}[-0.09,  0.52 ] & \cellcolor{lfd-green!50}{[ 0.11,  0.77 ]} & {}[-0.29,  0.64 ]\\
 & Grimoire & \cellcolor{lfd-lilac!60}{[ 0.06,  0.72 ]} & {}[-0.39,  0.41 ] & \cellcolor{lfd-lilac!60}{[ 0.10,  0.81 ]} & \cellcolor{lfd-lilac!60}{[ 0.01,  0.61 ]} & \cellcolor{lfd-lilac!60}{[ 0.06,  0.84 ]} & {}[-0.22,  0.68 ]\\
 & Kaiaulu & {}[-0.01,  0.65 ] & {}[-0.38,  0.43 ] & \cellcolor{lfd-blue!50}{[ 0.09,  0.81 ]} & {}[-0.08,  0.53 ] & \cellcolor{lfd-blue!50}{[ 0.04,  0.87 ]} & {}[-0.15,  0.72 ]\\
\midrule
\multirow{5}{*}{\rotatebox[origin=c]{90}{Ansible}} & Original & {}[-0.23,  0.73 ] & {}[-0.34,  0.64 ] & {}[-0.21,  0.67 ] & {}[-0.29,  0.72 ] & {}[-0.22,  0.75 ] & {}[-0.33,  0.73 ]\\
 & Codeface & -- & -- & -- & -- & -- & --\\
 & git2net & {}[-0.15,  0.75 ] & {}[-0.37,  0.66 ] & {}[-0.42,  0.68 ] & {}[-0.32,  0.70 ] & {}[-0.32,  0.69 ] & {}[-0.38,  0.73 ]\\
 & Grimoire & {}[-0.37,  0.71 ] & {}[-0.29,  0.64 ] & {}[-0.37,  0.68 ] & {}[-0.33,  0.69 ] & {}[-0.28,  0.72 ] & {}[-0.40,  0.72 ]\\
 & Kaiaulu & -- & -- & -- & -- & -- & --\\
\midrule
\multirow{5}{*}{\rotatebox[origin=c]{90}{Jenkins}} & Original & {}[-0.28,  0.30 ] & {}[-0.17,  0.42 ] & \cellcolor{lfd-gray!50}{[ 0.27,  0.74 ]} & {}[-0.07,  0.49 ] & \cellcolor{lfd-gray!50}{[ 0.08,  0.65 ]} & \cellcolor{lfd-gray!50}{[ 0.01,  0.63 ]}\\
 & Codeface & \cellcolor{lfd-yellow!50}{[ 0.07,  0.59 ]} & {}[-0.20,  0.43 ] & \cellcolor{lfd-yellow!50}{[ 0.02,  0.55 ]} & \cellcolor{lfd-yellow!50}{[ 0.40,  0.88 ]} & \cellcolor{lfd-yellow!50}{[ 0.30,  0.72 ]} & \cellcolor{lfd-yellow!50}{[ 0.37,  0.70 ]}\\
 & git2net & \cellcolor{lfd-green!50}{[ 0.14,  0.66 ]} & {}[-0.29,  0.32 ] & {}[-0.03,  0.55 ] & \cellcolor{lfd-green!50}{[ 0.30,  0.77 ]} & \cellcolor{lfd-green!50}{[ 0.07,  0.63 ]} & \cellcolor{lfd-green!50}{[ 0.13,  0.64 ]}\\
 & Grimoire & \cellcolor{lfd-lilac!60}{[ 0.05,  0.55 ]} & {}[-0.20,  0.38 ] & {}[-0.10,  0.55 ] & {}[-0.03,  0.55 ] & \cellcolor{lfd-lilac!60}{[ 0.08,  0.61 ]} & \cellcolor{lfd-lilac!60}{[ 0.13,  0.62 ]}\\
 & Kaiaulu & {}[-0.05,  0.50 ] & {}[-0.25,  0.33 ] & {}[-0.11,  0.55 ] & {}[-0.02,  0.55 ] & \cellcolor{lfd-blue!50}{[ 0.07,  0.62 ]} & \cellcolor{lfd-blue!50}{[ 0.07,  0.58 ]}\\
\midrule
\multirow{5}{*}{\rotatebox[origin=c]{90}{JQuery}} & Original & {}[-0.14,  0.67 ] & \cellcolor{lfd-gray!50}{[ 0.15,  0.81 ]} & \cellcolor{lfd-gray!50}{[ 0.02,  0.70 ]} & {}[-0.41,  0.46 ] & \cellcolor{lfd-gray!50}{[ 0.14,  0.85 ]} & \cellcolor{lfd-gray!50}{[ 0.07,  0.80 ]}\\
 & Codeface & {}[-0.59,  0.25 ] & {}[-0.19,  0.63 ] & {}[-0.24,  0.65 ] & {}[-0.52,  0.40 ] & {}[-0.23,  0.68 ] & {}[-0.27,  0.71 ]\\
 & git2net & {}[-0.53,  0.42 ] & \cellcolor{lfd-green!50}{[ 0.10,  0.75 ]} & {}[-0.11,  0.80 ] & {}[-0.28,  0.61 ] & {}[-0.20,  0.69 ] & {}[-0.12,  0.71 ]\\
 & Grimoire & {}[-0.44,  0.49 ] & \cellcolor{lfd-lilac!60}{[ 0.17,  0.80 ]} & {}[-0.06,  0.82 ] & {}[-0.31,  0.56 ] & {}[-0.23,  0.72 ] & {}[-0.15,  0.71 ]\\
 & Kaiaulu & {}[-0.55,  0.36 ] & \cellcolor{lfd-yellow!50}{[ 0.12,  0.81 ]} & {}[-0.16,  0.79 ] & {}[-0.42,  0.53 ] & {}[-0.35,  0.64 ] & {}[-0.27,  0.66 ]\\
\midrule
\multirow{5}{*}{\rotatebox[origin=c]{90}{Rails}} & Original & {}[-0.01,  0.53 ] & {}[-0.21,  0.30 ] & \cellcolor{lfd-gray!50}{[ 0.11,  0.58 ]} & {}[-0.18,  0.33 ] & \cellcolor{lfd-gray!50}{[ 0.16,  0.58 ]} & \cellcolor{lfd-gray!50}{[ 0.03,  0.55 ]}\\
 & Codeface & \cellcolor{lfd-yellow!50}{[ 0.04,  0.51 ]} & {}[-0.11,  0.42 ] & \cellcolor{lfd-yellow!50}{[ 0.10,  0.58 ]} & {}[-0.03,  0.49 ] & {}[-0.02,  0.47 ] & {}[-0.04,  0.50 ]\\
 & git2net & {}[-0.09,  0.39 ] & {}[-0.08,  0.44 ] & \cellcolor{lfd-green!50}{[ 0.14,  0.62 ]} & \cellcolor{lfd-green!50}{[ 0.06,  0.56 ]} & \cellcolor{lfd-green!50}{[ 0.10,  0.57 ]} & {}[-0.05,  0.49 ]\\
 & Grimoire & {}[-0.10,  0.40 ] & {}[-0.07,  0.46 ] & \cellcolor{lfd-lilac!60}{[ 0.08,  0.57 ]} & \cellcolor{lfd-lilac!60}{[ 0.09,  0.58 ]} & \cellcolor{lfd-lilac!60}{[ 0.09,  0.58 ]} & {}[-0.01,  0.51 ]\\
 & Kaiaulu & {}[-0.10,  0.39 ] & {}[-0.19,  0.40 ] & \cellcolor{lfd-blue!50}{[ 0.08,  0.55 ]} & \cellcolor{lfd-blue!50}{[ 0.09,  0.56 ]} & \cellcolor{lfd-blue!50}{[ 0.05,  0.55 ]} & {}[-0.02,  0.53 ]\\
\bottomrule
\end{tabular}}
\end{table}

\begin{table}[]
    \centering
    \caption{Result stability: Cross-tool range of the 95\% bootstrap CI bounds, $\max_{t \in T} \text{CI}_{95}^{b}(\rho_{m, \text{BD}, p, t}) - \min_{t \in T} \text{CI}_{95}^{b}(\rho_{m, \text{BD}, p, t})$ for $b \in \{\text{lo}, \text{hi}\}$, showing the larger of the two, for the Spearman correlations between each turnover metric and bug density. Deviations exceeding the threshold $\delta_\rho = 0.1$ are highlighted. Stability is only assessed where a CI could be calculated for all four tools.}
    \label{tab:foucault-result-stability}
    {\scriptsize

\begin{tabular}{lcccccc}
\toprule
\textbf{Project} & \textbf{INA} & \textbf{ILA} & \textbf{ENA} & \textbf{ELA} & \textbf{StA} & \textbf{Total A}\\
\midrule
AngularJS & \cellcolor{lfd-lilac!60}{0.10} & \cellcolor{lfd-lilac!60}{0.13} & \cellcolor{lfd-lilac!60}{0.14} & \cellcolor{lfd-lilac!60}{0.17} & 0.09 & \cellcolor{lfd-lilac!60}{0.14}\\
Ansible & -- & -- & -- & -- & -- & --\\
Jenkins & \cellcolor{lfd-lilac!60}{0.19} & \cellcolor{lfd-lilac!60}{0.11} & \cellcolor{lfd-lilac!60}{0.12} & \cellcolor{lfd-lilac!60}{0.43} & \cellcolor{lfd-lilac!60}{0.24} & \cellcolor{lfd-lilac!60}{0.30}\\
JQuery & \cellcolor{lfd-lilac!60}{0.24} & \cellcolor{lfd-lilac!60}{0.37} & \cellcolor{lfd-lilac!60}{0.18} & \cellcolor{lfd-lilac!60}{0.24} & \cellcolor{lfd-lilac!60}{0.14} & \cellcolor{lfd-lilac!60}{0.15}\\
Rails & \cellcolor{lfd-lilac!60}{0.13} & \cellcolor{lfd-lilac!60}{0.12} & 0.07 & \cellcolor{lfd-lilac!60}{0.12} & \cellcolor{lfd-lilac!60}{0.12} & 0.04\\
\bottomrule
\end{tabular}
}

\end{table}

\paragraph{Stability assessment:} 
The quantitative evaluation in Table~\ref{tab:foucault-result-stability} confirms that across nearly all project and metric combinations, the bounds of the bootstrap confidence intervals $\text{CI}_{95}(\rho_{m,\text{BD},p,t})$ vary between tools by more than $\delta_\rho$, most severely for Jenkins and JQuery. We therefore consider results \emph{not stable} across tools. 
For conclusion stability, cells fulfilling Foucault~\etal's significance criterion are highlighted in Table~\ref{tab:corr-table}. Only \textsc{Diggit} (original) and \textsc{Codeface} yield positive intervals for the majority of projects and support the verdict ($V_t=1$), whereas \textsc{git2net}, \textsc{GrimoireLab} and \textsc{Kaiaulu} indicate positive intervals for just two out of five projects, thus not supporting the verdict ($V_t=0$). Conclusions are therefore \emph{not stable} across tools either.

\begin{answerbox}
\textbf{Answer to RQ3b:}
The \emph{conclusion} that external turnover negatively impacts module quality is \emph{not stable} across all tools, and neither is the relation between turnover and quality in the underlying \emph{results}. 
This demonstrates that small differences in the baseline data can propagate through analyses and change even a central conclusion in complex scenarios.
\end{answerbox}

\subsection{Third replication: collaboration and coordination}

The study ``Classifying Developers into Core and Peripheral: An Empirical Study on Count and Network Metrics'' by Joblin~\etal~\citeyearpar{joblin_classifying_2017} evaluates the validity and agreement of established count-based metrics and novel network metrics to classify contributors into \emph{core} developers, responsible for coordination and major workload, and \emph{peripheral} developers as casual contributors. Their quantitative and qualitative results indicate that the level of agreement always exceeds random agreement, leading the authors to conclude that all proposed metrics are overall consistent. In addition, they find that core and peripheral developers exhibit different hierarchical positions in the network structure, which are consistent over time. From this study, we replicate two central analyses to answer \emph{\textbf{RQ3c:} Do count- and network-based core--peripheral operationalisations agree consistently across tools?}

\paragraph{Reproducibility:} The original study references a supplementary website linking to an archived version of the \textsc{Codeface} repository. Unfortunately, we could not determine which state of the repository was used. A manual code investigation alongside our replication showed that several functions changed in comparison to the replication version of \textsc{Codeface}. For instance, library versions and related function calls, time window splitting and commit parsing have been reworked multiple times, indicating that the behaviour from our \textsc{Codeface} instance differs from the one used in the original study. Although the possibility to calculate overlapping time windows is still implemented, we cannot verify that it still performs as intended. As the other tools provide no native support for overlapping time windows, we use non-overlapping windows throughout. The supplementary website also offers \emph{project data} to download. However, this consists of visualisations rather than the underlying raw data, which limits comparability with the original work. However, as explained in Section~\ref{sec:replication-stability}, this does not affect our stability evaluation across tools. 

\paragraph{Result stability:} As reasoned in Section~\ref{sec:replication-stability}, the following criteria illustrate our interpretative logic for assessing stability: We consider results across tools as \emph{stable} if pairwise Cohen's $\kappa_{m,p,t_a,t_b}$ for the actual developer classifications \emph{across tool pairs} $t_a, t_b \in T$ indicates at least substantial agreement (threshold $\delta_{\kappa,\text{R}}=0.61$) for all developer metrics $m \in M$, all projects $p \in P$, and all time windows $w \in W_p$:

\begin{equation*}
H_{\text{cR}}:
\forall m \in M,\;
\forall p \in P,\;
\forall w \in W_p,\;
\forall t_a, t_b \in T \text{ with } a \neq b:\\
\kappa_{m,p,w,t_a,t_b} \geq \delta_{\kappa,\text{R}}
\end{equation*}

The baseline data comparison in Section~\ref{sec:data-comparison} showed that the set of identified developers is not always consistent across tools in the first place. Therefore, we limit our stability evaluation to the group of developers that, based on their name, was found by all tools. As Cohen's $\kappa$ degenerates for very small sample sizes common in early time windows, but \emph{early} depends on the project, we further restrict our evaluation to time windows with at least $n_{\text{dev}}$ such developers.

We then evaluate $H_{\text{cR}}$ by constructing a 95\% bootstrap confidence interval (CI) for each $\kappa_{m,p,w,t_a,t_b}$ using $B = 10{,}000$ resamples of developers for each tuple $(m,p,w,t_a,t_b)$. We consider results as \emph{not stable} if the lower bound of the CI lies below $\delta_{\kappa,\text{R}}$ for at least one tuple. 

\paragraph{Conclusion stability:} Following Joblin~\etal, we consider the verdict as supported by a specific tool if the \emph{within-tool} pairwise agreement $\kappa_{m_c,m_n,p,w,t}$ between every count-based metric $m_c \in M_\text{count} \subset M$ and network-based metric $m_n \in M_\text{net} \subset M$ for core and peripheral developer classification exceeds random chance ($\delta_{\kappa,\text{C}}=0$) for the majority $\lceil k/2 \rceil$ of the $k$ metric, project and time window combinations $(m_c,m_n,p,w)$:

\begin{equation*}
V_t = 
\begin{cases}
1 & \text{if } \left| \{ (m_c,m_n,p,w) :\; \kappa_{m_c,m_n,p,w,t} >
    \delta_{\kappa,\text{C}} \} \right| \geq \left\lceil k/2 \right\rceil \\
0 & \text{otherwise}
\end{cases}
\end{equation*}

We then consider conclusions as \emph{stable} if all tools' verdicts agree unanimously, and \emph{not stable} if any tool yields $V_t = 0$:
\begin{equation*}
H_{\text{cC}}: \quad \forall\, t \in T :\quad V_t = 1
\end{equation*}

\paragraph{Agreement of count- and network-based metrics:} To study the agreement of developer metrics measured on VCS data, we aggregate the count-based metrics $M_\text{count}$ (LOC, commits) from the baseline data and compute the network-based metrics $M_\text{net}$ (node degree, eigenvector centrality, hierarchy centrality) from the adjacency matrices exported in a unified format for each tool.
Intuitively, core developers are expected to be more active and coordinate with more developers, thus showing higher counts, node degrees and centralities.

\begin{figure}[htbp]
  \centering
  \begin{subfigure}[b]{0.49\textwidth}
    \includegraphics[width=\linewidth]{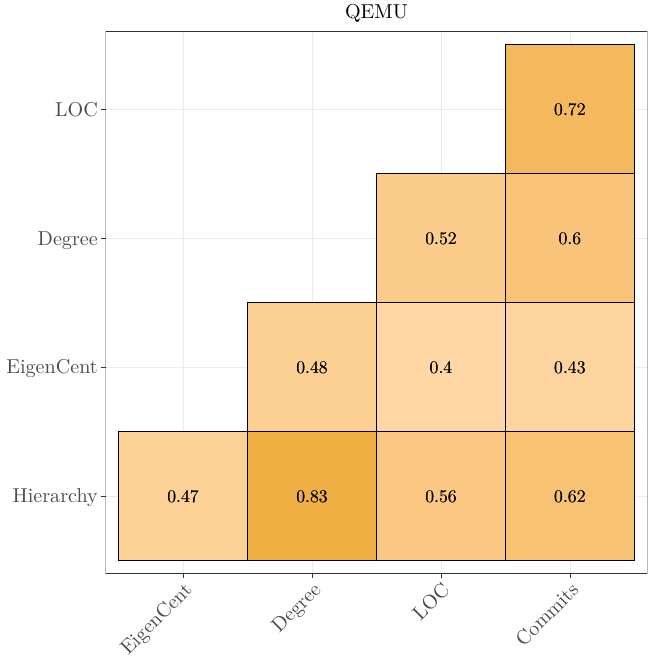}
  \end{subfigure}
  \begin{subfigure}[b]{0.49\textwidth}
    \includegraphics[width=\linewidth]{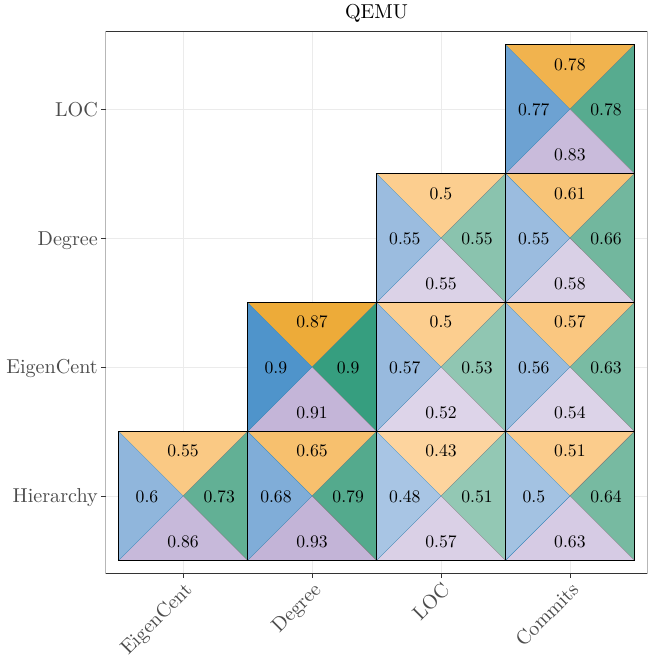}
  \end{subfigure}
  
  \vspace{1em}
  
  \begin{subfigure}[b]{0.49\textwidth}
    \includegraphics[width=\linewidth]{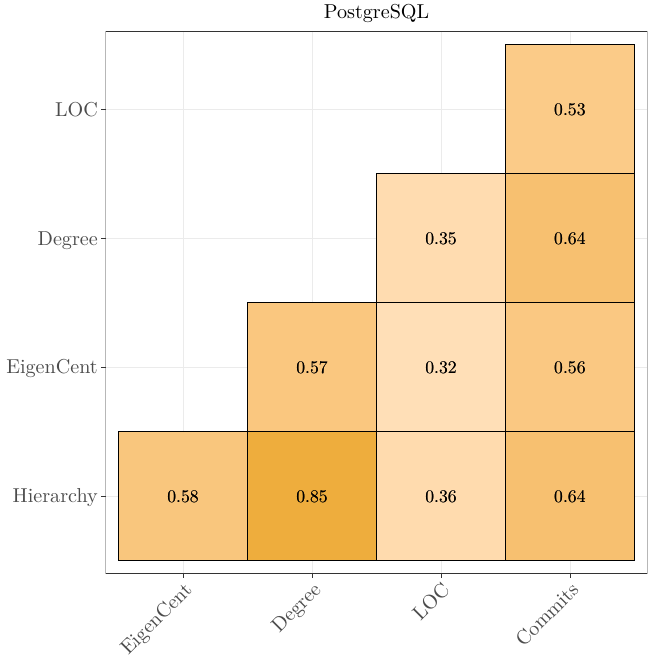}
  \end{subfigure}
  \begin{subfigure}[b]{0.49\textwidth}
    \includegraphics[width=\linewidth]{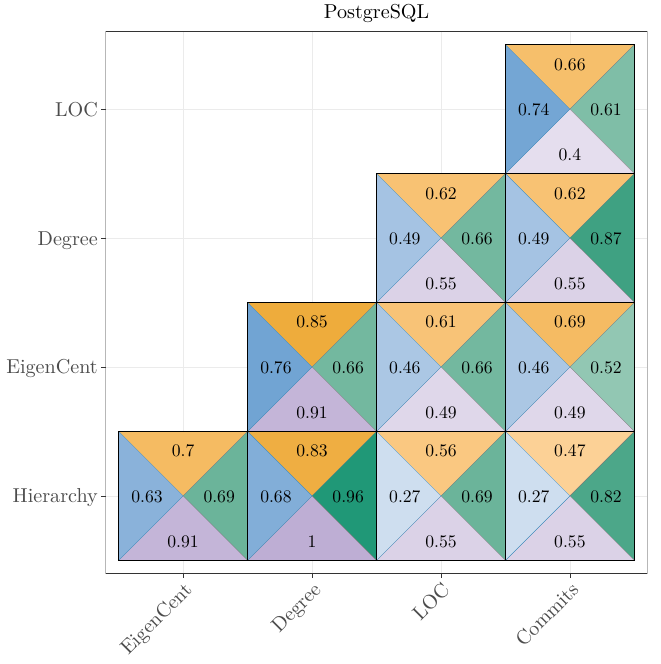}
  \end{subfigure}

  \caption{Time-averaged agreement in terms of Cohen's kappa for QEMU and PostgreSQL. The pairwise agreement is shown for the count-based metrics lines of code (LOC), number of commits and the network-based operationalisations node degree, eigenvector centrality and hierarchy centrality. The left column shows the agreement measured in the original study. The right column displays the agreement measured in the four replications using \textsc{Codeface} (yellow), \textsc{git2net} (green), \textsc{GrimoireLab} (lilac) and \textsc{Kaiaulu} (blue).}
  \label{fig:core-developer-agreement}
\end{figure}

Figure \ref{fig:core-developer-agreement} shows the metrics agreement for QEMU and PostgreSQL, with Cohen's $\kappa$ observed in the original study on the left and $\kappa_{m_c, m_n, p, t}$ observed in the replications with the same data aggregation method on the right.
In the original study, Joblin~\etal present the agreement averaged over one year of development but do not specify which; as their other analyses refer to the most recent development year, we present results for that interval.
As Joblin~\etal, we interpret the strength of agreement measured by $\kappa$ as follows: 0.81--1.00 almost perfect, 0.61--0.80 substantial, 0.41--0.60 moderate, 0.21--0.40 fair, 0.00--0.20 slight, and \(<\) 0.00 poor.

Although \textsc{Codeface} was used in the original and the replication study, we observe differences in the individual $\kappa$ values. As these differences likely stem from changes in our replication pipeline, which is applied identically to every tool, we accept this deviation from the original study in favour of highlighting tool-induced differences. For \textsc{git2net}, \textsc{GrimoireLab} and \textsc{Kaiaulu}, the strength of agreement mostly falls in the same class as \textsc{Codeface}, and at most one class apart. Hierarchy centrality, however, is a recurring exception. Most strikingly, the tools \textsc{git2net} and \textsc{Kaiaulu} disagree in PostgreSQL, where \textsc{git2net} yields almost perfect agreement between hierarchy centrality and commits, whereas \textsc{Kaiaulu} only finds fair agreement. Given that the tools follow different network construction techniques, it is surprising that all of them still indicate the validity of network-based metrics, as their agreement with count-based metrics exceeds random chance.

\begin{figure}[htbp]
  \centering
  \begin{subfigure}[b]{\textwidth}
    \includegraphics[width=\linewidth]{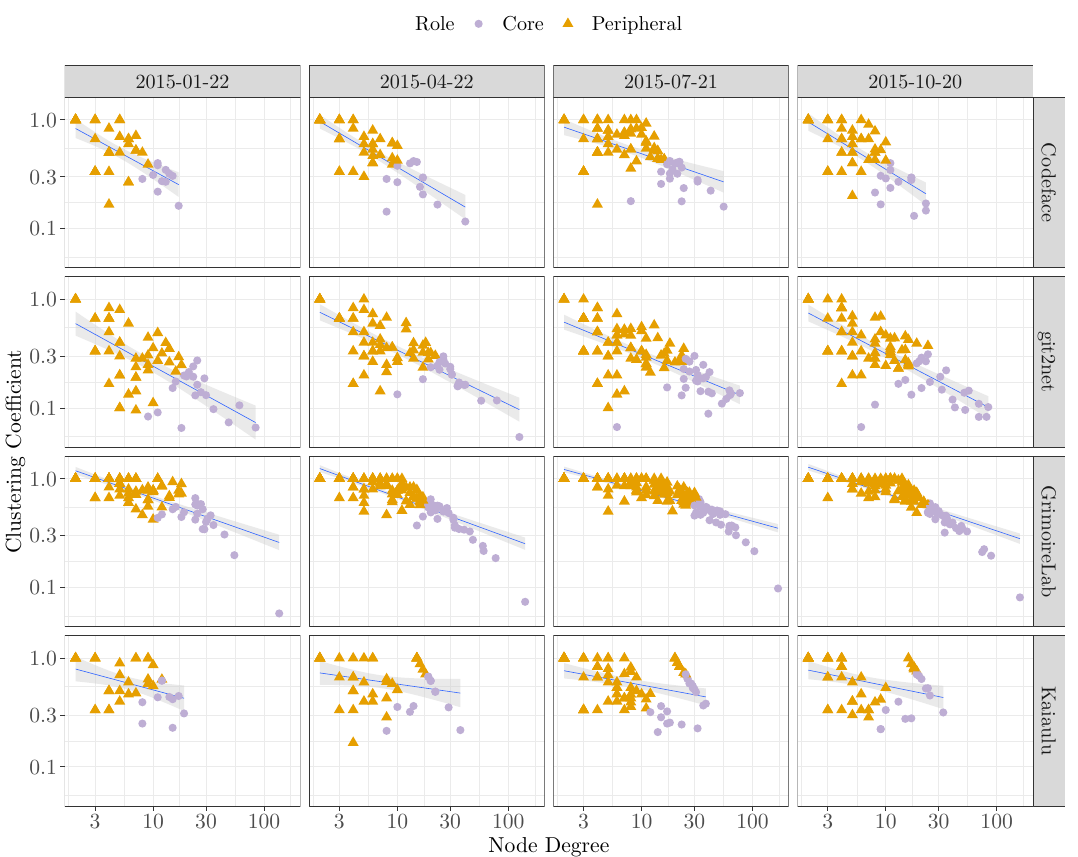}
  \end{subfigure}

  \vspace{1em}
  
  \begin{subfigure}[b]{\textwidth}
    \includegraphics[width=\linewidth]{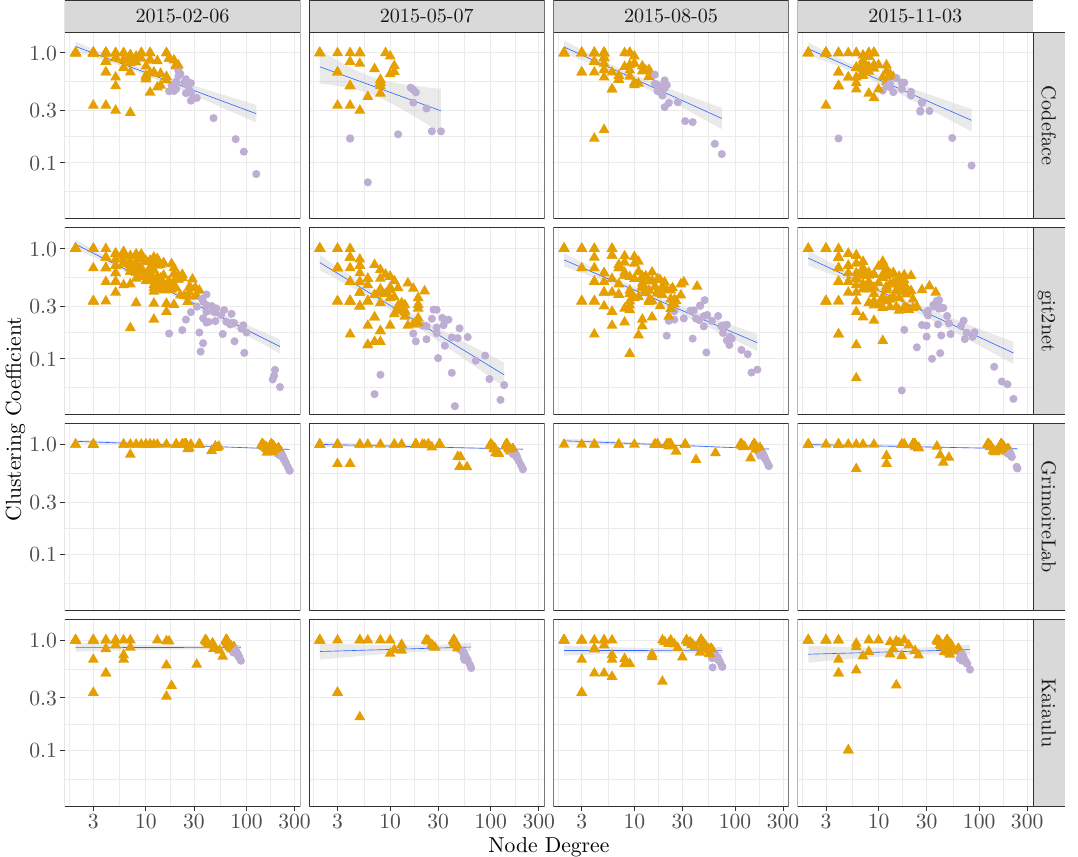}
  \end{subfigure}

  \caption{Hierarchy stability in projects QEMU (top graph) and GCC (bottom graph) during four development periods measured by \textsc{Codeface}, \textsc{git2net}, \textsc{GrimoireLab} and \textsc{Kaiaulu}. The linear dependence between clustering coefficient and degree expresses the hierarchy. In most cases, core developers appear clustered at the top of the hierarchy (bottom right region), while peripheral developers are clustered at the bottom of the hierarchy (upper left region).}
  \label{fig:hierarchy-stability}
\end{figure}

\paragraph{Hierarchical embedding:} Joblin~\etal further investigate the validity of the network-based operationalisations through relational differences between developers in the network hierarchy, which manifest as a mutual dependence between node degree and clustering coefficient. Intuitively, core developers are expected to occupy higher positions in the hierarchy. 

Figure~\ref{fig:hierarchy-stability} shows qualitative examples for QEMU and GCC. As we do not have the raw data from the original study, and since time window splitting in \textsc{Codeface} changed, we compare only the replication tools. 
For QEMU, we consistently see core developers with high node degree and low clustering coefficient clustered in the bottom-right region, and peripheral developers with lower node degree and high clustering coefficient clustered in the top-left region. The overall linear dependence between node degree and clustering coefficient appears to a similar degree with all tools, though slightly less pronounced with \textsc{Kaiaulu}. 
For GCC, trends differ more notably: While all tools yield a higher node degree of core developers, the difference in clustering coefficient is weaker with \textsc{GrimoireLab} and \textsc{Kaiaulu} than with \textsc{Codeface} and \textsc{git2net}. Several outliers even reverse the overall trend with \textsc{Kaiaulu} in some time intervals.

\paragraph{Stability assessment:} Table~\ref{tab:joblin-result-stability} presents the fraction of time windows fulfilling the result stability criterion $H_{\text{cR}}$, evaluated as the lower
bound of the bootstrap CI for $\kappa_{m,p,w,t_a,t_b}$ reaching
$\delta_{\kappa,\text{R}} = 0.61$ (substantial agreement), for several minimum developer-set sizes $n_{\text{dev}} \in \{2, 10, 20, 60\}$. For classifications based on network metrics, we observe at least substantial agreement in at most 13\% of time windows. Excluding small-sample windows ($n_{\text{dev}} = 60$) would be expected to increase the fraction, yet it decreases further to 3\%, indicating that the divergence stems from the tools' networks rather than from small-sample noise. The count-based metrics show at least substantial agreement in the majority of time windows, but not in \emph{all} of them. Therefore, we consider results \emph{not stable}.

Table~\ref{tab:joblin-conclusion-stability} shows that \emph{within} tool datasets, the level of agreement between count- and network-based metrics exceeds random chance in the large majority of time windows and projects. Our qualitative analysis shows a few negative values indicating agreement below chance mainly in the early years of development in subject projects including Django, GCC, LLVM, QEMU and U-Boot, likely due to the small developer set at this stage. As these form a minority, every tool yields $V_t = 1$ and we consider the conclusion \emph{stable} across tools.

\begin{table}[]
    \centering
    \caption{Result stability: fraction of time windows per metric in which the lower bound of the 95\% bootstrap CI for $\kappa_{m,p,w,t_a,t_b}$ reaches $\delta_{\kappa,\text{R}}$. Since Cohen's $\kappa$ can be degenerate for small numbers of developers common in early time windows, we show several thresholds for the minimum number of jointly identified developers $n_{\text{dev}}$ as a sensitivity analysis.}
    \label{tab:joblin-result-stability}
    {\scriptsize
\begin{tabular}{lcccc}
\toprule
\textbf{Metric} & \textbf{$n_{\text{dev}} \geq 2$} & \textbf{$n_{\text{dev}} \geq 10$} & \textbf{$n_{\text{dev}} \geq 20$} & \textbf{$n_{\text{dev}} \geq 60$}\\
\midrule
\multicolumn{5}{l}{\emph{Count-based}}\\
LOC & 0.59 & 0.57 & 0.59 & 0.62\\
Commits & 0.68 & 0.65 & 0.66 & 0.69\\
\midrule
\multicolumn{5}{l}{\emph{Network-based}}\\
Node degree & 0.13 & 0.06 & 0.03 & 0.03\\
Eigenvector & 0.12 & 0.05 & 0.03 & 0.03\\
Hierarchy & 0.12 & 0.06 & 0.03 & 0.04\\
\bottomrule
\end{tabular}
}

\end{table}

\begin{table}[]
    \centering
    \caption{Conclusion stability: per-tool verdicts $V_t$ on metrics agreement. A tool supports the verdict if the within-tool agreement $\kappa$ between count- and network-based metrics exceeds chance ($\delta_{\kappa,\text{C}} = 0$) for the majority of combinations. The number of assessed combinations $k$ differs
    slightly between tools because $\kappa$ cannot be computed for some combinations $(m_c,m_n,p,w)$.}
    \label{tab:joblin-conclusion-stability}
    {\scriptsize

\begin{tabular}{lcccc}
\toprule
 & \textbf{Codeface} & \textbf{git2net} & \textbf{GrimoireLab} & \textbf{Kaiaulu}\\
\midrule
$k$ & 2,405 & 2,310 & 2,437 & 2,317\\
$\lceil k/2 \rceil$ & 1,203 & 1,155 & 1,219 & 1,159\\
$\kappa > \delta_{\kappa,\text{C}}$ & 2,141 & 2,172 & 2,157 & 2,040\\
$(\kappa > \delta_{\kappa,\text{C}})/k$ & 0.89 & 0.94 & 0.89 & 0.88\\
\midrule
$V_t$ & 1 & 1 & 1 & 1\\
\bottomrule
\end{tabular}
}

\end{table}

\begin{answerbox}
\textbf{Answer to RQ3c:}
The \emph{conclusion} that count- and network-based metrics agree is reached by
every tool \emph{individually}, consistent with the original study. When comparing the actual developer classifications \emph{across} tools, however, agreement is substantially lower, indicating that \emph{results} are \emph{not stable}, even under a relaxed criterion. 
The network-based metrics, being more complex and further downstream in the pipeline, again prove more susceptible to tool-level differences than the simpler count-based metrics.
\end{answerbox}

\section{Discussion and lessons learned}
\label{discussion}

\paragraph{\textbf{Are the results of independent mining tools different?}} From RQ2 to RQ3a--c, we never observed \emph{exactly} matching metrics calculated with the same analysis algorithms applied to outputs of different tools. This is true for both the baseline data and subsequently derived metrics. Given the many implementation decisions along the tool pipeline, this is plausible and may be acceptable in some cases. However, under circumstances such as the existence of multiple active branches in a repository, we have seen tools indicate \emph{opposite} activity trends. The more fine-grained and downstream analyses are, the larger the differences in results become, as assumptions and uncertainties stack. 

\paragraph{\textbf{Are the differences critical?}} With differences in even the  most simple metrics, such as the number of commits, all of the discrepancies can impact decisions in practice, for instance by affecting  popular problems of monitoring open-source community health, its engagement and activity~\cite{claes_abnormal_2017, claes_towards_2018, izquierdo-cortazar_starting_2022, newton_leveraging_2023, robles_industrial_2024} or identifying the most suitable person to review a new code contribution~\citep{rahman_correct_2016, chen_code_2022}.

From a research perspective focusing on generalisable best practices and methods, this question cannot be answered conclusively. In two of three studies, conclusions remain consistent when switching tools. However, with this being the only change, our replications are as close to the original studies as possible, which often leads to confirming results~\citep{shepperd_role_2018}. While our replications show the \emph{minimal} impact of switching general-purpose mining tools, the impact when also switching subsequently applied tools is not evaluated, but expected to be even stronger. 

Although all three studies drew broad, relatively high-level conclusions, only the one investigating the more complex and uncertain relationships showed conclusion instability. This suggests that the more complex and uncertain a research question is, the more tool choice matters. With research questions becoming more and more complex and specific, we may not have a clear intuition on the direction of causes and effects, and instead rely heavily on data-driven approaches to validate possible assumptions. Data quality, in particular consistency and accuracy~\citep{sidi_data_2012}, should therefore be a major concern. Relatedly, research questions should therefore be scoped to the specific conditions the
chosen tools and data can actually satisfy, and claims should be restricted to the
contexts in which those conditions hold. Since tool development and application are highly relevant across empirical software engineering, as our literature review shows, many areas may face similar challenges, as exemplified by Lefever~\etal~\citeyearpar{lefever_lack_2021}. 

\paragraph{\textbf{What can be done about it?}} In our previous study, we addressed this question from a tool \emph{development} perspective and adjusted one tool to yield the same results as another. However, we also found a lack of standardisation, which is difficult to overcome, because there is often no clearly \emph{correct} implementation, as this highly depends on the specific study. The discrepancies we have identified are largely not due to bugs, but to different assumptions and decisions, all of which may be equally justified. Therefore, in Table~\ref{tab:lessons-learned}, we address this question by summarising our lessons learned and recommendations on aspects that should be considered during tool choice, empirical study design and practical application of general-purpose MSR tools. Several of these detection and handling mechanisms are already realised by mock data generators and test cases in~\textsc{Codeface} and~\textsc{Kaiaulu}, offering a reference for building similar checks into other tools.

\paragraph{\textbf{Future work}} The depth of technical detail behind the checks in Table~\ref{tab:lessons-learned} may give the impression that building a new, custom solution is easier than understanding and adjusting an existing tool and explain why many authors take this path. In many cases, however, it is preferable to build on established tools, whose authors have already encountered and addressed many of these pitfalls, rather than to reinvent the wheel. To support this, we suggest more dedicated studies~\citep{lefever_lack_2021, hoess_does_2025} analysing the effects of technical details on outcomes across mining tools with a shared purpose to help users trade off capabilities against customisation efforts. It often remains unclear which approach is best suited for a specific problem, for instance when constructing developer networks. Surveys with the target users of a method~\citep{joblin_developer_2015} can offer guidance. Relatedly, the community could curate a shared benchmark of mock repositories encoding the edge cases behind the checks in Table~\ref{tab:lessons-learned}, so that tool authors can validate their pipelines against a common ground truth instead of carefully constructing test cases themselves.

\begin{longtblr}[
  caption = {Checklist of threats related to MSR tools, organised by pipeline stage from Figure~\ref{fig:pipeline}. For each threat, we propose a \emph{white box} inspecting a tool's technical artefacts, a \emph{black box} check relying only on results, and handling strategies.},
  label = {tab:lessons-learned},
]{
  colspec = {X[0.9,l] X[2,l] X[2,l] X[2,l] X[2.1,l]},
  cells = {font=\scriptsize},
  row{1} = {bg=white, fg=black, font=\scriptsize\bfseries},
  rowhead = 1
}
\toprule
Stage & Threat & Tool check (white box) & Result check (black box) & Handling \\
\midrule

Commit analysis (1) 
& \textbf{T1}: Tools process different repository branches (e.g. all remote branches vs. single checked-out branch), which can reverse activity trends &
Check branch configuration options; search documentation or source code for implicit defaults, e.g. by keywords \texttt{checkout, branch, remote} 
& Run tool on a \emph{local} repository with $\geq$ two branches; verify whether a known side-branch commit is captured in the output 
& Configure branch scope; disable remote fetching for reproducibility; if branches are missing, run tool on multiple checkouts and merge data (e.g. commits, files)\\ \hline

Commit analysis (1)
& \textbf{T2}: Tools use different commit iteration strategies (remote API, \texttt{git log} or repeated updates via \texttt{git blame}) and filters (e.g. merges), resulting in different activity trends
& Check configuration options for commit filters; search documentation and source code for implicit defaults, e.g. by keywords \texttt{blame, git log, github, update}
& Run tool on a repository with known cherry-picked and merge commits; verify their presence in the results
& Configure commit filters, such as for merge commits, if possible; implement custom post-processing filters\\ \hline

Commit analysis (1) 
& \textbf{T3}: Tools filter different files (e.g. by file endings, programming language, and path substrings), which can provide an incomplete structural overview 
& Check file filter configuration options; search source code for implicit defaults, e.g. by keywords \texttt{filter, file ending} for the stage, \texttt{.exe, .xml, test, docs} for common exclusions and \texttt{.c, .java, .py} for common inclusions 
& Run tool on repositories with diverse programming languages (e.g. Scala, Ruby, Rust), structures (e.g. test folders, READMEs and configuration files); verify that respective files are included and excluded as desired
& Set filter options explicitly, if possible; if filters are missing, apply them by post-processing; if desired files are missing, adjust code to export intermediate data and implement post-processing \\ \hline

Commit analysis (1)
& \textbf{T4}: Tools aggregate time windows differently (e.g. fixed before extraction vs. configurable at analysis time; overlapping vs. non-overlapping), affecting data schemes and flexibility
& Check time window configuration options; search documentation or source code for implicit defaults, e.g. by keywords \texttt{time window, time interval, split, overlap, shift}
& Run tool on a small sample repository and check data scheme (interval tables or columns)
& Configure time intervals explicitly, if possible; if slicing is implicit, adjust code to export intermediate data for custom post-processing \\ \hline

Identity matching (2) 
& \textbf{T5}: Tools apply different identity matching algorithms (e.g. string matching, fuzzy matching, machine learning, white-listing), causing false positives and negatives in developer classifications 
& Determine identity matching settings by configuration options; search documentation or source code for implicit defaults, e.g. by keywords \texttt{levenshtein, person, developer} 
& Run tool on repository with edge cases, e.g. Jane Doe and Jana Doe (likely false positive in fuzzy matching); verify merged identities manually
& If not already provided by the tool, create a unified identity table; implement custom post-processing to merge or split failure cases; add a manual white list for domain-knowledge-based merges if necessary \\ \hline

Identity matching (2)
& \textbf{T6}: Tools merge identities across single or multiple projects, different tables (e.g. commits, mails) and columns (e.g. author, committer), resulting in inconsistencies
& Search documentation or source code for implicit defaults, e.g. by keywords from \textbf{T5} and usage of identity-related column names, e.g. \texttt{author, committer, email, username}
& Run tool on multiple related repositories and on a repository with different authors and committers; check schema and extract unique identities from all relevant tables and columns; verify them manually
& Join all identity-related columns from tables and projects; apply matching algorithm; store result as a dedicated identity table; replace identities in original data based on this table\\ \hline

Entity analysis (3) 
& \textbf{T7}: Tools use different parsers and options (e.g. granularity levels such as modules, classes, functions, namespaces), resulting in different structural and activity trends  
& Check configuration options for entity filters; search documentation and source code by keywords for tools such as \texttt{ctags, doxygen}, custom \texttt{parsers} and \texttt{filters} for languages or structures
& Run tool on repositories using procedural vs. object-oriented languages; inspect data scheme to verify captured structures manually
& Configure entity filter; if entities are missing, integrate additional third-party tools \\ \hline

Entity analysis (3)
& \textbf{T8}: Tools calculate metrics with the same name (e.g. \emph{complexity}) differently, for instance at different levels of granularity (repository, file, function, line block), making results incomparable
& Determine the metric's definition from documentation; search source code for the metric's name; add debugging statements and run tool on a small known repository to reconstruct the algorithm
& Compare metric values across tools or against a known formula computed manually from the extracted data
& If results differ from expectation, recompute metrics by an own algorithm or by integrating third-party MSR tools \\ \hline

Network construction (4) 
& \textbf{T9}: Tools aggregate, direct and weigh edges differently (e.g. bipartite vs. temporal, files vs. functions, single vs. cumulative contributions), affecting graphs and metrics
& Check configuration options for networks; search documentation and code by keywords, e.g. \texttt{edge, connect, link}; add logging and run tool on a mock repository to reconstruct algorithm
& Iteratively add commits and developers to a mock repository; run tool on each commit; manually track changes in the outcomes to reverse-engineer data aggregation and algorithm
& Configure entities and weight function for aggregation explicitly; otherwise export intermediate data and construct the network with a custom algorithm \\ \hline

Cross-stage
& \textbf{T10}: Tools use different data schema semantics, reflecting their original purpose, which can mislead users (e.g. expecting a column \texttt{file} in a \texttt{change} table to contain \emph{all} edited files, not only those with edited functions)
& Inspect data schema; search in documentation or source code under which circumstances data is written, e.g. by database-related keywords such as \texttt{insert, update, write}
& Run tool on a known repository; compare basic counts (e.g. files, commits) against another tool or plain \texttt{git} commands; large deviations not attributable to threats of stages 1--4 may indicate a schema misunderstanding
& If semantics differ, derive data from earlier pipeline stage and apply custom processing, possibly using a special-purpose MSR tool\\ \hline

Cross-stage 
& \textbf{T11}: Tools apply processing steps (e.g. filtering and storing data) in different order, causing non-obvious differences in the data
& Determine overall processing order from documentation or source code, e.g. by pipeline keywords such as \texttt{merge, save, store} 
& Run tool on a known repository; inspect data to verify that elements of interest are captured as expected
& Store intermediate data where possible during processing to aid debugging and adjustments; add informative log messages at multiple verbosity levels \\ \hline

Tool selection
& \textbf{T12}: Tools are configurable to different degrees; missing or implicitly hard-coded parameters limit adaptability, while many options increase the risk of misconfiguration
& Check configuration options for T1--T11; check documentation where unclear
& Run tool on a known repository; compare results under different configurations to assess each parameter's effect
& Prefer established, actively maintained OSS tools with documented defaults; prefer simpler tools if code changes may be needed; combine with special-purpose tools, if required\\ \hline

Empirical study
& \textbf{T13}: Studies choose different metrics; more expressive (e.g. network) metrics offer richer models but show lower cross-tool agreement, increasing sensitivity to tool choice
& Determine how metrics are calculated in tools (\textbf{T8}); compare different metrics for the same data
& Check whether conclusions hold for different metrics
& Compare additional metric variants; also report simpler metrics, which have fewer algorithmic choices and generalise better across tools and contexts (e.g. languages) \\ \hline

Empirical study
& \textbf{T14}: Studies generalise tool-driven results beyond their valid context
& Identify the technical assumptions and decisions underlying the results (check \textbf{T1--T11})
& Check the relevance of each assumption to the specific study context
& Define research questions for a specific context; restrict claims to contexts in which underlying assumptions hold \\ \hline

Empirical study
& \textbf{T15}: Studies report statistical summaries not capturing tool- or subject-induced variation, leading to overgeneralisation
& --
& Check that finer-grained metrics (e.g. project vs. entire data set) agree with summary statistics and thresholds, e.g. by graphs and manual inspection
& Report visualisations beyond statistical metrics for transparency; follow the ATOM principle (accept uncertainty; be thoughtful, open, and modest) \\ \hline

Empirical study
& \textbf{T16}: Studies provide insufficient technical details through analysis scripts or execution environments (e.g. missing code for metrics), hindering reproduction and reuse
& Ensure that the entire extraction and processing pipeline runs end-to-end within a defined environment; provide a minimal setup with example projects in large-scale studies
& Compare the pipeline results to the ones reported in the paper
& Provide a reproduction package as a Git repository \emph{and} long-term archive with execution environment (e.g. Docker image export), all data (raw, intermediate and results) and code (actual artefacts, \emph{not} volatile links or submodules)\\

\bottomrule
\end{longtblr}

On a methodological level, many studies explore relationships by measuring correlations and building regression models, despite being actually interested in \emph{causal} effects~\citep{graf-vlachy_cleaning_2024}, for instance when interested in improving software quality. As Dueñas~\etal~\citeyearpar{duenas_grimoirelab_2021} note, practitioners require explanations \emph{why} metrics are important and \emph{how} they change. Causal discovery is therefore a promising direction for MSR, although still limited by its own threats such as training instabilities~\citep{hulse_shaky_2025}.

Finally, tool choice is rarely reported as a threat in empirical studies, yet our results suggest that it \emph{can} become a relevant one. We therefore emphasise our recommendations on reproduction packages and transparent reporting following the ATOM principle (Table~\ref{tab:lessons-learned}). More broadly, we hope that replication studies continue to gain recognition as a means of assessing and strengthening the generalisability of empirical findings.

\section{Limitations and threats to validity}
\label{threats}

\paragraph{Construct validity:} Comparing tool outcomes and evaluating stability of conclusions can be done in many ways, based on a wide range of data sources, metrics and analyses. Therefore, there is a risk that we have chosen suboptimal tools and studies, representing outliers in the landscape. We reduce this threat by conducting a literature review to select representative studies and tools based on defined criteria. 

\paragraph{Internal validity:} When replicating studies, we focus on capturing tool-specific details to gain novel insights, rather than enforcing conformance to the original study. However, where a metric is \emph{not} readily comparable across tools, we adopt the original study's algorithm or, if the corresponding code is missing, replace it with a reconstructed component. These decisions may influence our results, but since we apply them consistently across all tools, they affect conformance to the original study rather than the cross-tool comparison (as explained in Section~\ref{sec:replication-stability}).

To compare the baseline data, we automate the calculation of statistics, but additional manual investigations of data schemas and the original Git log was needed to explore the actual causes of differences. Due to the large number of discrepancies and their interplay along the pipeline, we could not inspect all of them for each tool, so further hidden causes may exist. As the characteristics of differences resemble those in our previous work, and this extended paper focuses on conclusion stability, we accept this threat.

\paragraph{External validity:} A potential threat concerns the number of tools compared. Rather than sampling, we applied a systematic selection process with defined criteria alongside our literature review. The resulting set of four tools is complete with respect to these criteria, although it inherently excludes tools that are not actively maintained or do not support evolutionary socio-technical studies by means of commit, code entity and network analysis. Therefore, our findings only generalise to tools of a similar profile.

In addition, generalisation is limited by replicating only three studies. As motivated in Section~\ref{sec:study-selection}, three is the reasonable minimum to observe a pattern rather than an isolated result, and the studies were selected from distinct, yet highly prevalent fields following a systematic selection process to ensure representativeness. Evaluating more studies could reveal additional threats and lessons, but is beyond the scope of this study due to temporal, space and compute constraints.

As the literature review also emphasised, the field of MSR is broad. Having collected and annotated over 800 studies and more than 100 tools in our supplementary material, we encourage other researchers to extend and complement our lessons learned through similar tool comparisons in other fields.

\paragraph{Conclusion validity:} As explained in Section~\ref{sec:replication-stability}, drawing conclusions about tool and replication \emph{similarity} is inherently difficult. We deliberately avoid null-hypothesis significance testing and only define \emph{illustrative hypotheses} explaining our interpretative logic, while explicitly disclaiming statistical significance. Following the ATOM recommendation by Wasserstein~\etal~\citeyearpar{wasserstein_moving_2019}, we provide qualitative and visual evidence, ensuring that our conclusions and lessons learned stay transparent and open to the reader's own assessment.

\section{Conclusion}
\label{conclusion}

The lightweight literature review conducted in our study shows that empirical software engineering heavily relies on tools, and these tools often share a similar purpose. In this paper, we studied the agreement between four independent tools that help to analyse socio-technical software evolution. By replicating three carefully selected studies and quantitatively and qualitatively comparing result and conclusion stability across these tools, we find seemingly minor differences due to technical details and assumptions affecting
\begin{itemize}
    \item the extracted baseline data and their transformations;
    \item each extraction and processing step;
    \item subsequently calculated metrics and results of statistical analyses;
    \item the practical applicability of tools;
    \item and under specific conditions, the overall study conclusion. 
\end{itemize}

From these findings, we derive an actionable checklist of the technical details and assumptions that tool and study authors should verify, together with concrete handling strategies for each. Our results demonstrate that tools in this domain are not readily interchangeable, and that switching the tool can cause relevant distortions. We believe that the scope of studies and their related validity claims should be evaluated more explicitly (for which our checklist offers a concrete starting
point) and reported in the manner of a case study: tailored to the specific assumptions and context in which they hold, so that readers can judge whether a study's findings transfer to their own setting. Future work can conduct similar comparisons for tools in other fields, helping researchers and practitioners apply and implement tools with less uncertainty.

\section{Declarations}
\label{reproduction-package}

\subsection{Funding}

Nicole Hoess and Wolfgang Mauerer acknowledge support by the High-Tech Agenda of the Free State of Bavaria, Germany.

\subsection{Ethical approval} 

The authors confirm that this work complies with ethical standards.

\subsection{Informed consent}

All analyses were conducted with data already made public and reviewed by the community and do not involve additional interaction with human subjects.

\subsection{Author contributions}

\paragraph{Nicole Hoess:} Conceptualisation, Methodology, Tool software adjustment, Literature review, Data collection, Data analysis, Evaluation and Investigation, Discussion, Validation, Writing -- Original Draft

\paragraph{Carlos Paradis:} Conceptualisation, Methodology, Tool software adjustment, Discussion, Writing -- Review and Editing

\paragraph{Rick Kazman:} Conceptualisation, Methodology, Discussion, Writing -- Review and Editing

\paragraph{Wolfgang Mauerer:} Conceptualisation, Methodology, Discussion, Writing -- Review and Editing, Supervision

\subsection{Data availability}

The supplementary website\footnotemark[17] provides visualisations for all results not presented in the paper. 
For understandability and to facilitate future studies relying on the examined tools,
we also provide snapshots of the study tools and all analysis scripts in our GitHub
repository\footnotemark[18].
A full reproduction package with the exported Docker images for each tool and the entire analysis pipeline including code, data and visualisations is available at Zenodo\footnotemark[19].

\footnotetext[17]{Supplementary website: \href{https://lfd.github.io/emse2025.github.io/}{https://lfd.github.io/emse2025.github.io/}}
\footnotetext[18]{GitHub repository: \href{https://github.com/lfd/emse2025}{https://github.com/lfd/emse2025}}
\footnotetext[19]{Zenodo archive: \href{https://doi.org/10.5281/zenodo.17222931}{https://doi.org/10.5281/zenodo.17222931})
}

\subsection{Conflict of interest}
The authors declare that they have no conflicting interests.

\subsection{Clinical trial number}
Not applicable
\bigskip

\begin{acknowledgements}
We thank Christoph Gote (git2net), Georg Link and Daniel Izquierdo Cortázar (GrimoireLab) for their valuable input and feedback on setting up and improving the replication pipelines with their respective tools. We also thank the anonymous reviewers for their careful reading and constructive feedback, which helped us considerably strengthen the clarity and rigour of this paper.
\end{acknowledgements}

\newpage
\appendix
\section{Appendix} 

\subsection{Original studies and replication details} \label{sec:appendix}

\subsubsection{First replication: software maintenance}\label{sec:appendix-cs2}

\textbf{Context:} The study \emph{Big Data = Big Insights? Operationalising Brooks’ Law in a Massive GitHub Data Set} by Gote~\etal~\citeyearpar{gote_big_2022} examines the relationship between productivity and team size. In psychology, the Ringelmann effect describes the phenomenon of productivity linearly decreasing with team size. In software engineering, Brooks' law states that adding manpower to a late software project makes it later.\\

\noindent\textbf{Objective:} Productivity has been widely studied in software engineering research, but studies carried out on large data sources with similar empirical methods report disagreeing results. The authors examine threats in large-scale repository mining and study the validity of Brook's law based on multiple productivity and collaborative metrics to dissolve these conflicts.\\

\noindent\textbf{Method:} The original study employs a selection pipeline to systematically filter suitable subject projects based on criteria such as the number of developers, project size, activity and purpose from the GHTorrent dataset. The repositories are then mined with \textsc{git2net}. Based on this dataset, the authors calculate commit-based (number of commits, number of events measured by line changes, and number of modified characters) and code-based productivity metrics (changed LOC, changed number of code tokens, changed number of functions, change in McCabe cyclomatic complexity, Halstead effort to make changes) per time window of 42 weeks and respective team size. 
To explore the effect of collaborative aspects, the authors calculate metrics based on temporal line-granularity networks, including the number of nodes, number of edges, network density, network diameter, global clustering coefficient, mean indegree, mean foreign modification ratio and eigengap. 
The authors log- or square-root-transform metrics due to skewness.
To select a meaningful set of metrics to evaluate Brooks' law, the original study calculates Pearson's correlation coefficient between the productivity and network metrics among each other and cross-correlation. 
The authors then build multiple linear and quadratic regression models for different productivity target variables using team size and other network metrics to test the relationship between team size and productivity assumed by Brooks' law. This relationship is indicated by the regression coefficients of the model. Gote~\etal additionally analyse the existence of a global maximum of the quadratic functions to evaluate the existence of an optimal team size.\\

\noindent\textbf{Results:} The authors find a strong correlation between all productivity metrics, except for the number of commits and Halstead effort, which exhibit lower correlation coefficients. 
For the network metrics, Gote~\etal identify three clusters in the correlation matrix: The first cluster indicates strong positive relationships between team size, number of nodes and network diameter. The second cluster reports a positive correlation between clustering coefficient, mean indegree and eigengap, and the third cluster only consists of the mean foreign modification ratio. The authors pick one measure from each cluster to build their regression models.
Both the linear and quadratic models regressing different productivity metrics on team size consistently yield a negative relation, indicating that, on average, individual productivity decreases with higher team size. Furthermore, some of the models exploring the existence of an optimal team size suggest that individual productivity increases for small and decreases for larger teams, with an optimum of 7 or 19 members. The models with network metrics as control variables suggest a positive relationship between mean indegree and productivity, while the mean foreign modification ratio has a negative relationship with productivity. 
Models further exploring the effect of network structure indicate a positive relation between team size and the mean indegree of developers.\\

\noindent\textbf{Conclusions:} Since both, correlations and models expressing the relationship between team size and a diverse set of metrics capturing independent dimensions of productivity consistently indicate a negative trend, the original study concludes that the Ringelmann effect and Brooks' law apply to collaborative software development. The authors attribute contradictory results to threats in the method and interpretation of results in other studies.\\

\noindent\textbf{Scope of replication:} The replication of~\cite{gote_big_2022} focuses on the relationship of team size and productivity, while accounting for the effect of structural properties. We use a subset of expressive productivity and collaboration metrics to calculate correlations and build linear and quadratic regression models. 
Specifically, we include the number of commits and Halstead effort exhibiting unique characteristics. From the highly correlated productivity measures, we choose the change in the number of functions, as it has the lowest correlation with other productivity metrics. Among the network measures, we choose the ones used by the authors to build regression models.\\

\noindent\textbf{Limitations:} The authors analyse 201 projects with \textsc{git2net}, which took them over one million CPU-hours.
As resources are limited in our replication with four tools and three studies, we select a subset of ten representative subject projects. We reduce population threats by selecting projects with all different team size clusters investigated by the original study to approximate the distribution. 
As we do not know which version of the GHTorrent dataset 
was used for sampling, we select projects based on the team size in the latest time window in the reproduction data. This may lead to the selection of larger projects, as authors are already disambiguated in this dataset. To ensure comparability with the original results, we reduce the reproduction data to our set of projects and recalculate the originally observed metrics accordingly. 
Gote~\etal calculate all metrics based on time windows of 42 weeks. Since \textsc{Codeface} does not support weekly granularity, we would not be able to compare its results to those obtained by the other tools in a close replication. Therefore, we configure a time window of nine months for all tools and use \textsc{Codeface}'s time window boundaries as reference.
Unfortunately, the code to calculate metrics is missing in the reproduction package. We assume that code-level productivity metrics were calculated in the same way as by the official \textsc{git2net} repositories. However, even with help of the original author, we could not fully clarify the calculation of the network metrics. 
This leaves room for potential incorrect implementations, which we cannot fully overcome. Although this limits comparability with the original study, comparability across tools in the replication is not affected.\\

\noindent\textbf{Considerations:} The \textsc{git2net} tutorials repository~\citep{gote_git2net-tutorials_2022} indicates that the team size may be calculated after performing a left join between the complexity metrics and commits, meaning that all commits without complexity statistics are discarded. While this may appear as a technical detail, it influences the commit count as a unique productivity metric and the team size as the central collaboration metric. In our replication, we prioritise the impact of tool-specific variations over replication conformity. Therefore, we consider the measures extracted directly from each tool's data base whenever possible.
Productivity metrics calculated by \textsc{git2net}'s \emph{complexity} analysis depend on the tools \textsc{PyDriller} and \textsc{lizard}. Although these steps are part of \textsc{git2net}, we apply the same pipeline to data extracted by \textsc{Codeface}, \textsc{GrimoireLab} and \textsc{Kaiaulu}, as these tools do not implement comparable metrics calculations.

We contacted the lead author of the study and developer of the \textsc{git2net} tool regarding differing correlations for the network metrics. Although the original code was not available any more, he provided us with possible reasons explaining the discrepancies in the replication. For instance, in a previous version of the correlation matrix, which we included in the supplementary material, we exported the \textsc{git2net} developer networks in adjacency matrix format, which combined multiple edges into a single edge weight. This corrupted the multi-edge count used in the original study and led to positive correlations between the foreign modification ratio and the productivity metrics. When switching to the multi-edge edgelist, this issue was overcome, but at the same time changed the correlation between team size and mean in-degree to a negative value. Due to these side effects, a large number of possible combinations would have had to be explored to achieve an accurate replication in all metrics. However, this circumstance does not affect the agreement between the tools, which is of interest for this study.

\subsubsection{Second replication: software quality} \label{sec:appendix-cs3}

\noindent\textbf{Context:} The study \emph{Impact of Developer Turnover on Quality in Open-Source Software} by Foucault~\etal~\citeyearpar{foucault_impact_2015} investigates the effect of team dynamics on software quality. Developers may join and leave entire software projects or individual software module. Most theories assume that these \emph{turnover} phenomenons negatively impact software quality due to a loss of experience and knowledge. Others hypothesise positive or negative effects on team motivation and social interactions.\\

\noindent\textbf{Objective:} In an industrial setting, prior research found that developers who leave a software project have a negative impact on software quality, while newcomers who join a project have no impact on quality. The study aims to investigate whether the relationship between turnover and quality is similar in open-source software projects with extensive use and low entry barriers.\\

\noindent\textbf{Method:} The original study extracts data from five popular subject repositories using the tool \textsc{Diggit}. First, the authors evaluate whether turnover is an important phenomenon in open-source projects.
They measure turnover at module-granularity, with each module representing a finite set of files. 
To modularise the source code, files are clustered manually based on the directory structure. 
The study distinguishes between \emph{external turnover} affecting all modules of a project and \emph{internal turnover} affecting a specific module. 
At project level, turnover metrics are evaluated considering two subsequent time intervals of two weeks. Developers contributing in the second but not the first period are considered \emph{newcomers}. Developers contributing in the first, but not in the second period are considered \emph{leavers}. \emph{Stayers} contribute in both periods.
To detect patterns, the original study splits the project's history into development periods of six months. By measuring the groups' activity as the sum of code churn per developer and module, the authors consider the varying levels of involvement of individual developers. The study calculates bug density as the number of bug-fixing commits per module, normalised by its size in lines of code. Bug-fixing commits are identified manually from maintenance branches and include arithmetic and logic errors, security issues, requirements misunderstanding and design flaws. Finally, the authors conduct Spearman's rank correlation tests and estimate confidence intervals using bootstrapping to evaluate the relationship between each turnover metric and software quality.\\

\noindent\textbf{Results:} The study finds that at least \(80\%\) of developers are either newcomers or leavers.
Qualitatively analysing characteristics that cause developers to become stayers reveals that payment and consulting may be influencing factors.
Regarding contribution patterns, the authors find that no module is exclusively changed by only external newcomers. Besides that, the authors identify project-specific patterns. 
The confidence intervals for the correlation between turnover and quality indicate a positive correlation between external newcomer activity and bug density for the majority of subject projects. Foucault~\etal find no significant correlation between external leavers and quality. The confidence intervals for stayers majorly indicate a positive correlation with bug fixes. For the internal turnover metrics, results are unclear.\\

\noindent\textbf{Conclusions:} Finding high turnover in five highly successful projects leads the authors to conclude that the role of turnover in open-source projects differs from the one in industrial settings, disagreeing with the suggestion to control turnover.
Although contribution patterns are project-specific, the original study finds that external newcomers always work under supervision. 
Based on the confidence intervals, the study concludes that external turnover negatively impacts module quality, while internal turnover is not problematic.\\

\noindent\textbf{Scope of replication:} To address the domain of software quality, our replication of Foucault~\etal~\citeyearpar{foucault_impact_2015} focuses on the impact of turnover on quality. However, to answer this question in an informed manner, we calculate turnover metrics and evaluate patterns per project as done in the original study, which also indicate the general relevance of turnover in open-source projects. 
The original study evaluates different approaches for time interval splitting, project modularisation, bug fix identification and classification, but does not present results for all combinations in the paper and reproduction package. Therefore, we exclude these additional approaches from our replication.\\

\noindent\textbf{Limitations:} The comprehensive reproduction package allows for reuse of the same subject repositories and code logic.
Although the code to calculate confidence intervals was missing, we could recover it from an updated version in a git repository of the author~\citep{foucault_rdeveloperturnover_2016}.
An important difference in tools concerns the parsing of the git log. The original study tool \textsc{Diggit} traverses across all branches to identify commits, while \textsc{Codeface}, \textsc{git2net} and \textsc{Kaiaulu} only consider the linear history of the currently checked-out branch. While this technical detail is less relevant for the previous studies, it plays a major role when analysing the LTS branches containing bug fixing commits. 
This required us to perform separate analysis runs of multiple branches to simulate the behaviour of \textsc{Diggit} with the previously mentioned tools. Limitations remain, as this approach would actually require manual identity merging between the analysis runs. Another difference is the file filtering. \textsc{Diggit} supports complex filters of regular expressions to choose files. Most similarly, \textsc{Kaiaulu} provides configuration options for users, but since they are based on exact substring matching, we cannot accurately replicate all filters. The same is true for \textsc{Codeface} and \textsc{git2net} with internally defined filters and \textsc{GrimoireLab}, which would require custom post-processing.
All analyses performed in the original study rely directly on the extracted commit history, without more complex analyses such as network construction. This allows for an otherwise close replication.\\

\noindent\textbf{Considerations:} The authors modularise source code manually. As we are interested in comparing tool pipelines, we do not replicate the manual part but adopt the authors'
decomposition in all replications.
Similarly, we adopt the classification of bug-fixing commits. Although \textsc{Codeface} provides a keyword-based classification, its adoption would bias the comparison of quality measured by \textsc{Codeface} and by other tools without such capabilities. Therefore, we fixed the pipeline after data extraction to the original one for comparability.

\subsubsection{Third replication: collaboration and coordination}\label{sec:appendix-cs1}

\textbf{Context:} The study ``Classifying Developers into Core and Peripheral: An Empirical Study on Count and Network Metrics'' by Joblin~\etal~\citeyearpar{joblin_classifying_2017} addresses the organisational roles of developers in software projects.  
Classifying developers into core and peripheral is a common practice to understand a project's collaborative dynamics.\\

\noindent\textbf{Objective:} In addition to prevailing operationalisations based on simple counts of developer activities, the authors propose and evaluate novel operationalisations based on the organisational structure and collaboration behaviour derived from developer networks
to enrich information and overcome potential limitations. The study analyses the validity and agreement of these metrics.\\

\noindent\textbf{Method:} The original study calculates count- (commit, lines of code  and mail count) and network-based metrics (eigenvector centrality, hierarchy centrality, node degree) on version-control system and mailing-list data extracted from ten large open-source software projects using the tool \textsc{Codeface}. 
Developer operationalisations are calculated using overlapping time windows of three months. Cohen's kappa is used to measure the pairwise agreement of these operationalisations to evaluate whether the different operationalisations are statistically consistent across data sources and across count- and network-based metrics.
Assuming that core and peripheral developers exhibit different communication and coordination structures, the authors additionally analyse the position and temporal stability of developer roles. Therefore, the authors evaluate node degree and clustering coefficient over time and build Markov Chain models expressing the transition of developers across the roles \emph{core}, \emph{peripheral}, \emph{absent} and \emph{idle}. The authors additionally construct the core-periphery block model describing edge probabilities between core and peripheral nodes in the developer network to evaluate coordination preferences.
Finally, Joblin~\etal conduct a survey across 166 developers to evaluate to what extent the operationalisations agree with actual developer perceptions.\\ 

\noindent\textbf{Results:} Joblin~\etal find that the level of agreement between count-based metrics calculated on version-control system and mailing-list data is consistently substantial for the same data source and fair across data sources. For network-based metrics, the authors find substantial agreement on the same data source. The agreement across all metrics and data sources always exceeds random agreement. The time-resolved evaluation of structural and hierarchical properties demonstrates that core developers consistently exhibit a low clustering coefficient and high node degree in the network, while peripheral developers have the opposite. The Markov chain shows that core developers are less likely to transit to the absent or isolated state. The decreasing probability of core-core, core-peripheral and peripheral-peripheral edges in the core-periphery block model further confirms different coordination preferences. 
The developer survey reveals that among the metrics types, network-based metrics have a higher agreement with developer perception,
especially when constructed from mailing lists.\\

\noindent\textbf{Conclusions:} Based on the quantitative and qualitative study results, the authors conclude that all of the evaluated count- and network-based core-peripheral operationalisations are overall consistent
and agree with actual developer perception.
They find that network-based metrics can accurately discriminate core and peripheral developers due to their different structural and hierarchical positions and their communication and coordination behaviour.\\ 

\noindent\textbf{Scope of replication:} The replication of Joblin~\etal~\citeyearpar{joblin_classifying_2017} 
focuses on the agreement of core-peripheral developer operationalisations 
and the manifestation of different positions and role stability in the network structure. In particular, we focus on replicating the VCS metrics, their level of agreement and the structural and hierarchical properties as analysed by the original study.\\
In their results, Joblin~\etal present the agreement between metrics averaged over one year of development, but do not specify which year. As their other analyses refer to the most recent development year under study, we present qualitative results for the same interval. Results for all other time windows, and the agreement averaged over the entire studied interval, are available on our supplementary website.

\noindent\textbf{Limitations:} Our replication is further limited by several technical constraints. First, the original study constructs developer networks in overlapping time windows. This feature is not supported by \textsc{git2net}, \textsc{GrimoireLab} and \textsc{Kaiaulu} and would cause additional computational effort, which is beyond the scope of this study. Since the authors compare results pointwise and report temporal stability, we consider this threat as acceptable. In addition, as we extract data and compare results obtained by different tools for exactly the same time intervals, results should be consistent across tools regardless of whether overlapping time intervals were used or not.
Another limitation is caused by an anonymised subject project named \emph{Project X}, whose true identity could not be determined. Due to computing time constraints, we were also unable to complete the calculations for the Linux project with one of the tools. However, as shown in the original supplementary material, the findings for these projects are consistent with those of all other subject projects, indicating only minor information loss.\\

\noindent\textbf{Considerations:} The original study relies on \textsc{Codeface} and refers to its official repository, but as dependencies used at the time of study are outdated and several important bug fixes have been made, we conduct the replication using our own actively maintained version of \textsc{Codeface}. 
\textsc{Kaiaulu} was adjusted with new parameter options in previous work to support a very close replication of data obtained by \textsc{Codeface}, which we adopt in this study. 
\textsc{git2net} also allows for constructing temporal developer networks as used in the original study. However, the tool is only able to connect developers at line- instead on function-granularity for this type of network.
Similarly, file-level is the finest entity granularity supported by \textsc{GrimoireLab}. 
Although \textsc{GrimoireLab} allows for network construction through its Kibana Dashboard, the networks represent undirected graphs connecting arbitrary columns from its database. We evaluated different types of nodes and relations and found the connection of developers based on commonly edited files as the most similar representation to \textsc{Codeface}.  
To get the actual adjacency matrix for the visualisation,
we implemented a corresponding query and network construction with the help of \textsc{GrimoireLab}'s ElasticSearch API.\\

\begin{figure}[htbp]
  \centering
  \begin{subfigure}[b]{0.49\textwidth}
    \includegraphics[width=\linewidth]{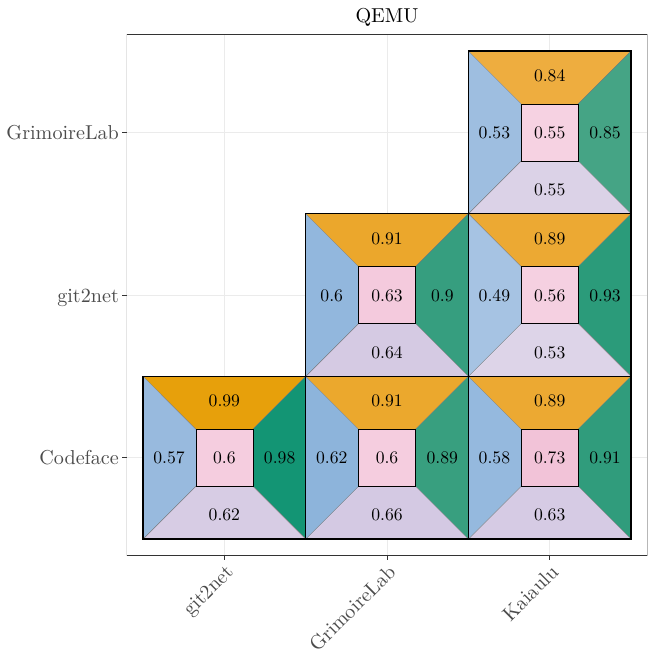}
  \end{subfigure}
  \begin{subfigure}[b]{0.49\textwidth}
    \includegraphics[width=\linewidth]{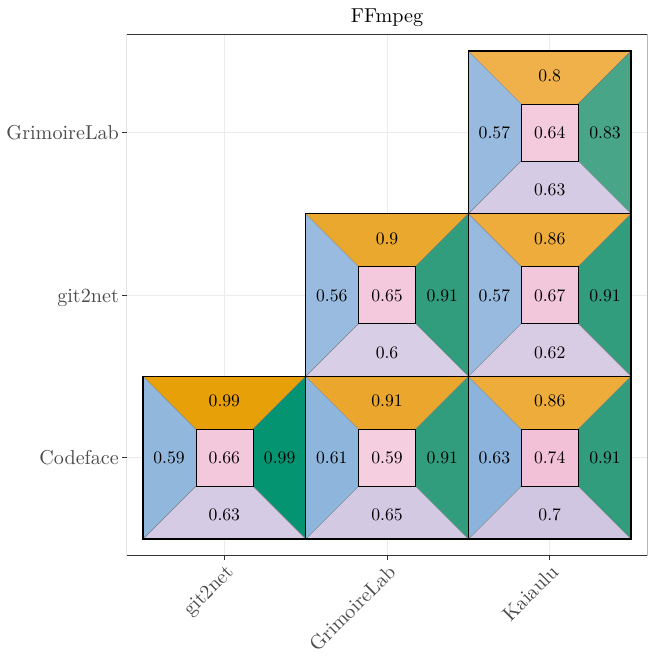}
  \end{subfigure}
  \vspace{1em}
  \begin{subfigure}[b]{0.49\textwidth}
    \includegraphics[width=\linewidth]{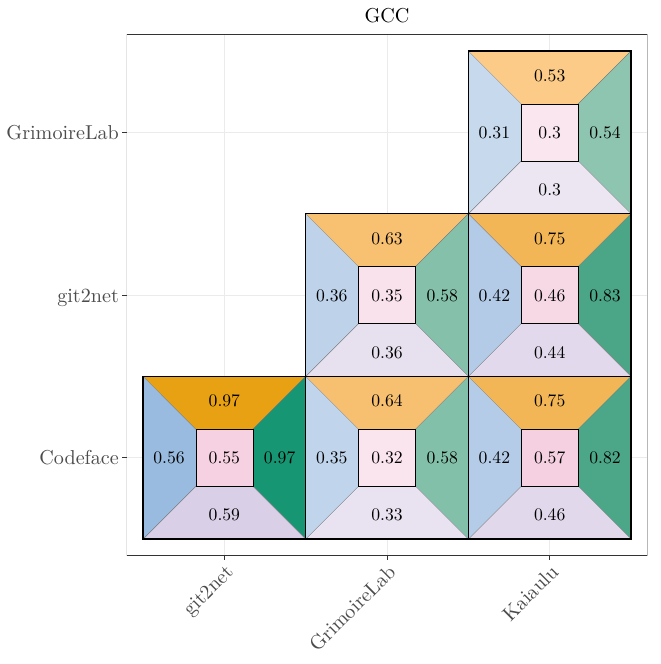}
  \end{subfigure}
  \begin{subfigure}[b]{0.49\textwidth}
    \includegraphics[width=\linewidth]{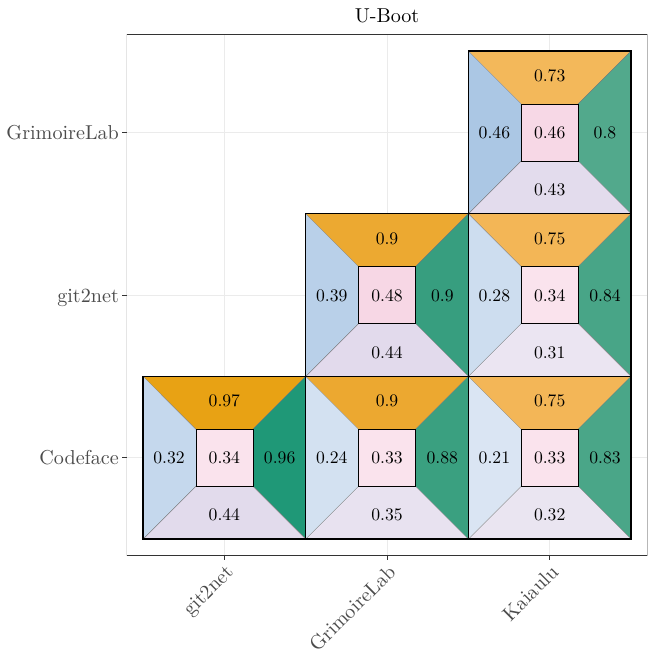}
  \end{subfigure}

  \caption{Time-averaged agreement of classifications of the same set of developers identified and classified by all four tools \textsc{Codeface}, \textsc{git2net}, \textsc{GrimoireLab} and \textsc{Kaiaulu}. The pairwise agreement on classifications between tools is shown for lines of code (LOC) count (yellow), commit count (green), node degree (lilac), eigenvector centrality (blue) and hierarchy centrality (pink).}
  \label{fig:tool-agreement}
\end{figure}

\noindent\textbf{Further investigations:} As an illustrative example of differences in developer classifications, Figure~\ref{fig:tool-agreement} shows the time-averaged pairwise agreement of cross-tool classifications per metric for QEMU, FFmpeg, GCC, and U-Boot, computed directly on the classifications as point estimates, \emph{without} bootstrap CIs. For QEMU and FFmpeg, the count-based metrics (LOC count, number of commits) reach the expected almost perfect agreement, whereas the agreement of classifications based on network metrics (node degree, eigenvector centrality and hierarchy centrality) is notably lower. For GCC and U-Boot, the agreement between tools is lower throughout, and for the network metrics often does not exceed the fair level. Given that inconsistently identified developers were already removed prior to calculating Cohen's $\kappa$, this points to substantial differences in the developer classifications tools produce, consistent with the formal stability assessment in Table~\ref{tab:joblin-result-stability}.

\newpage

\bibliographystyle{spbasic}      
\bibliography{references}   
\end{document}